\documentclass[pretext,two column,epic,coverpage]{svjour3} 
\pdfoutput=1
\usepackage{multirow}
\usepackage[british]{babel}
\usepackage{url}
\usepackage{cite}
\usepackage{atlasphysics}
\usepackage{subfigure}

\usepackage{color}

\newcommand{\trijet}{three-jet}
\newcommand{\Trijet}{Three-jet}
\newcommand{\antikt}{anti-\ensuremath{k_{t}}}

\newcommand{\rfour}{\ensuremath{R=0.4}}
\newcommand{\rfive}{\ensuremath{R=0.5}}
\newcommand{\rsix}{\ensuremath{R=0.6}}
\newcommand{\rseven}{\ensuremath{R=0.7}}

\newcommand{\topo} {topo-clust\-er}

\newcommand{\topos}{topo-clust\-ers}

\newcommand{\xs}{cross-sec\-tion}
\newcommand{\xss}{cross-sec\-tions}
\newcommand{\XS}{Cross-sec\-tion}
\newcommand{\XSS}{Cross-sec\-tion}

\def\ystar{\ensuremath{\left|Y^{*}\right|}}
\def\mjjj{\ensuremath{m_{jjj}}}

\def\avgmu{\ensuremath{\langle\mu\rangle}}
\def\npv{\ensuremath{N_{{\rm PV}}}}
\def\LUMI{\ensuremath{4.51\pm 0.08~\ifb}}
\def\LUMIE{\ensuremath{4.51~\ifb}}

\newcommand{\psAbstract}{
Double-differential  \trijet{} production \xss{} are measured in proton--proton colli\-sions at a centre-of-mass energy of $\sqrt{s} = 7\TeV{}$  using the ATLAS detector at the Large Hadron Collider. The measurements are presented
as a function of the \trijet{} mass $(\mjjj)$, in bins  of the sum of the absolute rapidity separations
between the three leading jets $(\ystar)$. Invariant masses extending up to 5~\TeV{} are reached for $8< \ystar < 10$.  These measurements use a  sample of data recorded using the ATLAS detector in 2011, which corresponds to an integrated luminosity of \LUMIE{}. Jets are identified using the \antikt{} algorithm with two different jet radius parameters, \rfour{} and \rsix. The dominant uncertainty in these measurements comes from the jet energy scale. Next-to-leading-order  QCD  calculations corrected to account for non-perturbative effects are compared to the measurements. Good agreement  is found  between the data and the theoretical predictions based on most of the available sets  of parton distribution functions, over the full kinematic range, covering almost seven orders of magnitude in the measured \xs{} values.
}

\newcommand{\psTitle}{Measurement of \trijet{} production \xss{} in \pp{} collisions at  7~\TeV{} centre-of-mass energy using the  ATLAS detector}

\RequirePackage[switch,mathlines]{lineno}

\usepackage{preprintcover}
\PreprintCoverPaperTitle{\psTitle}
\PreprintIdNumber{CERN-PH-EP-2014-157}
\PreprintCoverAbstract{\psAbstract}
\PreprintJournalName{Eur. Phys. J. C}

\clearpage

\RequirePackage{fix-cm}
\smartqed  
\RequirePackage{graphicx}
\DeclareGraphicsExtensions{.pdf}
\journalname{Eur. Phys. J. C}

\usepackage[hyperindex,breaklinks]{hyperref}

\begin{document}


\title{\psTitle}
\titlerunning{Measurement of \trijet{} production \xss{} using the ATLAS detector}

\author{The ATLAS Collaboration}

\date{Received: date / Accepted: date}

\maketitle

\begin{abstract}
\psAbstract{}

\keywords{QCD \and jet \and LHC \and PDF}
\end{abstract}

 \section{Introduction \label{sec:intro}}

Collimated jets of hadrons are a characteristic feature of high-energy particle interactions. In the theory of strong interactions, quantum chromodynamics (QCD), jets can be interpreted as the result of fragmentation of partons   produced in a scattering process.
In high-energy particle collisions two main phases can be distinguished.
In the perturbative phase,  partons with high-transverse momentum (\pt) are produced in a hard-scattering process at a scale $Q$. This phase is described by a perturbative expansion in QCD.
 In the transition to the second (non-perturbative) phase, these partons
 emit additional gluons and  produce quark--anti\-quark pairs. 
The non-per\-tur\-ba\-tive jet evolution is an interplay between the had\-ro\-ni\-sation process and  the underlying event. The hadronisation process governs the transition from partons to  hadrons
and the underlying event represents initial-state radiation, multiple parton interactions and colour-reconnection effects \cite{Field:2012jv}.
In spite of these phenomena, 
the  highly collimated sprays of particles, collectively identified as hadron jets, are observed in the final state.
The effects of both hadronisation and the underlying event vary strongly with   the jet radius parameter and are most pronounced at low \pt. They are accounted for using phenomenological models that are tuned to the data.

The ATLAS Collaboration has measured  the inclusive jet  \xss{}  at $7$~\TeV{} \cite{Aad:2011fc} and at $2.76$~\TeV{} \cite{Aad:2013lpa} centre-of-mass energies  in \pp{} collisions for jets defined by the \antikt{} algorithm \cite{Cacciari:2008gp} with two jet radius parameters, \rfour{} and \rsix. Recent inclusive jet \cite{Aad:2014vwa} and dijet \cite{Aad:2013tea} \xs{} measurements at $7$~\TeV{}  centre-of-mass energy in \pp{} collisions have exploited  improved jet energy calibration procedures  \cite{Aad:2014bia} leading to smaller systematic uncertainties compared to those achieved in Refs.~\cite{Aad:2011fc,Aad:2013lpa}.
Similar measurements at $7$~\TeV{}  centre-of-mass energy in \pp{} collisions \cite{Chatrchyan:2012bja,Chatrchyan:2012gwa}  have been carried out by the  CMS Collaboration.
These measurements test  perturbative QCD (pQCD) at very short distances and have provided constraints on the gluon momentum distribution within protons at large momentum fraction.
The impact of higher order effects on the inclusive jet \xs{} ratios of \antikt{} \rfive{} and \rseven{} jets has been studied  in \cite{Chatrchyan:2014gia}.  The inclusive  \trijet{} to two-jet ratio \cite{Chatrchyan:2013txa} is used to determine the strong coupling constant.
Theoretical predictions of the multi-jet \xss{}  in \pp{} collisions at $7$~\TeV{} centre-of-mass energy  have been tested in Refs.~\cite{Aad:2011tqa,Chatrchyan:2013qza}.

Previous measurements of \trijet{}  \xss{} in \ppbar{}  collisions were performed  by the $\textrm{D}{\emptyset}$ collaboration \cite{Abazov:2011ub}.
The measurements were compared to predictions, and agreement between data and theory was found within the uncertainties.

In this paper, measurements of  double-differential \trijet{} production \xss{}  are presented as a function of the \trijet{} mass  (\mjjj) and the sum of absolute rapidity separation between the three leading jets (\ystar). The measurements are corrected for experimental effects and reported at the particle level.
The \trijet{} mass distributions test the dynamics of the underlying $2\rightarrow 3$ scattering process. The distributions are sensitive to both the transverse momentum (\pt) spectra of the three leading jets and their angular correlations, since a massive \trijet{} system can be built either from high-\pt{} jets or from jets with large rapidity separation.  Binning in \ystar{} allows  events with \mjjj{} originating from these different regions of phase space to be separated.

The analysis presented in this paper  tests the description of multi-jet events in next-to-leading-order \\ (NLO) QCD and uses two different values of jet radius parameter, \rfour{} and \rsix{}, since  \trijet{} \xss{}
depend on the jet radius even at leading order (LO) in the perturbative expansion. The NLO QCD calculations corrected to account for non-perturbative effects are compared to the measured \xss.
The measurements also provide  constraints on the proton's parton distribution functions~(PDFs) beyond  those from inclusive and dijet \xss{}, since they  probe a different region of phase space in proton momentum fraction  and squared momentum transfer  $(x,Q^2)$  and  different combinations of initial-state partons.

The content of this paper is structured as follows. The ATLAS detector is briefly described in Sect.~\ref{sec:atlas}, followed by the definition of observables and description of Monte Carlo (MC) samples  in Sects.~\ref{sec:definition}~and~\ref{sec:mc}, respectively. The trigger, data selection and jet calibration are presented in Sect.~\ref{sec:data}.
Data unfolding and experimental uncertainties are described in Sects.~\ref{sec:unfold} and \ref{sec:sysunc}.
Section~\ref{sec:theory} describes the theoretical predictions for the measurements in this paper.
The \xs{} results are presented in Sect.~\ref{sec:results} and the conclusions are given in Sect.~\ref{sec:conclusions}.

\section{The ATLAS experiment \label{sec:atlas}}

The ATLAS detector is described in detail in Ref.~\cite{Aad:2008zzm}.
ATLAS uses a right-handed coordinate system with its origin at the nominal interaction point (IP) in the centre of the detector and the $z$-axis pointing along the beam axis. The $x$-axis points from the IP to the centre of the LHC ring, and the $y$-axis points upward. Cylindrical coordinates ($r$, $\phi$) are used in the transverse plane, $\phi$ being the azimuthal angle around the beam pipe.
The pseudorapidity is defined in terms of the polar angle $\theta$ as $\eta=-\ln\tan(\theta/2)$.
The rapidity is defined in terms of the energy $E$ and longitudinal to the beam pipe momentum $p_z$ as $y=1/2\ln{\left((E+p_z)/(E-p_z)\right)}$.
The transverse momentum \pt{} is defined as the component of the momentum transverse to the beam pipe.

The inner detector (ID) is used to measure the momenta and trajectories of charged particles. The ID has full coverage in the azimuthal angle $\phi$ and over the pseudorapidity range $|\eta|<2.5$. The ID is immersed in a 2~T magnetic field provided by a superconducting solenoid magnet.

The main detector system used for this analysis is the calorimeter.
The electromagnetic calorimeters use liquid argon (LAr) as the active detector medium. They employ accordion-shaped
electrodes and lead absorbers, and are divided into one barrel ($|\eta|<1.475$) and two end-cap components ($1.375<|\eta|<3.2$). The technology used for the hadronic calorimeters depends on $\eta$.
In the barrel region ($|\eta|<1.7$), the detector is made of scintillator tiles with steel absorbers.  In the end-cap region ($1.5<|\eta|<3.2$), the detector uses LAr and copper. A forward calorimeter consisting of LAr and tungsten/copper absorbers has both electromagnetic and hadronic sections, and extends the coverage to $|\eta| = 4.9$.

The muon spectrometer  has one barrel and two end-cap air-core toroid magnets.
Three layers of precision tracking stations  provide  muon momentum measurements over the range
$|\eta|<2.7$.

The ATLAS trigger system consists of three levels of event selection: a first level implemented using custom-made electronics, which selects events at a design rate of at most 75~kHz, followed by two successive software-based levels. The level-2 trigger uses fast online algorithms, and the final trigger stage, Event Filter~(EF), uses reconstruction software with algorithms similar to the offline versions.

\section{\XS{} definition \label{sec:definition}}

Jets are defined using  the \antikt{} algorithm as implemented in the FastJet~\cite{Cacciari:2011ma} package,
  with two different values of the radius parameter: \rfour{} and \rsix.

Events containing at least three jets  within the rapidity range $|y|<3.0$ with  $\pt>50$~\GeV{} are considered.
The leading, subleading and sub-subleading jets are required to have $\pt > 150$~\GeV, $\pt > 100$~\GeV{} and $\pt > 50$~\GeV, respectively.

\Trijet{} double-differential \xss{} are measured as a function of the \trijet{} mass $$\mjjj{}=\sqrt{\left(p_1+p_2+p_3\right)^2}$$ and  the summed absolute rapidity separation of the three leading jets $$\ystar = \left|y_1 - y_2\right| + \left|y_2 - y_3\right| + \left|y_1 - y_3\right|,$$ where $p_i(y_i)$ are the four-momenta (rapidities) of the three leading jets.
The measurements are made in five ranges of $\ystar < 10$, in equal steps of two.
In each range of \ystar, a lower limit on the \trijet{} mass is imposed  to avoid the region of phase space affected by the jet \pt{} cuts. The measurement starts at $\mjjj=380$~\GeV{} in the $\ystar{}<2$ bin, increasing to $1180$~\GeV{} for the  $8<\ystar<10$ bin.

The \trijet{} mass distributions are corrected for  detector effects, and  the measured \xss{} are defined at the particle level.
Here particle level refers to jets built using produced particles with a proper lifetime longer than $10~\mbox{ps}$, including muons and neutrinos from decaying hadrons~\cite{Buttar:2008jx}.

 \section{Monte Carlo samples\label{sec:mc}}

The default MC generator used to simulate  events is \pythia~6  
\cite{Sjostrand:2006za} with the \perugia~2011 tune \cite{Skands:2010akv4}  and the CTEQ5L PDFs~\cite{Lai:1999wy}. Usually, ``tune`` refers to a set of model parameters, which provide an optimal description of high-energy particle collisions. Data from previous colliders (LEP, TEVATRON, etc), as well as early LHC data are included in  the process of tuning the model parameters \cite{Skands:2010akv4,ATL-PHYS-PUB-2011-008,ATL-PHYS-PUB-2011-009}.
The \pythia~6 is a  generator with LO $2 \to 2$ matrix element calculations, supplemented by leading-logarithmic calculations of parton showers order\-ed in \pt. A simulation of the underlying event, including multiple parton interactions,  is also included. The Lund string model \cite{lundString,Andersson:1979ij} is used to simulate the fragmentation process.
The signal reconstruction is affected by multiple proton--proton interactions occurring during the same bunch crossing and  by remnants of electronic signals from previous bunch crossings in the detectors (pileup).
To simulate  pileup, inelastic \pp{} events
are generated using \pythia~8 \cite{Sjostrand:2007gs} with the 4C tune \cite{Corke:2010yf} and MRST~LO$^{**}$ proton PDF set \cite{Sherstnev:2008dm}.
The number of minimum-bias events overlaid on each signal event is chosen to reproduce the distribution of the average number of simultaneous \pp{} collisions \avgmu{} in an event.  During the 2011 data-taking period \avgmu{}  changed from 5 to 18 with increasing  instantaneous luminosity.

To estimate the uncertainties in the modelling of the hard scattering, hadronisation, the underlying event and of parton showers, events are also simulated using  \alpgen{} \cite{Mangano:2002ea}, a multi-leg LO MC simulation,  with up to six final-state partons in the matrix element calculations, interfaced to \herwig~6.5.10 \cite{Corcella:2002jc,Marchesini:1991ch,Corcella:2000bw} using the AUET2 tune \cite{ATL-PHYS-PUB-2011-008} with the CTEQ6L1 PDF set~\cite{Pumplin:2002vw} for parton showers and \jimmy~4.31 \cite{Butterworth:1996zw} for the underlying event.

The outputs from these event generators are passed to the detector simulation \cite{Aad:2010ah}, based on \geant{} \cite{Agostinelli:2002hh}.
Simulated events are digitised \cite{ATLAS:1999uwa,ATLAS:1999vwa} to model the detector responses, and then reconstructed using the same software as used to process the data.

 \section{Data selection and jet calibration\label{sec:data}}

This analysis is based on data collected with the ATLAS detector in the year 2011 during periods with stable \pp{} collisions at $\sqrt{s}=7~\TeV$ in which all relevant detector components were  operational. The resulting data sample corresponds to an integrated luminosity of \LUMI~\cite{Aad:2013ucp}.

The presence of at least one primary vertex (compatible with the position of the beam spot), reconstructed using two or more tracks with $\pt>500$~\MeV,  is required  to reject cosmic ray events and beam-related backgrounds.
The primary vertex with the largest sum of squared transverse momenta of associated tracks is used as the interaction point for the analysis.

Due to the high instantaneous luminosity and a limited detector readout bandwidth, a set of single-jet triggers with increasing transverse energy (\et) thresholds  is used to collect data events with jets.
Only a fraction of the events that fired the trigger are actually recorded. The reciprocal of this fraction is the prescale factor of the trigger considered.
The triggers with  lower \et{} thresholds were prescaled with higher factors and only the trigger with the highest \et{} threshold remained unprescaled during the whole data-taking period. The pre\-scale factors are adjusted to keep the jet yield  approximately constant as a function of \et.

An event must pass all three levels of the jet trigger system. The trigger is based on the  \et{} of jet-like objects.
Level-1 provides a fast hardware decision based on the summed \et{} of calorimeter towers using a sliding-window algorithm.
Level-2 performs a simple jet reconstruction in a  geometric region around the object that fired the Level-1 trigger.
Finally, a full  jet reconstruction using the \antikt{} algorithm with \rfour{} is performed over the entire detector by the third level trigger.

The trigger efficiencies are determined as a function of \mjjj{} in each bin of \ystar{} separately for \rfour{} and \rsix{} jet radius  parameters. They are evaluated using an unbiased sample of events that fired the jet trigger with a $\pt=30$~\GeV{} threshold at the EF level.  This trigger is fully efficient in events with a leading jet passing the \trijet{} analysis requirements.
For every \ystar{} bin, the full range of \trijet{} mass is divided into subranges, each filled by only one of the several single-jet triggers. Triggers are used only where the trigger efficiency is above $99\%$.
Moreover, the lower \mjjj{} bound for each trigger is shifted up by $15\%$ from the $99\%$ efficiency point to avoid any possible biases from the  trigger strategy chosen for this measurement. This shift leads to a negligible increase in the statistical error on the measured \xss{}, compared to the total uncertainty.

Since the EF reconstructs  jets with a radius parameter \rfour, the \pt{} threshold at which the trigger for  jets defined with \rsix{} becomes fully efficient is significantly higher  than for \rfour{} jets. Using the same trigger subranges for both jet sizes would reduce the number of events with \antikt{} \rfour{} jets. To take advantage of the lower \pt{} at which  triggers are fully efficient for \rfour{} jets,
 different assignments between triggers and \mjjj{} ranges are considered  for these jets and jets  reconstructed with \rsix.

After events are selected by the trigger system, they are fully reconstructed offline.
The input objects to the jet algorithm are three-dimensional \topos{} \cite{Lampl:1099735}. Each \topo{} is constructed from a seed calorimeter cell with energy $|E_{\rm cell}| > 4\sigma$, where $\sigma$ is the width of the total noise distribution of the cell from both the electronics and pileup sources. Neighbouring cells are added to the \topo{} if they have $|E_{\rm cell}| > 2\sigma$. At the last step, all neighbouring cells are added. A local hadronic calibration~(LC) that accounts for inactive material, out-of-cluster losses for pions, and calorimeter response  is applied to clusters identified as hadronic by their energy density distribution \cite{Barillari:1112035}. The LC improves the \topo{} energy resolution, and the jet clustering algorithm  propagates this improvement to the jet level.  The LC  is validated using single pions in the combined test-beam \cite{Barillari:1112035}.

Each \topo{} is considered as a massless particle with an energy $E=\sum E_{\rm cell}$, and a direction given by the energy-weighted barycentre of the cells in the cluster with respect to the geometrical centre of the ATLAS detector.
The four-momentum of an uncalibrated jet is defined as the sum of the four-momenta of the clusters making up the jet.
The jet is then calibrated in four steps:
\begin{enumerate}
\item
  An estimated mean additional energy due to pileup is subtracted using a correction
  derived from MC simulation and validated \insitu{}  using  track-jets in dijet events and photons in $\gamma$--jet events as a function of the average number of \pp{} collisions in the same
  bunch crossing, \avgmu,
  the number of primary vertices, \npv,  and jet $\eta$ \cite{ATLAS:2012lla}.
  Here,  track-jets are reconstructed from all tracks associated to the primary vertex using the \antikt{} jet algorithm. 

\item
  The direction of the jet is corrected such that the jet
  originates from the selected hard-scatter vertex of the event instead of the geometrical centre of ATLAS.
\item
  The energy and the position of the jet are corrected for
  instrumental effects (calorimeter non-com\-pen\-sa\-tion, additional inactive
  material, effects due to the magnetic field) using correction factors obtained from MC simulation. The jet energy scale is restored
  on average to that of the particle-level jet.
  For the calibration, the particle-level jet does not include muons and non-interacting particles.
\item
  An additional \insitu{} calibration is applied to correct for
  residual differences between the MC simulation and data, derived by combining the
  results of dijet, $\gamma$--jet, $Z$--jet, and multi-jet momentum balance techniques.
\end{enumerate}
The full calibration procedure is described in detail in Ref.~\cite{Aad:2014bia}.

Data-taking in the year 2011 was affected by a read-out problem in a region of the LAr calorimeter, causing jets in this region to be poorly reconstructed. In order to avoid a bias in the spectra, events with any of the three leading jets falling in the region $-0.88 < \phi < -0.5$ were rejected. Approximately $ 15\%$ of events are removed by this requirement. This inefficiency is  corrected for using  MC simulation~(cf. Sect.~\ref{sec:unfold}).

The three leading jets  are required to satisfy the ``me\-dium'' quality criteria as described in Ref.~\cite{ATLAS-CONF-2012-020}, designed to reject cosmic-rays, beam-halo particles, and detector noise.  More than $5.3(2.5)\times 10^6$   \trijet{} events are selected   with radius parameter  \rfour$(0.6)$.

 \section{Data unfolding   \label{sec:unfold}}

The \trijet{} \xss{} as a function of \mjjj{} are obtained by unfolding the data distributions, and correcting for detector resolutions and inefficiencies. 
This procedure includes  a correction for the undetected presence of muons and neutrinos from hadron decays in jets.
The unfolding procedure is based on the iterative, dynamically stabilised (IDS) unfolding me\-thod \cite{Malaescu:2011yg}. Further details can be found in  Ref.~\cite{Aad:2011fc}.
To account for bin-to-bin migrations, a transfer matrix is built from the MC simulation, relating the particle-level and recon\-struction-level \trijet{} masses. The re\-construc\-tion-level to parti\-cle-level event association is done in the \mjjj--\ystar{} plane, such that only a requirement on the presence of a \trijet{} system is made.
Since bin-to-bin migrations  are usually due to jet energy  smearing of the \trijet{} mass, and less often due to jet angular resolution, the migrations across \ystar{} bins are negligible and the unfolding is performed separately in each  \ystar{} bin.

The data are unfolded to the particle level using a three-step procedure 
\begin{equation}
N_i^\mathrm{\cal P} = \frac{1}{\epsilon^\mathrm{\cal P}_i} \sum_{(j)} N_j^\mathrm{\cal R} \cdot \epsilon_j^\mathrm{\cal R}A_{ij},
\end{equation}
where $i$ ($j$) is the particle-level (re\-construc\-tion-level) bin index, and $N_i^\mathrm{\cal P}$ ($N_i^\mathrm{\cal R}$) is the number of particle-level (recon\-struction-level) events in bin $i$.
The quantities $\epsilon_i^\mathrm{\cal R}$ ($\epsilon_i^\mathrm{\cal P}$) are the fractions of recon\-struction-level (particle-level) events matching (associated with) particle-level (recon\-struction-level) events in each bin $i$.
These efficiencies are used to correct  for the matching inefficiency at the recon\-struction and particle level, respectively.
The element $A_{ij}$ of the transfer matrix is the probability for a recon\-struction-level event in bin $j$ to be associated with a particle-level event in bin $i$. It is used to unfold the re\-construc\-tion-level spectrum for detector effects.

A data-driven closure test is used to evaluate the bias in the unfolded data spectrum shape due to mis-modelling of the recon\-struction-level spectrum shape in the MC simulation.
The transfer matrix is improved through a series of iterations, where the pa\-rticle-level distribution from simulation is re-weighted  such that the re\-construc\-tion-level distribution from simulation  ma\-tches the data distribution.
The modified recon\-struction-level MC simulation is unfolded using the original transfer matrix, and the result is compared with the modified pa\-rticle-level spectrum. 
The resulting bias is considered as a systematic uncertainty.
For the analyses in this paper, one iteration is used, which leads to a bias in closure tests of less than one percent.

The statistical uncertainties in the unfolded results are estimated using pseudo-experiments.
Each event in the data and in the MC simulation is counted $n$ times, where $n$ is sampled from a Poisson distribution with a mean of one.
A fluctuated transfer matrix and efficiency corrections are calculated as the average over these pseudo-experiments in MC simulation.
Then, each resulting pseudo-experiment of the data spectrum is unfolded using the fluctuated transfer matrix and efficiency corrections.
Finally, the covariance matrix bet\-ween bins of measured \mjjj{} \xs{} is calculated using the set of unfolded pseudo-experiments of the data. The random numbers for the pseudo-experiments are generated  using unique seeds.
 The  dijet \cite{Aad:2013tea} and inclusive jet  \cite{Aad:2014vwa} \xs{} measurements use the same unique seeds to evaluate the statistical uncertainties.
In this way, the statistical uncertainty and bin-to-bin correlations in both the data and the MC simulation are encoded in the covariance matrix and the statistical correlation between different measurements can be taken into account in  combined fits.

\section{Experimental uncertainties
\label{sec:sysunc}}

The uncertainty in the jet energy scale (JES) calibration is the dominant uncertainty in this measurement.
The uncertainties in the central region are determined using a combination of the transverse momentum balance techniques,
such as   $Z$--jet, $\gamma$--jet and multi-jet balance measurements performed \insitu{}. In each of the methods, the uncertainties in the energy of  the well-measured objects, e.g. $Z$/photon or system of low-\pt{} jets, are propagated to the energy of the balancing jet.
The JES uncertainty in the central region is propagated to the forward region using transverse momentum balance between a central and a forward jet in events with two jets. The difference in the balance observed between MC simulation samples generated with \pythia{} and \herwig{}  is treated as  an additional uncertainty in the forward region.
The JES uncertainty in the high-\pt{} range is evaluated using the \insitu{} measurement of the single isolated hadron response~\cite{Aad:2012vm}.
The total JES uncertainty is described by the set of fully correlated in \pt{} independent uncertainty sources. Complete details of the JES derivation and its uncertainties can be found in Ref.~\cite{Aad:2014bia}.

The uncertainty  in the \pt{} of each individual jet due to the JES calibration is between $1\%$ and $4\%$ in the central region $(|\eta|<1.8)$, and increases  to $5\%$ in the forward region $(1.8<|\eta|<4.5)$.

The uncertainties due to the JES calibration are propagated to the measured \xss{} using the MC simulation. The energy and \pt{} of each jet in the \trijet{} sample are scaled up or down by one standard deviation of a given uncertainty component, after which the luminosity-nor\-ma\-lised \trijet{} event yield is measured from the resulting sample. The yields from the nominal sample and the samples where all jets were scaled up and down are unfolded, and the difference between each of these variations and the nominal result is taken as the uncertainty due to that JES uncertainty component.
 For example, the uncertainty in the \trijet{} \xs{} in the $8<\ystar<10(\ystar<2)$ bin  due to the LAr electromagnetic energy scale uncertainty increases from $2(3)\%$ to $10(8)\%$ with the \mjjj{} increasing from $1(0.4)$~\TeV{} to $4(3)$~\TeV{}.  In the same \ystar{} bins, the  uncertainty in the \trijet{} \xs{} due to  the uncertainty in the jet energy measurements in the forward region varies from $15(4)\%$ to $30(0.5)\%$, as a function of \mjjj{}.
Since the sources of JES calibration uncertainty are  uncorrelated with each other by construction, the corresponding uncertainty components in the \xs{} are also taken as uncorrelated.

Each jet is affected by the additional energy deposited in the calorimeters due to pileup effects.
Additional energy due to pileup  is  subtracted during the jet energy calibration procedure \cite{Aad:2014bia}.
To check for any residual pileup effects in the measured \xss, the luminosity-normalised \trijet{} yields in all \trijet{} mass and rapidity-separation bins are split into bins of different pileup conditions under which the data were collected. No statistically significant deviation from the nominal result is observed.

The jet energy resolution (JER) is measured in the data using the bisector method in dijet events \cite{Aad:2012ag}, where good agreement  with the MC simulation is observed. The uncertainty in the JER is affected by selection parameters for jets, such as the amount of nearby jet activity, and depends on both jet \pt{} and jet $\eta$.

Jet  angular resolution (JAR) is studied by matching  particle-level jets to reconstruction-level jets in simulation. Jets are matched by requiring  that  the angular distance $\Delta R= \sqrt{\left(\Delta\phi\right)^2+\left(\Delta y\right)^2}$ between the particle-level and re\-con\-struc\-tion-level jet  is less than the jet radius parameter. The angular resolution is obtained from a Gaussian fit to the distribution of the  difference of reconstruction-level and particle-level jet rapidity.

\begin{figure}[htp!]
\centering
 \subfigure[$\ystar<2$]{\includegraphics[width=\linewidth]{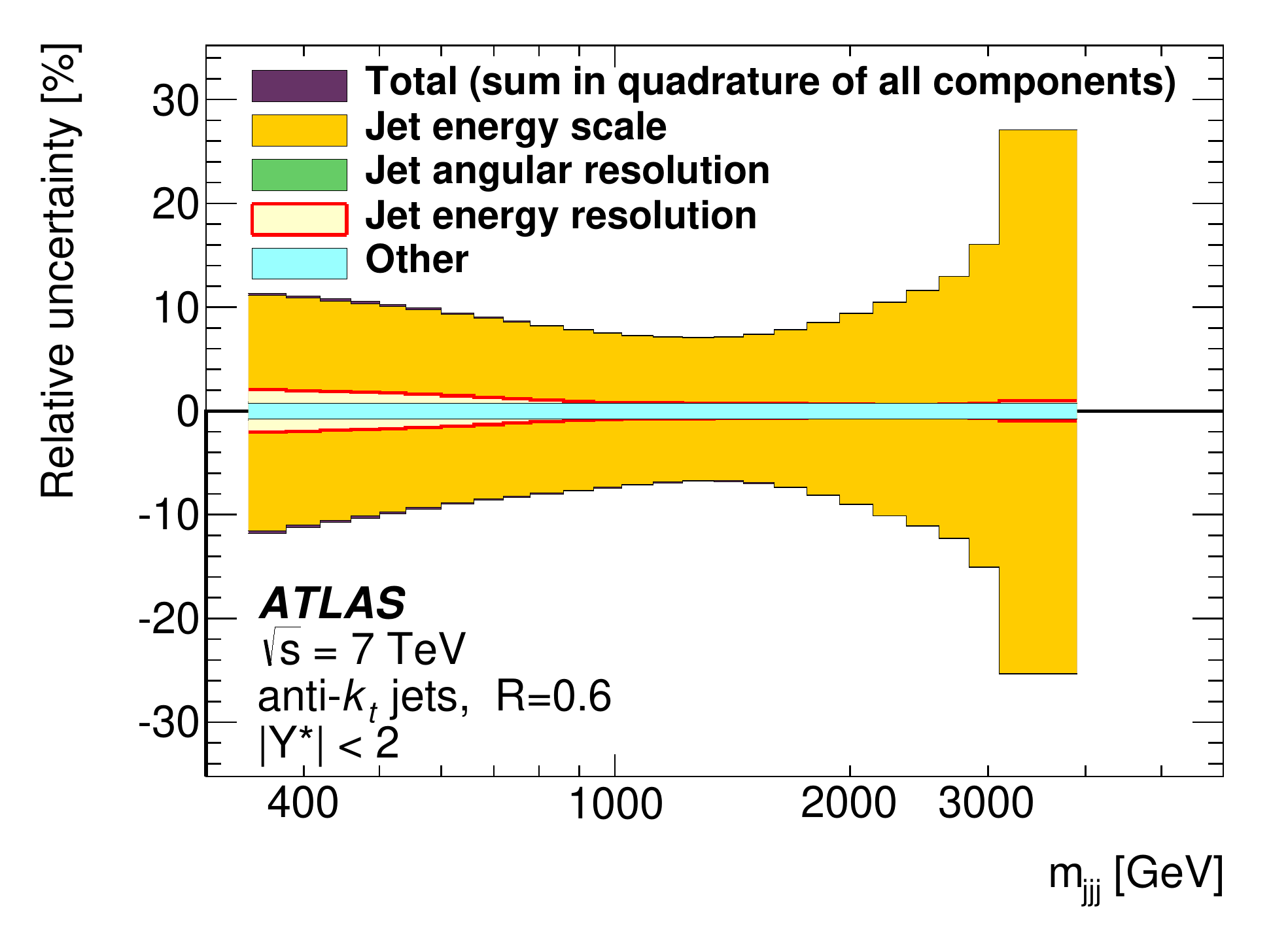}}\\
 \subfigure[$8<\ystar<10$]{\includegraphics[width=\linewidth]{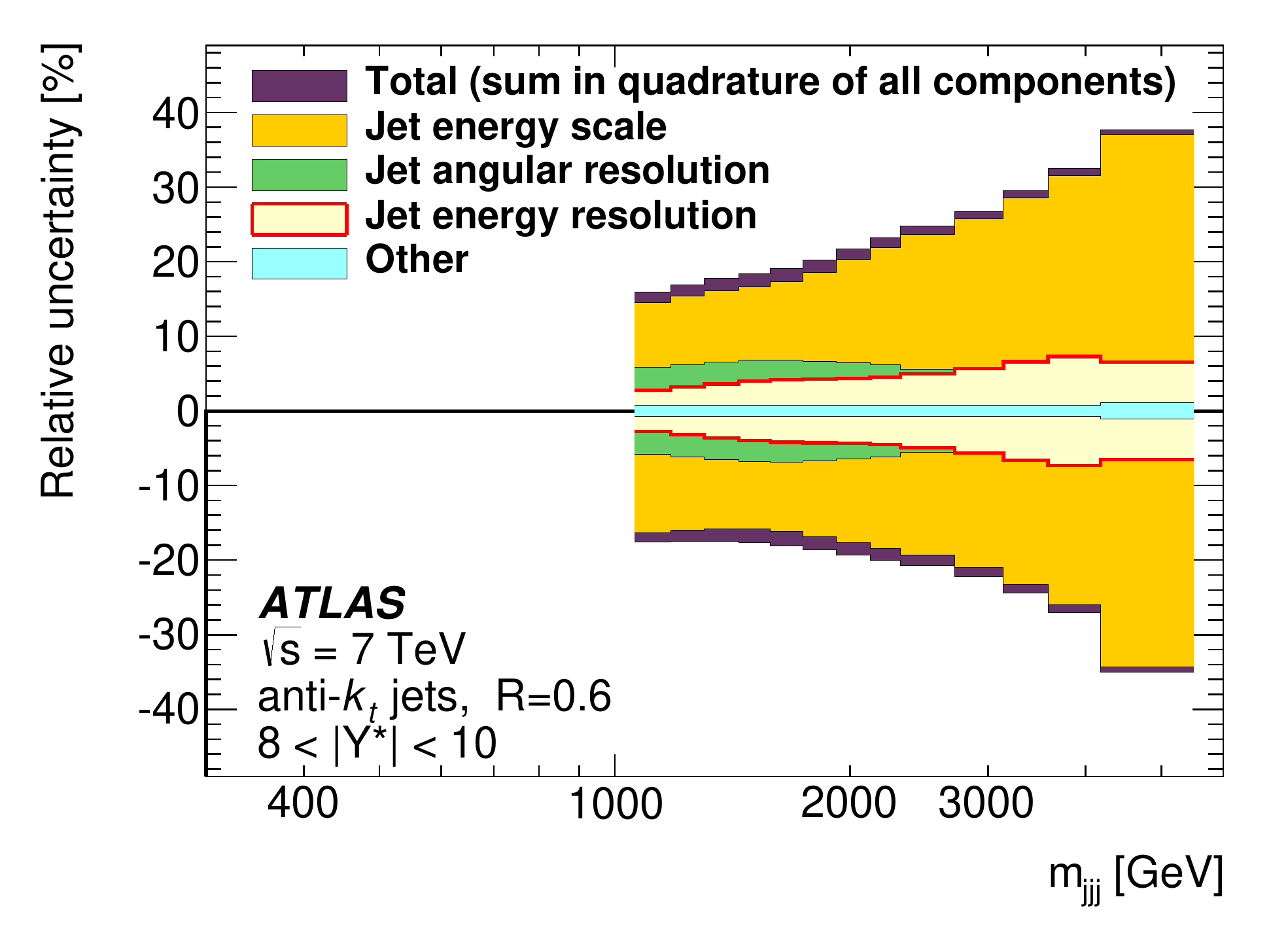}}
\caption{
Total systematic uncertainty in the \trijet{} \xs{}  for \antikt{} \rsix{} jets as a function of \mjjj{} (a) in  $\ystar<2$ and (b) $8<\ystar<10$ bins. The  bands shows the uncertainties due to jet energy scale, jet angular resolution, jet energy resolution and the combined uncertainty due to jet quality selection and unfolding. The outer band represents the total experimental uncertainty.
\label{fig:systematics}
}
\end{figure}

The difference between the JAR determined from the nominal MC simulation and that from the \alpgen{} sample is taken as a systematic uncertainty. The resolution varies between $0.005$ radians and $0.03$ radians depending on the jet $\eta$ and \pt{} values. The JAR uncertainty is about $10\mbox{--}15\%$ for $\pt{}<150$~\GeV{} and decreases to $\sim 1\%$ for $\pt>400$~\GeV.
 The jet angular bias is found to be negligible.

The JER and JAR uncertainties are propagated to the measured \xs{} through the unfolding transfer matrix. The energy and direction of each jet in the MC sample are smeared according to their uncertainties.
To avoid being limited by statistical fluctuations 
this procedure is repeated  1000 times in each event. The average transfer matrix derived from these pseudo-experiments is used to unfold the  \trijet{} yields, and the deviation from the \trijet{}  yield unfolded using the nominal transfer matrix is taken as a symmetrised systematic uncertainty.

The uncertainty due to the jet reconstruction inefficiency as a function of jet \pt{} is estimated by comparing the efficiency for reconstructing a calorimeter jet, given the  presence of an independently measured track-jet of the same radius, in data and in MC simulation \cite{ATLAS-CONF-2010-054,Aad:2014bia}.  Since this method relies on tracking,  its application is restricted to jets with $|\eta| < 1.9$ to ensure that both the \rfour{} and \rsix{} jets are fully within the tracker acceptance. For jets with $\pt > 50$~\GeV, relevant for this analysis, the reconstruction efficiency in both the data and the MC simulation is found to be $100\%$ for this rapidity region, leading to no additional uncertainty. The same efficiency is assumed for the forward region, where jets of a given \pt{}  are more energetic and, therefore, their reconstruction efficiency is expected to be at least as good as that of jets in the central region.

The efficiencies for single-jet selection using the ``me\-dium'' criteria agree within $0.25\%$ in data and MC simulation~\cite{ATLAS-CONF-2012-020}. Because three jets are considered for each event selected for the analysis, a $0.75\%$ systematic uncertainty in the \xs{} is assigned.

The impact of a possible mis-modelling of the  shape of \mjjj{} spectra in MC simulation, introduced through the unfolding as described in Sect.~\ref{sec:unfold}, is also included. The luminosity uncertainty is $1.8\%$ \cite{Aad:2013ucp} and is fully correlated between all data points.

The total experimental uncertainty in the  \trijet{} \xs{} is summarised in Fig.~\ref{fig:systematics}.
The total uncertainty ranges from $8\mbox{--}10\%$ at low \trijet{} mass  to $28\%$ at high \trijet{} mass for the range $\ystar < 6$ (see Appendix), and  increases slightly for larger \ystar{} bins.  In the $8<\ystar<10$ bin the total uncertainty ranges from $18\%$ to $38\%$, where it is dominated by the   jet energy scale uncertainty component for forward jets.

 \section{Theoretical predictions and uncertainties}
\label{sec:theory}

The  NLO QCD predictions by the parton-level MC \xs{} calculator \nlojetpp~\cite{Nagy:2003tz}, corrected for hadronisation effects  and  underlying-event activity using Monte Carlo simulation
with \perugia~2011 tune~\cite{Skands:2010akv4} of \pythia~6, are compared to the measured \trijet{} \xss{}.

\subsection{Fixed-order  predictions}
\label{subsec:nlo}

The fixed-order  QCD calculations are performed with the \nlojetpp{}  program interfaced to \applgrid{} \cite{applgrid:2009} for fast convolution with various PDF sets.
The renormalisation ($Q_\mathrm{R}$) and factorisation ($Q_\mathrm{F}$) scales are set to the mass of the \trijet{} system,
$
  Q = Q_\mathrm{R} = Q_\mathrm{F} = \mjjj.
$
The following proton PDF sets are considered for the theoretical predictions:
 CT~10~\cite{Lai:2010vv}, GJR~08~\cite{Gluck:2008gs},  MSTW~2008~\cite{Martin:2009iq}, 
 NNPDF~2.3~\cite{Ball:2012cx}, HERA\-PDF~1.5~\cite{HERAPDF15},  and ABM~11~\cite{Alekhin:2012ig}.

To estimate the uncertainty due to missing higher-order terms in the fixed-order perturbative expansion, the renormalisation scale is varied up and down by a factor of two. The uncertainty due to the dependence of the theoretical predictions on the factorisation scale, which specifies the separation between the short-distance hard scattering and long-distance non-pertur\-ba\-tive dynamics, is  estimated by varying the factorisation scale up and down by a factor of two. All permutations of these two scale choices are considered, except the cases where  the  scales are shifted in opposite directions. The maximum deviations from the nominal prediction are taken as the scale uncertainty. The scale uncertainty is generally  $10\mbox{--}20\%$ depending on the \mjjj.

The multiple uncorrelated uncertainty components of each PDF set, as provided by the various PDF analyses, are also propagated through the theoretical calculations. The PDF groups generally derive these from the experimental uncertainties in the data used in the fits.
For the results shown in Sect.~\ref{sec:results}, the standard Hessian sum in quadrature \cite{Pumplin:2001ct} of the various independent components is calculated taking into account asymmetries of the uncertainty components. The NNPDF~2.3 PDF set is an exception, where uncertainties are expressed in terms of \emph{replicas} instead of independent components. These replicas represent a collection of equally likely PDF sets, where the data used in the PDF fit were fluctuated within their experimental uncertainties. For the plots shown in Sect.~\ref{sec:results}, the uncertainties in the NNPDF~2.3 PDF set are evaluated as the RMS of the replicas in each bin of \mjjj, producing equivalent PDF uncertainties in the theoretical predictions. These uncertainties are symmetric by construction.
Where needed, the uncertainties of PDF sets are rescaled to the $68\%$ confidence level (CL).
HERA\-PDF provides three types of uncertainties: experimental, model and parameterisation. The three uncertainty sour\-ces are added in quadrature to get a total PDF uncertain\-ty.

The uncertainties in the \xss{} due to  the strong coupling, $\alphas$, are estimated using two additional proton PDF sets, for which different values of $\alphas$ are assumed in the fits, such that the effect of the strong coupling value on the PDFs  is included. This follows Ref.~\cite{Lai:2010nw}. The resulting uncertainty is approximately $3\%$ across all \trijet{} mass and \ystar{} ranges considered.

The scale uncertainties are dominant in low and intermediate \trijet{} mass regions, while the PDF uncertainties become dominant at high \mjjj. The uncertainties in the theoretical predictions due to those on the PDFs range from $5\%$ at low \mjjj{}  to $30\%$ at high \trijet{} mass for the range of  \ystar{} values up to four. For the  values of \ystar{} between four and ten, the PDF uncertainties reach $40\mbox{--}80\%$ at high \trijet{} mass, depending on the PDF set and the \ystar{} value.

\subsection{Non-perturbative effects \label{subsec:npc}}

\begin{figure}[htp!]
\centering
\subfigure[\rfour{} jets]{
  \includegraphics[width=\linewidth]{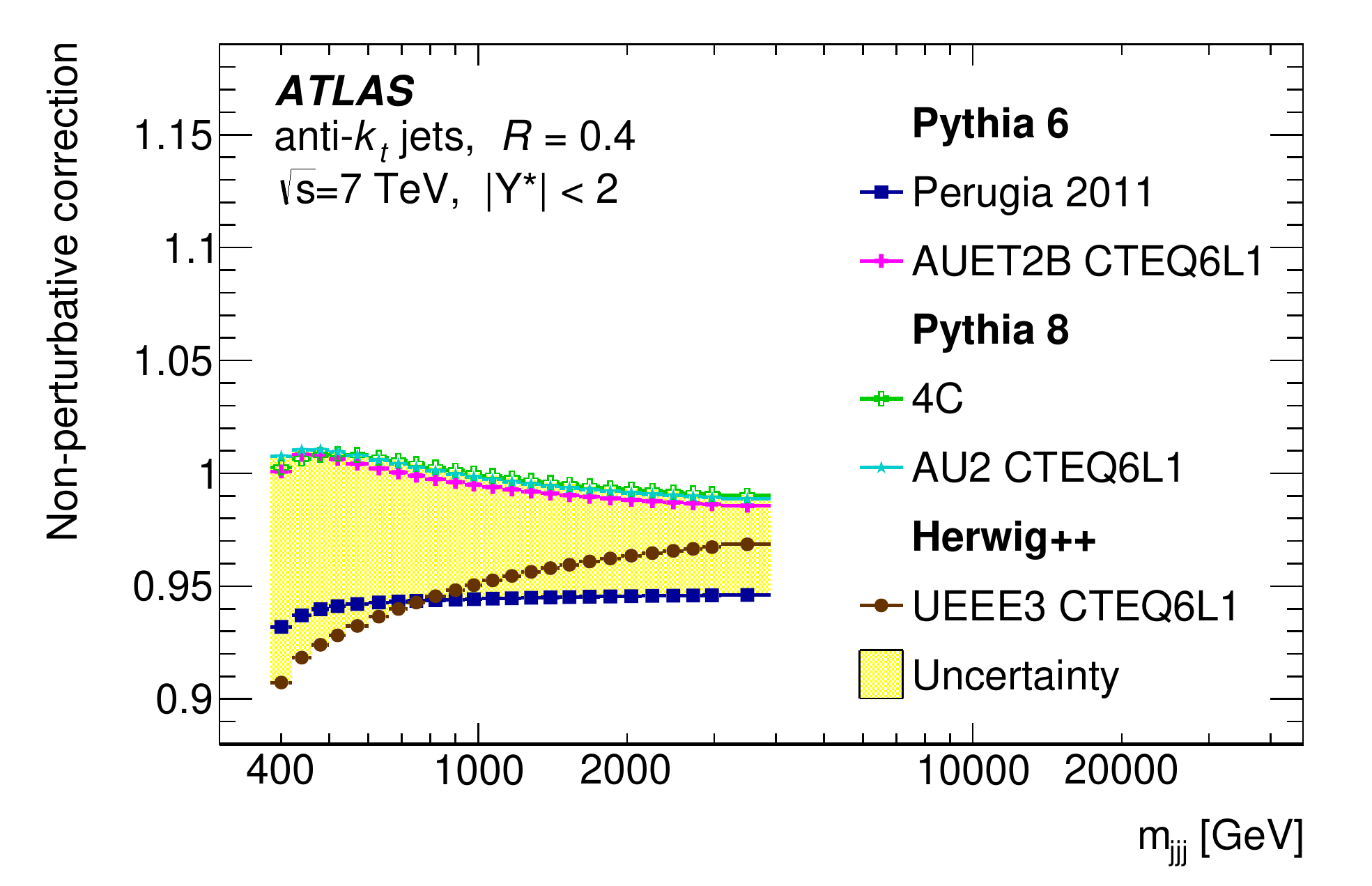}
}
\subfigure[ \rsix{} jets]{
  \includegraphics[width=\linewidth]{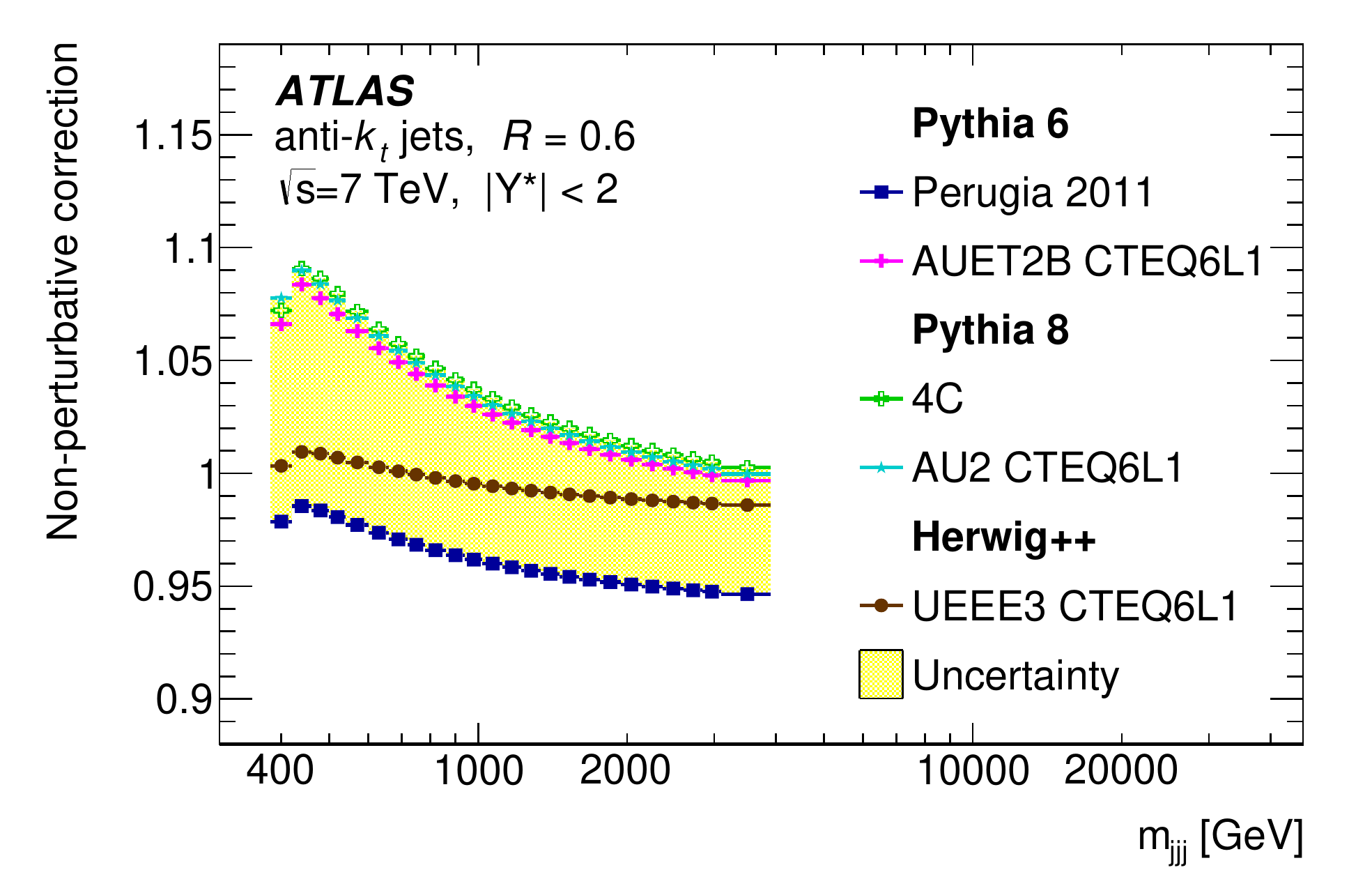}
}
\caption{
  \label{fig:npcorr}
Non-perturbative corrections  obtained using various MC generators and tunes   for the differential \trijet{} \xs{} as a function of \trijet{} mass in the range $\ystar<2$ for \antikt{} jet (a) \rfour{} and (b)~\rsix.
}
\end{figure}

Non-perturbative corrections (NPC) are evaluated using leading-logarithmic parton-shower generators, separately for each value of the jet radius parameter.
The corrections are calculated as bin-by-bin ratios of the \trijet{} differential \xs{} at the particle level, including hadronisation and underlying-event effects, to that at  parton-level after the parton shower (before the hadronisation process starts) with the underlying-event simulation switched off.
The nominal corrections are calculated using \pythia~6 
 with the \perugia{}~2011 tune.
The non-perturbative corrections as a function of \trijet{} mass are shown in Fig.~\ref{fig:npcorr} for the range $\ystar<2$ for \rfour{} and \rsix{} jets.
The NPC are  smaller than $10\%$ in all \mjjj{} and \ystar{} bins.

The uncertainties in the non-perturbative corrections, arising from the modelling of the hadronisation process and the underlying event, are estimated as the maximum deviations of the corrections from the nominal ones, using the following  configurations:
\pythia~8 
 with the 4C~\cite{Corke:2010yf} and AU2~\cite{ATL-PHYS-PUB-2011-008} tunes using the CTEQ6L1 PDF set~\cite{Pumplin:2002vw};
\pythia~6  with the AUET2B~\cite{ATL-PHYS-PUB-2011-009} tune with  CTEQ6L1; 
and \herwigpp~2.6.3~\cite{Bahr:2008pv,Gieseke:2011na} with the UE-EE-3 tune~\cite{Gieseke:2012ft} using the CTEQ6L1 set. 
The uncertainty in the non-per\-tur\-ba\-tive corrections ran\-ges  up to $\sim 10\%$ depending on the \trijet{} mass in all  \ystar{} bins.

The total theoretical uncertainty is calculated as a sum in quadrature of PDF, scale, $\alphas$ and NPC uncertainties.

 \section{\XSS{} results \label{sec:results}}

Measurements of the  double-differential \trijet{} \xss{} as  a function of the \trijet{} mass in various ranges of \ystar{} are shown in Figs.~\ref{fig:MassSummary04}~and~\ref{fig:MassSummary06} for \antikt{} jets with values of the radius parameter \rfour{} and \rsix, respectively.
The \xs{} decreases rapidly as a function of  the \trijet{} mass. The NLO QCD calculations using \nlojetpp{} with the CT~10 PDF set corrected for non-perturbative effects are compared to the measured \xss. Good agreement between the data and the theoretical predictions is found over the full kinematic range, covering almost seven orders of magnitude in the measured \xs{} values.

The ratios of the theoretical predictions calculated with various PDF sets to the measured \xss{} are presented in Figs.~\ref{fig:MassRatio04}~and~\ref{fig:MassRatio04_2}  for \rfour{} jets and   in Figs.~\ref{fig:MassRatio06}~and~\ref{fig:MassRatio06_2} for \rsix{} jets.
Theoretical calculations that use  CT~10, MSTW~2008 and GJR~08 PDFs are compared to data in Figs.~\ref{fig:MassRatio04}~and~\ref{fig:MassRatio06} and  comparisons to other  global PDFs, namely  NNPDF~2.3, ABM~11 and  HERAPDF~1.5.
are presented in Figs.~\ref{fig:MassRatio04_2}~and~\ref{fig:MassRatio06_2}.

The \trijet{} \xss{} are well described by the calculations that use  CT~10, NNPDF~2.3, GJR~08, MSTW~2008 and HERAPDF~1.5 PDFs. Disagreement between data and the predictions using ABM~11 PDFs is observed for most of the \xss{} measured with both jet radius parameters.

For all PDF sets, the predictions for \antikt{} \rfour{} jets agree well with measured \xss{}, while the calculations that use the  ABM~11 PDF set are systematically below  all other theory curves.
Theory predictions for \antikt{} \rsix{} jets underestimate the data
 across the full \mjjj--\ystar{} plane. This shift is within the  experimental and theoretical uncertainties. The jet radius dependence of theory-to-data ratios is similar for all PDF sets considered, demonstrating that this tendency is independent of the assumptions made in different PDF determinations.

\newcommand{\psRatioCaption}
{The experimental error bands are centered at one and designate the relative statistical  (thin dashed line) and total (statistical and systematic uncertainties added in quadrature) experimental uncertainties (thick solid line).
The theoretical predictions are represented by thick lines with the hatched or filled band around it. The line show the central values and the  band represent the total theory uncertainty.
}

\newcommand{\psXSCaption}
{
 For convenience, the \xss{} are multiplied by the factors indicated in the
legend. Also shown is the comparison with the \nlojetpp{} prediction with the CT~10 PDF set corrected for non-perturbative effects.
The statistical uncertainties are smaller than the size of the symbols. Where visible, the sum in quadrature of the statistical and experimental systematic uncertainties is plotted.
}

\begin{figure*}
\begin{center}
\includegraphics[width=0.85\textwidth]{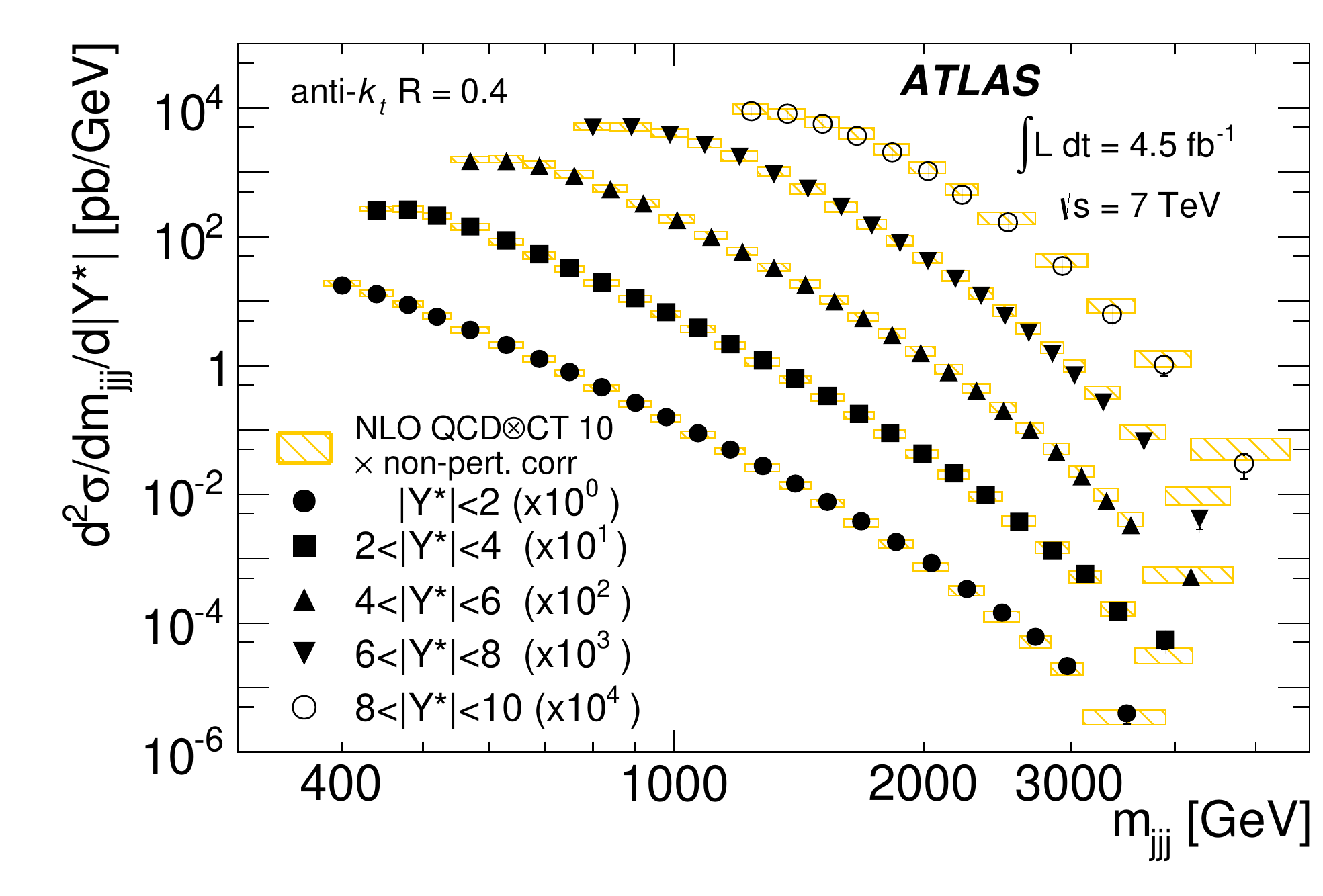}
\caption{
\label{fig:MassSummary04}
The \trijet{} double-differential \xs{} as a function of \mjjj{} in bins \ystar, as denoted in the legend. The jets are identified using the \antikt{} algorithm with \rfour.
\psXSCaption
}
\includegraphics[width=0.85\textwidth]{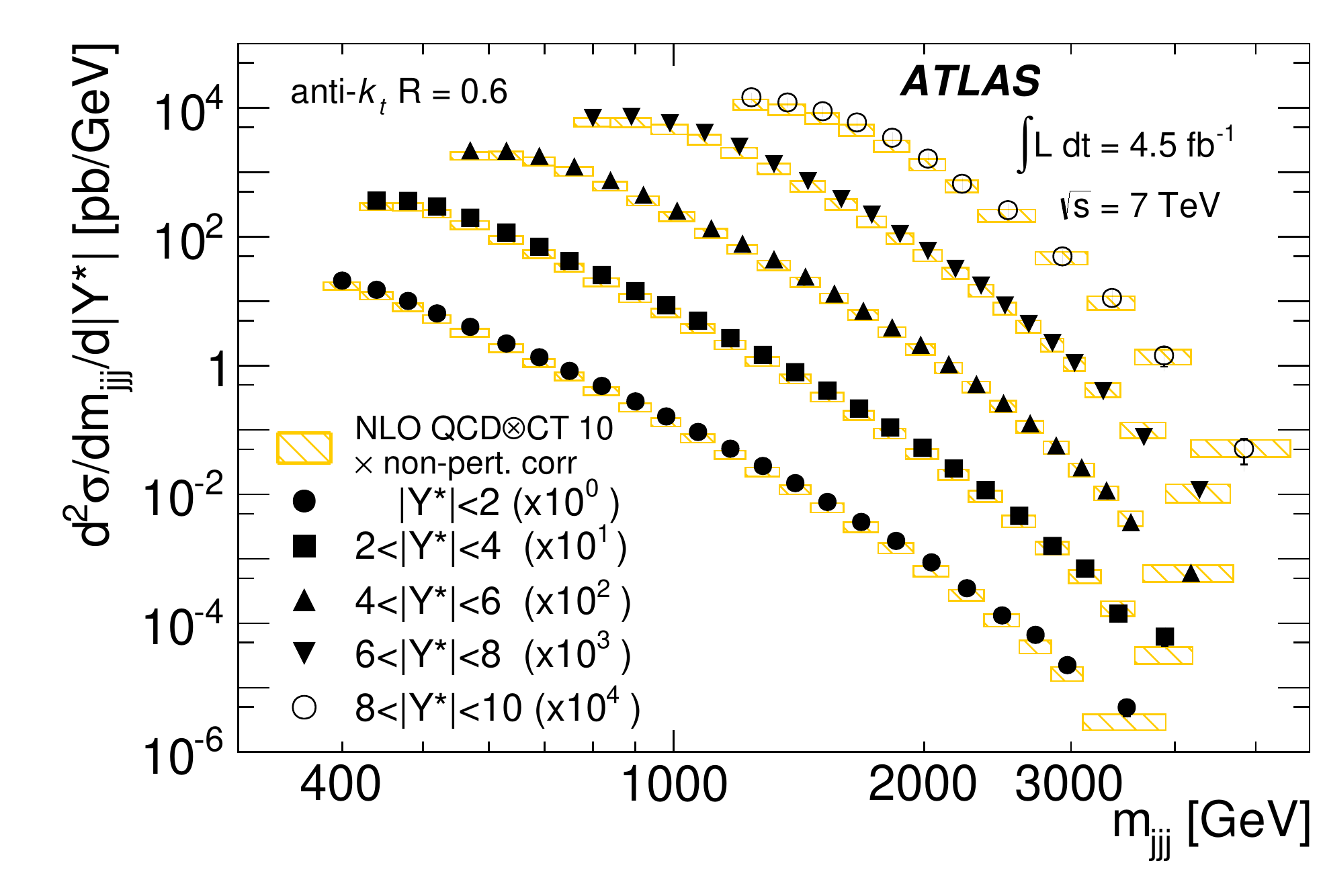}
\caption{
\label{fig:MassSummary06}
The \trijet{} double-differential \xs{} as a function of \mjjj{} in bins \ystar, as denoted in the legend. The jets are identified using the \antikt{} algorithm with \rsix.
\psXSCaption
}
\end{center}
\end{figure*}

\begin{figure*}[h]
\begin{center}
\includegraphics[width=0.98\textwidth]{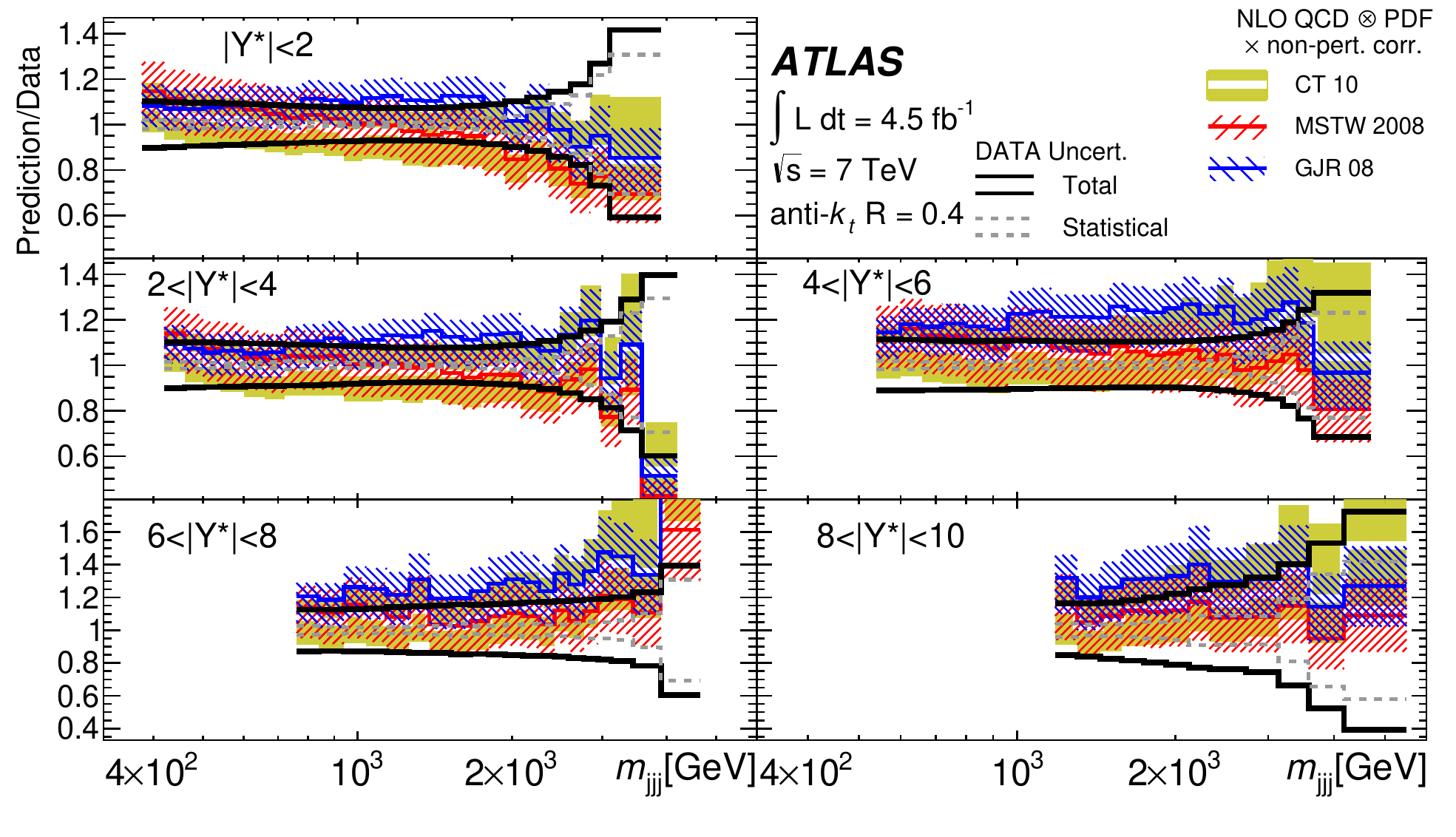}
\caption{
\label{fig:MassRatio04}
The ratio of  NLO QCD predictions, obtained by using \nlojetpp{} with different PDF sets (CT~10, MSTW~2008,  GJR~08) and corrected for non-perturbative effects, to data  as a function of \mjjj{} in bins of \ystar, as denoted in the legend.
The ratios are  for  jets identified using the \antikt{} algorithm with \rfour.
\psRatioCaption
}
\end{center}
\begin{center}
\includegraphics[width=0.98\textwidth]{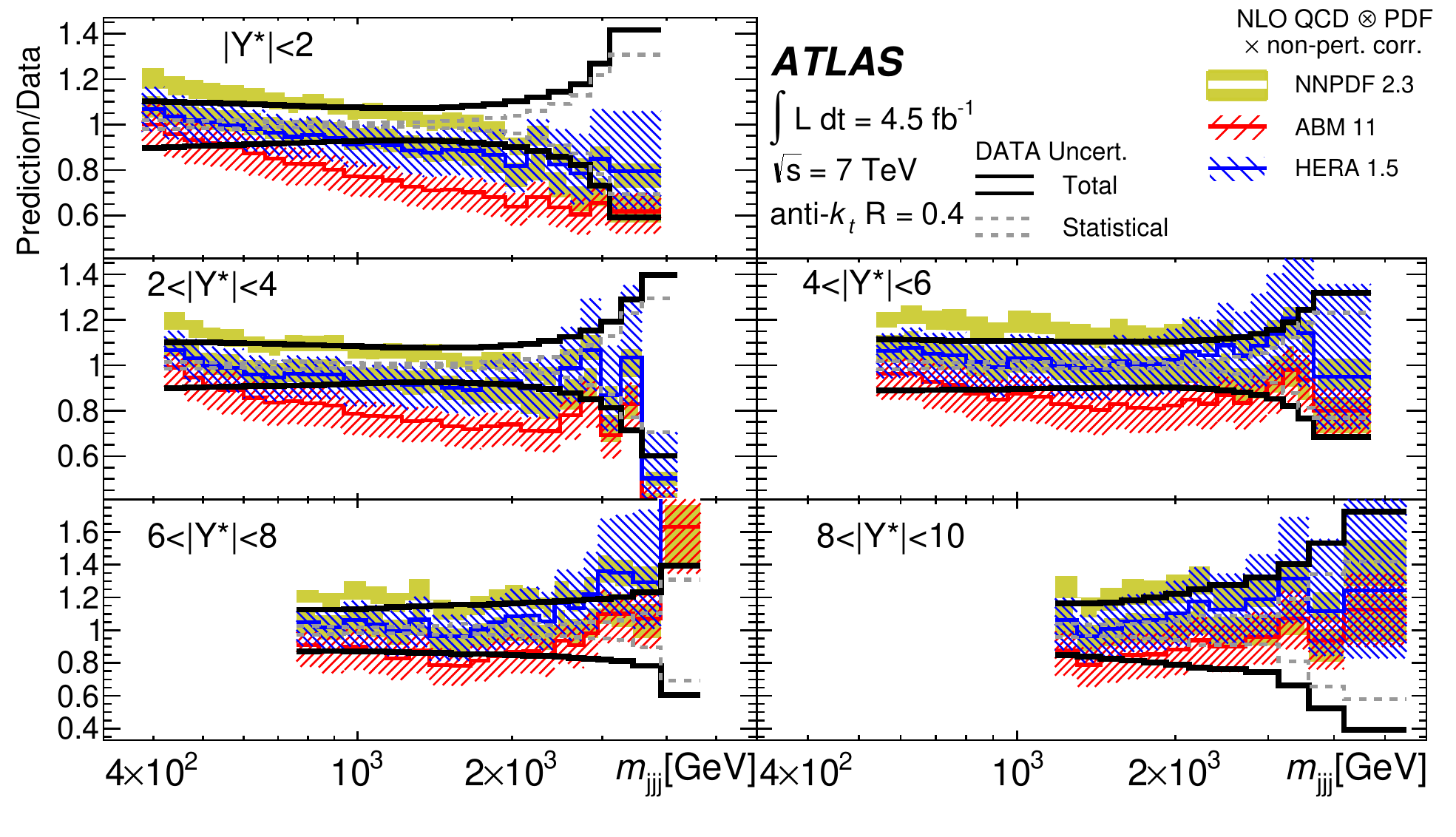}
\caption{
\label{fig:MassRatio04_2}
The ratio of  NLO QCD predictions, obtained by using \nlojetpp{} with different PDF sets ( NNPDF~2.3, ABM~11, HERAPDF~1.5) and corrected for non-perturbative effects, to data  as a function of \mjjj{} in bins of \ystar, as denoted in the legend.
The ratios are  for  jets identified using the \antikt{} algorithm with \rfour.
\psRatioCaption
}

 \end{center}
\end{figure*}

\begin{figure*}[h]
\begin{center}
\includegraphics[width=0.98\textwidth]{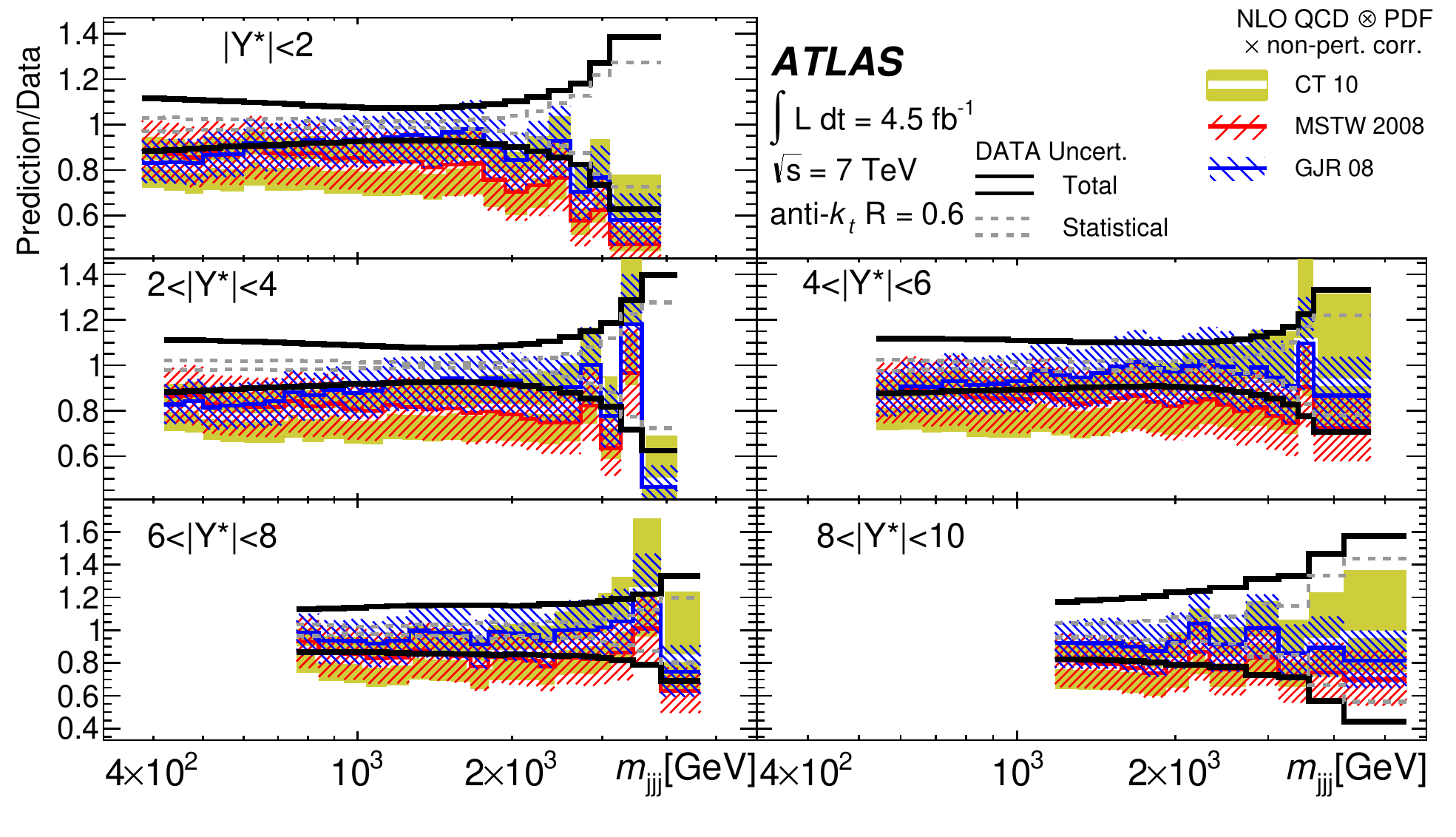}
\caption{
\label{fig:MassRatio06}
The ratio of  NLO QCD predictions, obtained by using \nlojetpp{} with different PDF sets (CT~10, MSTW~2008,  GJR~08) and corrected for non-perturbative effects, to data  as a function of \mjjj{} in bins of \ystar, as denoted in the legend.
The ratios are  for  jets identified using the \antikt{} algorithm with \rsix.
\psRatioCaption
}
\end{center}
\begin{center}
\includegraphics[width=0.98\textwidth]{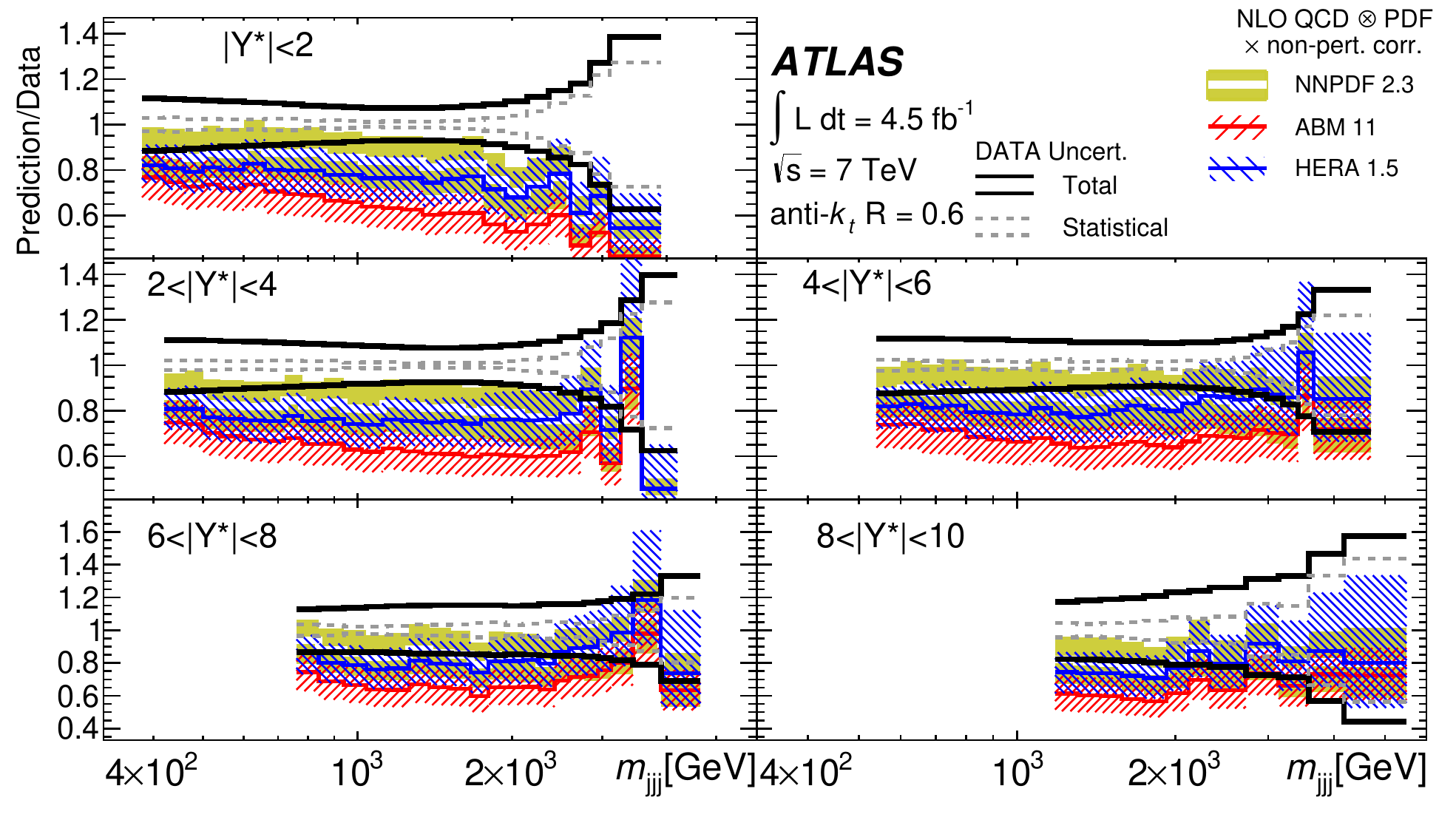}
\caption{
\label{fig:MassRatio06_2}
The ratio of  NLO QCD predictions, obtained by using \nlojetpp{} with different PDF sets ( NNPDF~2.3, ABM~11, HERAPDF~1.5) and corrected for non-perturbative effects, to data  as a function of \mjjj{} in bins of \ystar, as denoted in the legend.
The ratios are  for  jets identified using the \antikt{} algorithm with \rsix.
\psRatioCaption
}
\end{center}
\end{figure*}

\clearpage

 \section{Conclusions\label{sec:conclusions}}

\XSS{} measurements of   \trijet{} production in \pp{} collisions at 7~\TeV{} centre-of-mass energy  as a function of the \trijet{} mass, in bins of the sum of the absolute rapidity separations between the three leading jets are presented. Jets are reconstructed with the \antikt{} algorithm using two values of the radius parameter, \rfour{} and \rsix. The measurements are based on the full data set collected with the ATLAS detector during 2011 data-taking at the LHC, corresponding to an integrated luminosity of \LUMIE.
The measurements are corrected for detector effects and reported at the particle level. The total experimental uncertainty in these measurements is dominated by the jet energy scale calibration uncertainty. The measurement  uncertainties  are smaller than, or  similar to,  those in the theoretical predictions.

The measurements probe  \trijet{} masses up to \\ $\sim 5$~\TeV{} and are well described by perturbative QCD at NLO accuracy across the full \mjjj--\ystar{} plane. The comparison of NLO QCD predictions corrected for non-perturbative effects to the measured \xss{} is performed using several modern PDF sets.
The data are well described by the theoretical predictions when using  CT~10, NNPDF~2.3, HERAPDF~1.5, GJR~08 and MSTW~2008  PDFs.
 The  theoretical calculations based on the ABM~11 PDFs are systematically below all the other predictions.

Comparison of measured \xss{} to theoretical predictions for two different jet radius parameters shows good agreement for \rfour{} jets but shifted theory-to-data ratios for \rsix{} jets.
This shift is covered by the experimental and theoretical uncertainty bands and it has only a minor dependence on the PDF set used.


We thank CERN for the very successful operation of the LHC, as well as the
support staff from our institutions without whom ATLAS could not be
operated efficiently.

We acknowledge the support of ANPCyT, Argentina; YerPhI, Armenia; ARC,
Australia; BMWFW and FWF, Austria; ANAS, Azerbaijan; SSTC, Belarus; CNPq and FAPESP,
Brazil; NSERC, NRC and CFI, Canada; CERN; CONICYT, Chile; CAS, MOST and NSFC,
China; COLCIENCIAS, Colombia; MSMT CR, MPO CR and VSC CR, Czech Republic;
DNRF, DNSRC and Lundbeck Foundation, Denmark; EPLANET, ERC and NSRF, European Union;
IN2P3-CNRS, CEA-DSM/IRFU, France; GNSF, Georgia; BMBF, DFG, HGF, MPG and AvH
Foundation, Germany; GSRT and NSRF, Greece; RGC, Hong Kong SAR, China; ISF, MINERVA, GIF, I-CORE and Benoziyo Center, Israel; INFN, Italy; MEXT and JSPS, Japan; CNRST, Morocco; FOM and NWO, Netherlands; BRF and RCN, Norway; MNiSW and NCN, Poland; GRICES and FCT, Portugal; MNE/IFA, Romania; MES of Russia and NRC KI, Russian Federation; JINR; MSTD,
Serbia; MSSR, Slovakia; ARRS and MIZ\v{S}, Slovenia; DST/NRF, South Africa;
MINECO, Spain; SRC and Wallenberg Foundation, Sweden; SER, SNSF and Cantons of
Bern and Geneva, Switzerland; NSC, Taiwan; TAEK, Turkey; STFC, the Royal
Society and Leverhulme Trust, United Kingdom; DOE and NSF, United States of
America.

The crucial computing support from all WLCG partners is acknowledged
gratefully, in particular from CERN and the ATLAS Tier-1 facilities at
TRIUMF (Canada), NDGF (Denmark, Norway, Sweden), CC-IN2P3 (France),
KIT/GridKA (Germany), INFN-CNAF (Italy), NL-T1 (Netherlands), PIC (Spain),
ASGC (Taiwan), RAL (UK) and BNL (USA) and in the Tier-2 facilities
worldwide.

\bibliographystyle{atlasBibStyleWithTitle}
\bibliography{threejetPaper}

\onecolumn
\appendix

\section{Tables of measured \xss{} \label{sec:appendixTables}}

\begin{table*}[!ht]\tiny\centering\begin{tabular}{@{}c@{}@{}c@{}@{}c@{}@{}c@{}@{}r@{}@{}r@{}@{}r@{}@{}r@{}@{}r@{}@{}r@{}@{}r@{}@{}r@{}@{}r@{}@{}r@{}}\hline\hline \\[-2mm]\ \ $m_{jjj}$ \ \ & \ \ $m_{jjj}$-range \ \ & \ \ $\sigma$ \ \ & \ \ $\delta_{\textrm{stat}}^{\textrm{data}}$ \ \ & \ \ $\delta_{\textrm{stat}}^{\textrm{MC}}$ \ \ & \ \ $\gamma_{\insituT}$ \ \ & \ \ $\gamma_{\textrm{pileup}}$ \ \ & \ \ $\gamma_{\textrm{close-by}}$ \ \ & \ \ $\gamma_{\textrm{flavour}}$ \ \ & \ \ $u_{\textrm{JER}}$ \ \ & \ \ $u_{\textrm{JAR}}$ \ \ & \ \ $u_{\textrm{unfold}}$ \ \ & \ \ $u_{\textrm{qual.}}$ \ \ & \ \ $u_{\textrm{lumi}}$ \ \ \\ \relax
\ \ bin \# \ \ & \ \ [TeV] \ \ & \ \ [pb/GeV] \ \ & \ \ [\%] \ \ & \ \ [\%] \ \ & \ \ [\%] \ \ & \ \ [\%] \ \ & \ \ [\%] \ \ & \ \ [\%] \ \ & \ \ [\%] \ \ & \ \ [\%] \ \ & \ \ [\%] \ \ & \ \ [\%] \ \ & \ \ [\%] \\
\hline \\[-1.5mm]
\ \ $1$ \ \ & \ \ $ 0.38- 0.42$ \ \ & \ \ 17.5 \ \ & \ \ 1.8 \ \ & \ \ 0.73 \ \  & \ \ $^{+6.4}_{-6.3}$ \ \ & \ \ $^{+0.4}_{-2.7}$ \ \ & \ \ $^{+3.3}_{-3.4}$ \ \ & \ \ $^{+7.0}_{-6.6}$ \ \ & \ \ 0.7 \ \ & \ \ 0.0 \ \ & \ \ 0.0 \ \ & \ \ 0.75 \ \ & \ \ 1.8 \ \ \\[1.5mm]
\ \ $2$ \ \ & \ \ $ 0.42- 0.46$ \ \ & \ \ 12.8 \ \ & \ \ 2.0 \ \ & \ \ 0.62 \ \  & \ \ $^{+6.2}_{-6.1}$ \ \ & \ \ $^{+0.3}_{-1.9}$ \ \ & \ \ $^{+3.2}_{-3.2}$ \ \ & \ \ $^{+6.7}_{-6.3}$ \ \ & \ \ 0.8 \ \ & \ \ 0.0 \ \ & \ \ 0.0 \ \ & \ \ 0.75 \ \ & \ \ 1.8 \ \ \\[1.5mm]
\ \ $3$ \ \ & \ \ $ 0.46- 0.50$ \ \ & \ \ 8.75 \ \ & \ \ 1.3 \ \ & \ \ 0.50 \ \  & \ \ $^{+6.1}_{-6.0}$ \ \ & \ \ $^{+0.2}_{-1.4}$ \ \ & \ \ $^{+3.2}_{-3.2}$ \ \ & \ \ $^{+6.5}_{-6.2}$ \ \ & \ \ 0.8 \ \ & \ \ 0.0 \ \ & \ \ 0.0 \ \ & \ \ 0.75 \ \ & \ \ 1.8 \ \ \\[1.5mm]
\ \ $4$ \ \ & \ \ $ 0.50- 0.54$ \ \ & \ \ 5.72 \ \ & \ \ 1.5 \ \ & \ \ 0.55 \ \  & \ \ $^{+6.0}_{-5.9}$ \ \ & \ \ $^{+0.1}_{-1.2}$ \ \ & \ \ $^{+3.2}_{-3.2}$ \ \ & \ \ $^{+6.3}_{-6.0}$ \ \ & \ \ 0.8 \ \ & \ \ 0.0 \ \ & \ \ 0.0 \ \ & \ \ 0.75 \ \ & \ \ 1.8 \ \ \\[1.5mm]
\ \ $5$ \ \ & \ \ $ 0.54- 0.60$ \ \ & \ \ 3.57 \ \ & \ \ 1.8 \ \ & \ \ 0.49 \ \  & \ \ $^{+5.9}_{-5.7}$ \ \ & \ \ $^{+0.1}_{-1.1}$ \ \ & \ \ $^{+3.2}_{-3.2}$ \ \ & \ \ $^{+6.0}_{-5.7}$ \ \ & \ \ 0.8 \ \ & \ \ 0.0 \ \ & \ \ 0.0 \ \ & \ \ 0.75 \ \ & \ \ 1.8 \ \ \\[1.5mm]
\ \ $6$ \ \ & \ \ $ 0.60- 0.66$ \ \ & \ \ 2.09 \ \ & \ \ 1.6 \ \ & \ \ 0.49 \ \  & \ \ $^{+5.7}_{-5.6}$ \ \ & \ \ $^{+0.3}_{-1.1}$ \ \ & \ \ $^{+3.3}_{-3.2}$ \ \ & \ \ $^{+5.7}_{-5.4}$ \ \ & \ \ 0.7 \ \ & \ \ 0.0 \ \ & \ \ 0.0 \ \ & \ \ 0.75 \ \ & \ \ 1.8 \ \ \\[1.5mm]
\ \ $7$ \ \ & \ \ $ 0.66- 0.72$ \ \ & \ \ 1.27 \ \ & \ \ 1.0 \ \ & \ \ 0.55 \ \  & \ \ $^{+5.5}_{-5.4}$ \ \ & \ \ $^{+0.4}_{-1.1}$ \ \ & \ \ $^{+3.3}_{-3.3}$ \ \ & \ \ $^{+5.4}_{-5.2}$ \ \ & \ \ 0.7 \ \ & \ \ 0.0 \ \ & \ \ 0.0 \ \ & \ \ 0.75 \ \ & \ \ 1.8 \ \ \\[1.5mm]
\ \ $8$ \ \ & \ \ $ 0.72- 0.78$ \ \ & \ \ $ 7.93\cdot 10^{-1}$ \ \ &\ \ 1.1 \ \ & \ \ 0.53 \ \  & \ \ $^{+5.4}_{-5.3}$ \ \ & \ \ $^{+0.4}_{-1.1}$ \ \ & \ \ $^{+3.3}_{-3.2}$ \ \ & \ \ $^{+5.1}_{-4.9}$ \ \ & \ \ 0.6 \ \ & \ \ 0.0 \ \ & \ \ 0.0 \ \ & \ \ 0.75 \ \ & \ \ 1.8 \ \ \\[1.5mm]
\ \ $9$ \ \ & \ \ $ 0.78- 0.86$ \ \ & \ \ $ 4.61\cdot 10^{-1}$ \ \ &\ \ 0.91 \ \ & \ \ 0.42 \ \  & \ \ $^{+5.3}_{-5.2}$ \ \ & \ \ $^{+0.5}_{-1.0}$ \ \ & \ \ $^{+3.2}_{-3.1}$ \ \ & \ \ $^{+4.9}_{-4.7}$ \ \ & \ \ 0.6 \ \ & \ \ 0.0 \ \ & \ \ 0.0 \ \ & \ \ 0.75 \ \ & \ \ 1.8 \ \ \\[1.5mm]
\ \ $10$ \ \ & \ \ $ 0.86- 0.94$ \ \ & \ \ $ 2.64\cdot 10^{-1}$ \ \ &\ \ 0.69 \ \ & \ \ 0.33 \ \  & \ \ $^{+5.2}_{-5.1}$ \ \ & \ \ $^{+0.4}_{-0.9}$ \ \ & \ \ $^{+3.0}_{-2.9}$ \ \ & \ \ $^{+4.7}_{-4.5}$ \ \ & \ \ 0.6 \ \ & \ \ 0.0 \ \ & \ \ 0.0 \ \ & \ \ 0.75 \ \ & \ \ 1.8 \ \ \\[1.5mm]
\ \ $11$ \ \ & \ \ $ 0.94- 1.02$ \ \ & \ \ $ 1.58\cdot 10^{-1}$ \ \ &\ \ 0.82 \ \ & \ \ 0.32 \ \  & \ \ $^{+5.2}_{-5.1}$ \ \ & \ \ $^{+0.3}_{-0.8}$ \ \ & \ \ $^{+2.8}_{-2.7}$ \ \ & \ \ $^{+4.4}_{-4.3}$ \ \ & \ \ 0.6 \ \ & \ \ 0.0 \ \ & \ \ 0.0 \ \ & \ \ 0.75 \ \ & \ \ 1.8 \ \ \\[1.5mm]
\ \ $12$ \ \ & \ \ $ 1.02- 1.12$ \ \ & \ \ $ 8.91\cdot 10^{-2}$ \ \ &\ \ 0.58 \ \ & \ \ 0.34 \ \  & \ \ $^{+5.2}_{-5.1}$ \ \ & \ \ $^{+0.2}_{-0.6}$ \ \ & \ \ $^{+2.4}_{-2.4}$ \ \ & \ \ $^{+4.2}_{-4.1}$ \ \ & \ \ 0.6 \ \ & \ \ 0.0 \ \ & \ \ 0.0 \ \ & \ \ 0.75 \ \ & \ \ 1.8 \ \ \\[1.5mm]
\ \ $13$ \ \ & \ \ $ 1.12- 1.22$ \ \ & \ \ $ 4.96\cdot 10^{-2}$ \ \ &\ \ 0.71 \ \ & \ \ 0.42 \ \  & \ \ $^{+5.4}_{-5.2}$ \ \ & \ \ $^{+0.2}_{-0.5}$ \ \ & \ \ $^{+2.1}_{-2.0}$ \ \ & \ \ $^{+4.0}_{-3.9}$ \ \ & \ \ 0.6 \ \ & \ \ 0.0 \ \ & \ \ 0.0 \ \ & \ \ 0.75 \ \ & \ \ 1.8 \ \ \\[1.5mm]
\ \ $14$ \ \ & \ \ $ 1.22- 1.34$ \ \ & \ \ $ 2.76\cdot 10^{-2}$ \ \ &\ \ 0.92 \ \ & \ \ 0.39 \ \  & \ \ $^{+5.6}_{-5.4}$ \ \ & \ \ $^{+0.2}_{-0.4}$ \ \ & \ \ $^{+1.8}_{-1.8}$ \ \ & \ \ $^{+3.9}_{-3.7}$ \ \ & \ \ 0.6 \ \ & \ \ 0.0 \ \ & \ \ 0.0 \ \ & \ \ 0.75 \ \ & \ \ 1.8 \ \ \\[1.5mm]
\ \ $15$ \ \ & \ \ $ 1.34- 1.46$ \ \ & \ \ $ 1.48\cdot 10^{-2}$ \ \ &\ \ 1.2 \ \ & \ \ 0.46 \ \  & \ \ $^{+5.9}_{-5.7}$ \ \ & \ \ $^{+0.2}_{-0.4}$ \ \ & \ \ $^{+1.6}_{-1.5}$ \ \ & \ \ $^{+3.7}_{-3.6}$ \ \ & \ \ 0.6 \ \ & \ \ 0.0 \ \ & \ \ 0.0 \ \ & \ \ 0.75 \ \ & \ \ 1.8 \ \ \\[1.5mm]
\ \ $16$ \ \ & \ \ $ 1.46- 1.60$ \ \ & \ \ $ 7.63\cdot 10^{-3}$ \ \ &\ \ 1.6 \ \ & \ \ 0.39 \ \  & \ \ $^{+6.3}_{-6.1}$ \ \ & \ \ $^{+0.2}_{-0.4}$ \ \ & \ \ $^{+1.3}_{-1.3}$ \ \ & \ \ $^{+3.6}_{-3.5}$ \ \ & \ \ 0.5 \ \ & \ \ 0.0 \ \ & \ \ 0.0 \ \ & \ \ 0.75 \ \ & \ \ 1.8 \ \ \\[1.5mm]
\ \ $17$ \ \ & \ \ $ 1.60- 1.76$ \ \ & \ \ $ 3.83\cdot 10^{-3}$ \ \ &\ \ 2.1 \ \ & \ \ 0.38 \ \  & \ \ $^{+6.9}_{-6.7}$ \ \ & \ \ $^{+0.1}_{-0.4}$ \ \ & \ \ $^{+1.2}_{-1.2}$ \ \ & \ \ $^{+3.5}_{-3.4}$ \ \ & \ \ 0.6 \ \ & \ \ 0.0 \ \ & \ \ 0.0 \ \ & \ \ 0.75 \ \ & \ \ 1.8 \ \ \\[1.5mm]
\ \ $18$ \ \ & \ \ $ 1.76- 1.94$ \ \ & \ \ $ 1.82\cdot 10^{-3}$ \ \ &\ \ 2.9 \ \ & \ \ 0.38 \ \  & \ \ $^{+7.7}_{-7.5}$ \ \ & \ \ $^{+0.1}_{-0.3}$ \ \ & \ \ $^{+1.0}_{-1.0}$ \ \ & \ \ $^{+3.4}_{-3.3}$ \ \ & \ \ 0.6 \ \ & \ \ 0.0 \ \ & \ \ 0.0 \ \ & \ \ 0.75 \ \ & \ \ 1.8 \ \ \\[1.5mm]
\ \ $19$ \ \ & \ \ $ 1.94- 2.14$ \ \ & \ \ $ 8.60\cdot 10^{-4}$ \ \ &\ \ 4.0 \ \ & \ \ 0.37 \ \  & \ \ $^{+8.7}_{-8.4}$ \ \ & \ \ $^{+0.0}_{-0.2}$ \ \ & \ \ $^{+0.9}_{-0.9}$ \ \ & \ \ $^{+3.4}_{-3.2}$ \ \ & \ \ 0.6 \ \ & \ \ 0.0 \ \ & \ \ 0.0 \ \ & \ \ 0.75 \ \ & \ \ 1.8 \ \ \\[1.5mm]
\ \ $20$ \ \ & \ \ $ 2.14- 2.36$ \ \ & \ \ $ 3.40\cdot 10^{-4}$ \ \ &\ \ 6.0 \ \ & \ \ 0.54 \ \  & \ \ $^{+9.8}_{-9.4}$ \ \ & \ \ $^{+0.0}_{-0.1}$ \ \ & \ \ $^{+0.9}_{-0.8}$ \ \ & \ \ $^{+3.3}_{-3.1}$ \ \ & \ \ 0.6 \ \ & \ \ 0.0 \ \ & \ \ 0.0 \ \ & \ \ 0.75 \ \ & \ \ 1.8 \ \ \\[1.5mm]
\ \ $21$ \ \ & \ \ $ 2.36- 2.60$ \ \ & \ \ $ 1.46\cdot 10^{-4}$ \ \ &\ \ 9.1 \ \ & \ \ 0.70 \ \  & \ \ $^{+10.8}_{-10.4}$ \ \ & \ \ $^{+0.0}_{-0.1}$ \ \ & \ \ $^{+0.8}_{-0.8}$ \ \ & \ \ $^{+3.2}_{-3.1}$ \ \ & \ \ 0.7 \ \ & \ \ 0.0 \ \ & \ \ 0.0 \ \ & \ \ 0.75 \ \ & \ \ 1.8 \ \ \\[1.5mm]
\ \ $22$ \ \ & \ \ $ 2.60- 2.84$ \ \ & \ \ $ 6.16\cdot 10^{-5}$ \ \ &\ \ 13 \ \ & \ \ 0.79 \ \  & \ \ $^{+11.9}_{-11.8}$ \ \ & \ \ $^{+0.0}_{-0.1}$ \ \ & \ \ $^{+0.8}_{-0.8}$ \ \ & \ \ $^{+3.1}_{-3.1}$ \ \ & \ \ 0.7 \ \ & \ \ 0.0 \ \ & \ \ 0.0 \ \ & \ \ 0.75 \ \ & \ \ 1.8 \ \ \\[1.5mm]
\ \ $23$ \ \ & \ \ $ 2.84- 3.10$ \ \ & \ \ $ 2.17\cdot 10^{-5}$ \ \ &\ \ 22 \ \ & \ \ 1.1 \ \  & \ \ $^{+15.4}_{-15.5}$ \ \ & \ \ $^{+0.0}_{-0.1}$ \ \ & \ \ $^{+0.8}_{-0.7}$ \ \ & \ \ $^{+3.0}_{-3.1}$ \ \ & \ \ 0.8 \ \ & \ \ 0.0 \ \ & \ \ 0.0 \ \ & \ \ 0.75 \ \ & \ \ 1.8 \ \ \\[1.5mm]
\ \ $24$ \ \ & \ \ $ 3.10- 3.90$ \ \ & \ \ $ 4.00\cdot 10^{-6}$ \ \ &\ \ 31 \ \ & \ \ 0.87 \ \  & \ \ $^{+27.9}_{-26.9}$ \ \ & \ \ $^{+0.0}_{-0.1}$ \ \ & \ \ $^{+0.7}_{-0.7}$ \ \ & \ \ $^{+3.0}_{-3.1}$ \ \ & \ \ 1.3 \ \ & \ \ 0.0 \ \ & \ \ 0.3 \ \ & \ \ 0.75 \ \ & \ \ 1.8 \ \ \\[1.5mm]
\hline
\hline
\end{tabular}
\caption{
   Measured double-differential \trijet{} \xs{}, $\sigma$, for $\rfour{}$~jets and $\ystar{}<2$, along with uncertainties in the measurement. 
   All uncertainties are given in \%, where 
   $\delta_\textrm{stat}^\textrm{data}$ ($\delta_\textrm{stat}^\textrm{MC}$) are the statistical uncertainties in the data (MC simulation). 
   The $\gamma$ components are the uncertainty in the jet energy calibration from the \insitu{}, the pileup, the close-by jet, and flavour components. 
   The $u$ components show the uncertainty for the jet energy and angular resolution, the unfolding, the quality selection, and the luminosity. 
   While all columns are uncorrelated with each other, the \insitu{}, pileup, and flavour uncertainties shown here are the sum in quadrature of multiple uncorrelated components. 
}
\label{tab:sysunc_r00_ystar0}
\end{table*}

\begin{table*}[!ht]\tiny\centering\begin{tabular}{@{}c@{}@{}c@{}@{}c@{}@{}c@{}@{}r@{}@{}r@{}@{}r@{}@{}r@{}@{}r@{}@{}r@{}@{}r@{}@{}r@{}@{}r@{}@{}r@{}}\hline\hline \\[-2mm]\ \ $m_{jjj}$ \ \ & \ \ $m_{jjj}$-range \ \ & \ \ $\sigma$ \ \ & \ \ $\delta_{\textrm{stat}}^{\textrm{data}}$ \ \ & \ \ $\delta_{\textrm{stat}}^{\textrm{MC}}$ \ \ & \ \ $\gamma_{\insituT}$ \ \ & \ \ $\gamma_{\textrm{pileup}}$ \ \ & \ \ $\gamma_{\textrm{close-by}}$ \ \ & \ \ $\gamma_{\textrm{flavour}}$ \ \ & \ \ $u_{\textrm{JER}}$ \ \ & \ \ $u_{\textrm{JAR}}$ \ \ & \ \ $u_{\textrm{unfold}}$ \ \ & \ \ $u_{\textrm{qual.}}$ \ \ & \ \ $u_{\textrm{lumi}}$ \ \ \\ \relax
\ \ bin \# \ \ & \ \ [TeV] \ \ & \ \ [pb/GeV] \ \ & \ \ [\%] \ \ & \ \ [\%] \ \ & \ \ [\%] \ \ & \ \ [\%] \ \ & \ \ [\%] \ \ & \ \ [\%] \ \ & \ \ [\%] \ \ & \ \ [\%] \ \ & \ \ [\%] \ \ & \ \ [\%] \ \ & \ \ [\%] \\
\hline \\[-1.5mm]
\ \ $1$ \ \ & \ \ $ 0.38- 0.42$ \ \ & \ \ 20.8 \ \ & \ \ 2.9 \ \ & \ \ 0.91 \ \  & \ \ $^{+6.6}_{-6.6}$ \ \ & \ \ $^{+0.1}_{-3.4}$ \ \ & \ \ $^{+5.0}_{-4.6}$ \ \ & \ \ $^{+7.0}_{-6.6}$ \ \ & \ \ 2.0 \ \ & \ \ 0.7 \ \ & \ \ 0.0 \ \ & \ \ 0.75 \ \ & \ \ 1.8 \ \ \\[1.5mm]
\ \ $2$ \ \ & \ \ $ 0.42- 0.46$ \ \ & \ \ 15.0 \ \ & \ \ 3.1 \ \ & \ \ 0.81 \ \  & \ \ $^{+6.5}_{-6.5}$ \ \ & \ \ $^{+0.1}_{-2.6}$ \ \ & \ \ $^{+4.8}_{-4.4}$ \ \ & \ \ $^{+6.8}_{-6.5}$ \ \ & \ \ 1.9 \ \ & \ \ 0.6 \ \ & \ \ 0.0 \ \ & \ \ 0.75 \ \ & \ \ 1.8 \ \ \\[1.5mm]
\ \ $3$ \ \ & \ \ $ 0.46- 0.50$ \ \ & \ \ 10.1 \ \ & \ \ 2.1 \ \ & \ \ 0.60 \ \  & \ \ $^{+6.4}_{-6.4}$ \ \ & \ \ $^{+0.1}_{-2.2}$ \ \ & \ \ $^{+4.6}_{-4.3}$ \ \ & \ \ $^{+6.7}_{-6.3}$ \ \ & \ \ 1.8 \ \ & \ \ 0.5 \ \ & \ \ 0.0 \ \ & \ \ 0.75 \ \ & \ \ 1.8 \ \ \\[1.5mm]
\ \ $4$ \ \ & \ \ $ 0.50- 0.54$ \ \ & \ \ 6.44 \ \ & \ \ 2.4 \ \ & \ \ 0.59 \ \  & \ \ $^{+6.3}_{-6.2}$ \ \ & \ \ $^{+0.1}_{-1.9}$ \ \ & \ \ $^{+4.4}_{-4.1}$ \ \ & \ \ $^{+6.5}_{-6.0}$ \ \ & \ \ 1.7 \ \ & \ \ 0.4 \ \ & \ \ 0.0 \ \ & \ \ 0.75 \ \ & \ \ 1.8 \ \ \\[1.5mm]
\ \ $5$ \ \ & \ \ $ 0.54- 0.60$ \ \ & \ \ 3.99 \ \ & \ \ 2.2 \ \ & \ \ 0.49 \ \  & \ \ $^{+6.2}_{-5.9}$ \ \ & \ \ $^{+0.2}_{-1.6}$ \ \ & \ \ $^{+4.3}_{-3.9}$ \ \ & \ \ $^{+6.2}_{-5.8}$ \ \ & \ \ 1.6 \ \ & \ \ 0.3 \ \ & \ \ 0.0 \ \ & \ \ 0.75 \ \ & \ \ 1.8 \ \ \\[1.5mm]
\ \ $6$ \ \ & \ \ $ 0.60- 0.66$ \ \ & \ \ 2.20 \ \ & \ \ 2.1 \ \ & \ \ 0.53 \ \  & \ \ $^{+6.0}_{-5.7}$ \ \ & \ \ $^{+0.4}_{-1.3}$ \ \ & \ \ $^{+4.1}_{-3.8}$ \ \ & \ \ $^{+5.9}_{-5.4}$ \ \ & \ \ 1.5 \ \ & \ \ 0.2 \ \ & \ \ 0.0 \ \ & \ \ 0.75 \ \ & \ \ 1.8 \ \ \\[1.5mm]
\ \ $7$ \ \ & \ \ $ 0.66- 0.72$ \ \ & \ \ 1.35 \ \ & \ \ 2.6 \ \ & \ \ 0.63 \ \  & \ \ $^{+5.8}_{-5.5}$ \ \ & \ \ $^{+0.5}_{-1.1}$ \ \ & \ \ $^{+3.9}_{-3.7}$ \ \ & \ \ $^{+5.5}_{-5.2}$ \ \ & \ \ 1.3 \ \ & \ \ 0.1 \ \ & \ \ 0.0 \ \ & \ \ 0.75 \ \ & \ \ 1.8 \ \ \\[1.5mm]
\ \ $8$ \ \ & \ \ $ 0.72- 0.78$ \ \ & \ \ $ 8.27\cdot 10^{-1}$ \ \ &\ \ 2.4 \ \ & \ \ 0.67 \ \  & \ \ $^{+5.6}_{-5.4}$ \ \ & \ \ $^{+0.6}_{-1.1}$ \ \ & \ \ $^{+3.7}_{-3.6}$ \ \ & \ \ $^{+5.2}_{-4.9}$ \ \ & \ \ 1.2 \ \ & \ \ 0.1 \ \ & \ \ 0.0 \ \ & \ \ 0.75 \ \ & \ \ 1.8 \ \ \\[1.5mm]
\ \ $9$ \ \ & \ \ $ 0.78- 0.86$ \ \ & \ \ $ 4.83\cdot 10^{-1}$ \ \ &\ \ 1.4 \ \ & \ \ 0.57 \ \  & \ \ $^{+5.4}_{-5.3}$ \ \ & \ \ $^{+0.6}_{-1.1}$ \ \ & \ \ $^{+3.5}_{-3.4}$ \ \ & \ \ $^{+4.9}_{-4.7}$ \ \ & \ \ 1.0 \ \ & \ \ 0.0 \ \ & \ \ 0.0 \ \ & \ \ 0.75 \ \ & \ \ 1.8 \ \ \\[1.5mm]
\ \ $10$ \ \ & \ \ $ 0.86- 0.94$ \ \ & \ \ $ 2.78\cdot 10^{-1}$ \ \ &\ \ 1.8 \ \ & \ \ 0.45 \ \  & \ \ $^{+5.3}_{-5.3}$ \ \ & \ \ $^{+0.6}_{-1.1}$ \ \ & \ \ $^{+3.2}_{-3.1}$ \ \ & \ \ $^{+4.6}_{-4.4}$ \ \ & \ \ 0.9 \ \ & \ \ 0.0 \ \ & \ \ 0.0 \ \ & \ \ 0.75 \ \ & \ \ 1.8 \ \ \\[1.5mm]
\ \ $11$ \ \ & \ \ $ 0.94- 1.02$ \ \ & \ \ $ 1.62\cdot 10^{-1}$ \ \ &\ \ 1.5 \ \ & \ \ 0.43 \ \  & \ \ $^{+5.3}_{-5.2}$ \ \ & \ \ $^{+0.6}_{-1.1}$ \ \ & \ \ $^{+2.9}_{-2.9}$ \ \ & \ \ $^{+4.4}_{-4.2}$ \ \ & \ \ 0.8 \ \ & \ \ 0.0 \ \ & \ \ 0.0 \ \ & \ \ 0.75 \ \ & \ \ 1.8 \ \ \\[1.5mm]
\ \ $12$ \ \ & \ \ $ 1.02- 1.12$ \ \ & \ \ $ 9.31\cdot 10^{-2}$ \ \ &\ \ 1.0 \ \ & \ \ 0.38 \ \  & \ \ $^{+5.3}_{-5.2}$ \ \ & \ \ $^{+0.5}_{-1.1}$ \ \ & \ \ $^{+2.6}_{-2.5}$ \ \ & \ \ $^{+4.1}_{-3.9}$ \ \ & \ \ 0.8 \ \ & \ \ 0.0 \ \ & \ \ 0.0 \ \ & \ \ 0.75 \ \ & \ \ 1.8 \ \ \\[1.5mm]
\ \ $13$ \ \ & \ \ $ 1.12- 1.22$ \ \ & \ \ $ 5.12\cdot 10^{-2}$ \ \ &\ \ 1.4 \ \ & \ \ 0.44 \ \  & \ \ $^{+5.4}_{-5.2}$ \ \ & \ \ $^{+0.3}_{-0.9}$ \ \ & \ \ $^{+2.3}_{-2.2}$ \ \ & \ \ $^{+3.9}_{-3.7}$ \ \ & \ \ 0.8 \ \ & \ \ 0.0 \ \ & \ \ 0.0 \ \ & \ \ 0.75 \ \ & \ \ 1.8 \ \ \\[1.5mm]
\ \ $14$ \ \ & \ \ $ 1.22- 1.34$ \ \ & \ \ $ 2.77\cdot 10^{-2}$ \ \ &\ \ 1.3 \ \ & \ \ 0.47 \ \  & \ \ $^{+5.7}_{-5.4}$ \ \ & \ \ $^{+0.2}_{-0.7}$ \ \ & \ \ $^{+2.0}_{-1.9}$ \ \ & \ \ $^{+3.6}_{-3.5}$ \ \ & \ \ 0.8 \ \ & \ \ 0.0 \ \ & \ \ 0.0 \ \ & \ \ 0.75 \ \ & \ \ 1.8 \ \ \\[1.5mm]
\ \ $15$ \ \ & \ \ $ 1.34- 1.46$ \ \ & \ \ $ 1.50\cdot 10^{-2}$ \ \ &\ \ 1.2 \ \ & \ \ 0.49 \ \  & \ \ $^{+5.9}_{-5.6}$ \ \ & \ \ $^{+0.1}_{-0.5}$ \ \ & \ \ $^{+1.7}_{-1.7}$ \ \ & \ \ $^{+3.4}_{-3.2}$ \ \ & \ \ 0.7 \ \ & \ \ 0.0 \ \ & \ \ 0.0 \ \ & \ \ 0.75 \ \ & \ \ 1.8 \ \ \\[1.5mm]
\ \ $16$ \ \ & \ \ $ 1.46- 1.60$ \ \ & \ \ $ 7.63\cdot 10^{-3}$ \ \ &\ \ 1.6 \ \ & \ \ 0.50 \ \  & \ \ $^{+6.4}_{-6.0}$ \ \ & \ \ $^{+0.2}_{-0.3}$ \ \ & \ \ $^{+1.6}_{-1.5}$ \ \ & \ \ $^{+3.3}_{-3.1}$ \ \ & \ \ 0.7 \ \ & \ \ 0.0 \ \ & \ \ 0.0 \ \ & \ \ 0.75 \ \ & \ \ 1.8 \ \ \\[1.5mm]
\ \ $17$ \ \ & \ \ $ 1.60- 1.76$ \ \ & \ \ $ 3.73\cdot 10^{-3}$ \ \ &\ \ 2.0 \ \ & \ \ 0.44 \ \  & \ \ $^{+7.0}_{-6.6}$ \ \ & \ \ $^{+0.2}_{-0.3}$ \ \ & \ \ $^{+1.4}_{-1.4}$ \ \ & \ \ $^{+3.1}_{-2.9}$ \ \ & \ \ 0.7 \ \ & \ \ 0.0 \ \ & \ \ 0.0 \ \ & \ \ 0.75 \ \ & \ \ 1.8 \ \ \\[1.5mm]
\ \ $18$ \ \ & \ \ $ 1.76- 1.94$ \ \ & \ \ $ 1.90\cdot 10^{-3}$ \ \ &\ \ 2.8 \ \ & \ \ 0.38 \ \  & \ \ $^{+7.8}_{-7.4}$ \ \ & \ \ $^{+0.3}_{-0.2}$ \ \ & \ \ $^{+1.3}_{-1.2}$ \ \ & \ \ $^{+3.0}_{-2.8}$ \ \ & \ \ 0.7 \ \ & \ \ 0.0 \ \ & \ \ 0.0 \ \ & \ \ 0.75 \ \ & \ \ 1.8 \ \ \\[1.5mm]
\ \ $19$ \ \ & \ \ $ 1.94- 2.14$ \ \ & \ \ $ 8.81\cdot 10^{-4}$ \ \ &\ \ 3.9 \ \ & \ \ 0.40 \ \  & \ \ $^{+8.8}_{-8.5}$ \ \ & \ \ $^{+0.4}_{-0.2}$ \ \ & \ \ $^{+1.2}_{-1.1}$ \ \ & \ \ $^{+2.9}_{-2.8}$ \ \ & \ \ 0.6 \ \ & \ \ 0.0 \ \ & \ \ 0.0 \ \ & \ \ 0.75 \ \ & \ \ 1.8 \ \ \\[1.5mm]
\ \ $20$ \ \ & \ \ $ 2.14- 2.36$ \ \ & \ \ $ 3.50\cdot 10^{-4}$ \ \ &\ \ 5.9 \ \ & \ \ 0.58 \ \  & \ \ $^{+10.0}_{-9.7}$ \ \ & \ \ $^{+0.4}_{-0.2}$ \ \ & \ \ $^{+1.1}_{-1.1}$ \ \ & \ \ $^{+2.9}_{-2.7}$ \ \ & \ \ 0.6 \ \ & \ \ 0.0 \ \ & \ \ 0.0 \ \ & \ \ 0.75 \ \ & \ \ 1.8 \ \ \\[1.5mm]
\ \ $21$ \ \ & \ \ $ 2.36- 2.60$ \ \ & \ \ $ 1.32\cdot 10^{-4}$ \ \ &\ \ 9.4 \ \ & \ \ 0.70 \ \  & \ \ $^{+11.2}_{-10.7}$ \ \ & \ \ $^{+0.4}_{-0.2}$ \ \ & \ \ $^{+1.0}_{-1.0}$ \ \ & \ \ $^{+2.8}_{-2.6}$ \ \ & \ \ 0.6 \ \ & \ \ 0.0 \ \ & \ \ 0.0 \ \ & \ \ 0.75 \ \ & \ \ 1.8 \ \ \\[1.5mm]
\ \ $22$ \ \ & \ \ $ 2.60- 2.84$ \ \ & \ \ $ 6.60\cdot 10^{-5}$ \ \ &\ \ 13 \ \ & \ \ 0.83 \ \  & \ \ $^{+12.6}_{-11.9}$ \ \ & \ \ $^{+0.4}_{-0.2}$ \ \ & \ \ $^{+1.0}_{-1.0}$ \ \ & \ \ $^{+2.8}_{-2.6}$ \ \ & \ \ 0.7 \ \ & \ \ 0.0 \ \ & \ \ 0.0 \ \ & \ \ 0.75 \ \ & \ \ 1.8 \ \ \\[1.5mm]
\ \ $23$ \ \ & \ \ $ 2.84- 3.10$ \ \ & \ \ $ 2.24\cdot 10^{-5}$ \ \ &\ \ 22 \ \ & \ \ 1.2 \ \  & \ \ $^{+15.8}_{-14.8}$ \ \ & \ \ $^{+0.4}_{-0.2}$ \ \ & \ \ $^{+1.0}_{-1.0}$ \ \ & \ \ $^{+2.8}_{-2.5}$ \ \ & \ \ 0.7 \ \ & \ \ 0.0 \ \ & \ \ 0.0 \ \ & \ \ 0.75 \ \ & \ \ 1.8 \ \ \\[1.5mm]
\ \ $24$ \ \ & \ \ $ 3.10- 3.90$ \ \ & \ \ $ 4.95\cdot 10^{-6}$ \ \ &\ \ 27 \ \ & \ \ 0.93 \ \  & \ \ $^{+26.9}_{-25.2}$ \ \ & \ \ $^{+0.4}_{-0.2}$ \ \ & \ \ $^{+1.0}_{-0.9}$ \ \ & \ \ $^{+2.6}_{-2.4}$ \ \ & \ \ 1.0 \ \ & \ \ 0.0 \ \ & \ \ 0.0 \ \ & \ \ 0.75 \ \ & \ \ 1.8 \ \ \\[1.5mm]
\hline
\hline
\end{tabular}
\caption{
   Measured double-differential \trijet{} \xs{}, $\sigma$, for $\rsix{}$~jets and $\ystar{}<2$, along with uncertainties in the measurement. 
   All uncertainties are given in \%, where 
   $\delta_\textrm{stat}^\textrm{data}$ ($\delta_\textrm{stat}^\textrm{MC}$) are the statistical uncertainties in the data (MC simulation). 
   The $\gamma$ components are the uncertainty in the jet energy calibration from the \insitu{}, the pileup, the close-by jet, and flavour components. 
   The $u$ components show the uncertainty for the jet energy and angular resolution, the unfolding, the quality selection, and the luminosity. 
   While all columns are uncorrelated with each other, the \insitu{}, pileup, and flavour uncertainties shown here are the sum in quadrature of multiple uncorrelated components. 
}
\label{tab:sysunc_r01_ystar0}
\end{table*}

\begin{table*}[!ht]\tiny\centering\begin{tabular}{@{}c@{}@{}c@{}@{}c@{}@{}c@{}@{}r@{}@{}r@{}@{}r@{}@{}r@{}@{}r@{}@{}r@{}@{}r@{}@{}r@{}@{}r@{}@{}r@{}}\hline\hline \\[-2mm]\ \ $m_{jjj}$ \ \ & \ \ $m_{jjj}$-range \ \ & \ \ $\sigma$ \ \ & \ \ $\delta_{\textrm{stat}}^{\textrm{data}}$ \ \ & \ \ $\delta_{\textrm{stat}}^{\textrm{MC}}$ \ \ & \ \ $\gamma_{\insituT}$ \ \ & \ \ $\gamma_{\textrm{pileup}}$ \ \ & \ \ $\gamma_{\textrm{close-by}}$ \ \ & \ \ $\gamma_{\textrm{flavour}}$ \ \ & \ \ $u_{\textrm{JER}}$ \ \ & \ \ $u_{\textrm{JAR}}$ \ \ & \ \ $u_{\textrm{unfold}}$ \ \ & \ \ $u_{\textrm{qual.}}$ \ \ & \ \ $u_{\textrm{lumi}}$ \ \ \\ \relax
\ \ bin \# \ \ & \ \ [TeV] \ \ & \ \ [pb/GeV] \ \ & \ \ [\%] \ \ & \ \ [\%] \ \ & \ \ [\%] \ \ & \ \ [\%] \ \ & \ \ [\%] \ \ & \ \ [\%] \ \ & \ \ [\%] \ \ & \ \ [\%] \ \ & \ \ [\%] \ \ & \ \ [\%] \ \ & \ \ [\%] \\
\hline \\[-1.5mm]
\ \ $1$ \ \ & \ \ $ 0.42- 0.46$ \ \ & \ \ 25.5 \ \ & \ \ 1.4 \ \ & \ \ 0.83 \ \  & \ \ $^{+6.6}_{-6.5}$ \ \ & \ \ $^{+0.2}_{-4.1}$ \ \ & \ \ $^{+2.9}_{-2.8}$ \ \ & \ \ $^{+7.0}_{-6.2}$ \ \ & \ \ 0.6 \ \ & \ \ 0.2 \ \ & \ \ 0.0 \ \ & \ \ 0.75 \ \ & \ \ 1.8 \ \ \\[1.5mm]
\ \ $2$ \ \ & \ \ $ 0.46- 0.50$ \ \ & \ \ 26.3 \ \ & \ \ 1.4 \ \ & \ \ 0.66 \ \  & \ \ $^{+6.6}_{-6.5}$ \ \ & \ \ $^{+0.2}_{-3.2}$ \ \ & \ \ $^{+2.8}_{-2.7}$ \ \ & \ \ $^{+6.9}_{-6.2}$ \ \ & \ \ 0.7 \ \ & \ \ 0.2 \ \ & \ \ 0.0 \ \ & \ \ 0.75 \ \ & \ \ 1.8 \ \ \\[1.5mm]
\ \ $3$ \ \ & \ \ $ 0.50- 0.54$ \ \ & \ \ 21.3 \ \ & \ \ 0.90 \ \ & \ \ 0.60 \ \  & \ \ $^{+6.7}_{-6.4}$ \ \ & \ \ $^{+0.2}_{-2.4}$ \ \ & \ \ $^{+2.8}_{-2.7}$ \ \ & \ \ $^{+6.7}_{-6.1}$ \ \ & \ \ 0.7 \ \ & \ \ 0.2 \ \ & \ \ 0.0 \ \ & \ \ 0.75 \ \ & \ \ 1.8 \ \ \\[1.5mm]
\ \ $4$ \ \ & \ \ $ 0.54- 0.60$ \ \ & \ \ 14.4 \ \ & \ \ 0.85 \ \ & \ \ 0.43 \ \  & \ \ $^{+6.7}_{-6.4}$ \ \ & \ \ $^{+0.2}_{-1.8}$ \ \ & \ \ $^{+2.8}_{-2.7}$ \ \ & \ \ $^{+6.5}_{-5.9}$ \ \ & \ \ 0.8 \ \ & \ \ 0.1 \ \ & \ \ 0.0 \ \ & \ \ 0.75 \ \ & \ \ 1.8 \ \ \\[1.5mm]
\ \ $5$ \ \ & \ \ $ 0.60- 0.66$ \ \ & \ \ 8.76 \ \ & \ \ 1.0 \ \ & \ \ 0.38 \ \  & \ \ $^{+6.6}_{-6.4}$ \ \ & \ \ $^{+0.1}_{-1.4}$ \ \ & \ \ $^{+2.8}_{-2.7}$ \ \ & \ \ $^{+6.2}_{-5.8}$ \ \ & \ \ 0.8 \ \ & \ \ 0.1 \ \ & \ \ 0.0 \ \ & \ \ 0.75 \ \ & \ \ 1.8 \ \ \\[1.5mm]
\ \ $6$ \ \ & \ \ $ 0.66- 0.72$ \ \ & \ \ 5.35 \ \ & \ \ 1.3 \ \ & \ \ 0.39 \ \  & \ \ $^{+6.5}_{-6.3}$ \ \ & \ \ $^{+0.1}_{-1.2}$ \ \ & \ \ $^{+2.9}_{-2.7}$ \ \ & \ \ $^{+5.9}_{-5.6}$ \ \ & \ \ 0.8 \ \ & \ \ 0.1 \ \ & \ \ 0.0 \ \ & \ \ 0.75 \ \ & \ \ 1.8 \ \ \\[1.5mm]
\ \ $7$ \ \ & \ \ $ 0.72- 0.78$ \ \ & \ \ 3.27 \ \ & \ \ 1.6 \ \ & \ \ 0.46 \ \  & \ \ $^{+6.4}_{-6.2}$ \ \ & \ \ $^{+0.1}_{-1.1}$ \ \ & \ \ $^{+2.9}_{-2.8}$ \ \ & \ \ $^{+5.6}_{-5.3}$ \ \ & \ \ 0.8 \ \ & \ \ 0.1 \ \ & \ \ 0.0 \ \ & \ \ 0.75 \ \ & \ \ 1.8 \ \ \\[1.5mm]
\ \ $8$ \ \ & \ \ $ 0.78- 0.86$ \ \ & \ \ 1.95 \ \ & \ \ 1.3 \ \ & \ \ 0.45 \ \  & \ \ $^{+6.3}_{-6.1}$ \ \ & \ \ $^{+0.2}_{-1.1}$ \ \ & \ \ $^{+2.9}_{-2.8}$ \ \ & \ \ $^{+5.4}_{-5.1}$ \ \ & \ \ 0.8 \ \ & \ \ 0.0 \ \ & \ \ 0.0 \ \ & \ \ 0.75 \ \ & \ \ 1.8 \ \ \\[1.5mm]
\ \ $9$ \ \ & \ \ $ 0.86- 0.94$ \ \ & \ \ 1.11 \ \ & \ \ 0.96 \ \ & \ \ 0.47 \ \  & \ \ $^{+6.1}_{-6.0}$ \ \ & \ \ $^{+0.4}_{-1.1}$ \ \ & \ \ $^{+2.8}_{-2.8}$ \ \ & \ \ $^{+5.1}_{-4.9}$ \ \ & \ \ 0.7 \ \ & \ \ 0.0 \ \ & \ \ 0.0 \ \ & \ \ 0.75 \ \ & \ \ 1.8 \ \ \\[1.5mm]
\ \ $10$ \ \ & \ \ $ 0.94- 1.02$ \ \ & \ \ $ 6.73\cdot 10^{-1}$ \ \ &\ \ 1.1 \ \ & \ \ 0.44 \ \  & \ \ $^{+6.0}_{-5.9}$ \ \ & \ \ $^{+0.5}_{-1.2}$ \ \ & \ \ $^{+2.8}_{-2.7}$ \ \ & \ \ $^{+4.9}_{-4.7}$ \ \ & \ \ 0.7 \ \ & \ \ 0.0 \ \ & \ \ 0.0 \ \ & \ \ 0.75 \ \ & \ \ 1.8 \ \ \\[1.5mm]
\ \ $11$ \ \ & \ \ $ 1.02- 1.12$ \ \ & \ \ $ 3.87\cdot 10^{-1}$ \ \ &\ \ 0.57 \ \ & \ \ 0.39 \ \  & \ \ $^{+5.9}_{-5.8}$ \ \ & \ \ $^{+0.6}_{-1.1}$ \ \ & \ \ $^{+2.6}_{-2.6}$ \ \ & \ \ $^{+4.7}_{-4.4}$ \ \ & \ \ 0.7 \ \ & \ \ 0.0 \ \ & \ \ 0.0 \ \ & \ \ 0.75 \ \ & \ \ 1.8 \ \ \\[1.5mm]
\ \ $12$ \ \ & \ \ $ 1.12- 1.22$ \ \ & \ \ $ 2.14\cdot 10^{-1}$ \ \ &\ \ 0.65 \ \ & \ \ 0.34 \ \  & \ \ $^{+5.9}_{-5.7}$ \ \ & \ \ $^{+0.5}_{-1.0}$ \ \ & \ \ $^{+2.4}_{-2.4}$ \ \ & \ \ $^{+4.5}_{-4.2}$ \ \ & \ \ 0.7 \ \ & \ \ 0.0 \ \ & \ \ 0.0 \ \ & \ \ 0.75 \ \ & \ \ 1.8 \ \ \\[1.5mm]
\ \ $13$ \ \ & \ \ $ 1.22- 1.34$ \ \ & \ \ $ 1.20\cdot 10^{-1}$ \ \ &\ \ 0.75 \ \ & \ \ 0.30 \ \  & \ \ $^{+6.0}_{-5.7}$ \ \ & \ \ $^{+0.4}_{-0.7}$ \ \ & \ \ $^{+2.2}_{-2.1}$ \ \ & \ \ $^{+4.3}_{-4.0}$ \ \ & \ \ 0.7 \ \ & \ \ 0.0 \ \ & \ \ 0.0 \ \ & \ \ 0.75 \ \ & \ \ 1.8 \ \ \\[1.5mm]
\ \ $14$ \ \ & \ \ $ 1.34- 1.46$ \ \ & \ \ $ 6.32\cdot 10^{-2}$ \ \ &\ \ 0.59 \ \ & \ \ 0.33 \ \  & \ \ $^{+6.1}_{-5.9}$ \ \ & \ \ $^{+0.3}_{-0.5}$ \ \ & \ \ $^{+1.9}_{-1.9}$ \ \ & \ \ $^{+4.1}_{-3.9}$ \ \ & \ \ 0.7 \ \ & \ \ 0.0 \ \ & \ \ 0.0 \ \ & \ \ 0.75 \ \ & \ \ 1.8 \ \ \\[1.5mm]
\ \ $15$ \ \ & \ \ $ 1.46- 1.60$ \ \ & \ \ $ 3.37\cdot 10^{-2}$ \ \ &\ \ 0.77 \ \ & \ \ 0.34 \ \  & \ \ $^{+6.3}_{-6.1}$ \ \ & \ \ $^{+0.3}_{-0.4}$ \ \ & \ \ $^{+1.7}_{-1.6}$ \ \ & \ \ $^{+4.0}_{-3.8}$ \ \ & \ \ 0.6 \ \ & \ \ 0.0 \ \ & \ \ 0.0 \ \ & \ \ 0.75 \ \ & \ \ 1.8 \ \ \\[1.5mm]
\ \ $16$ \ \ & \ \ $ 1.60- 1.74$ \ \ & \ \ $ 1.78\cdot 10^{-2}$ \ \ &\ \ 0.98 \ \ & \ \ 0.41 \ \  & \ \ $^{+6.6}_{-6.4}$ \ \ & \ \ $^{+0.3}_{-0.4}$ \ \ & \ \ $^{+1.4}_{-1.4}$ \ \ & \ \ $^{+3.8}_{-3.6}$ \ \ & \ \ 0.6 \ \ & \ \ 0.0 \ \ & \ \ 0.0 \ \ & \ \ 0.75 \ \ & \ \ 1.8 \ \ \\[1.5mm]
\ \ $17$ \ \ & \ \ $ 1.74- 1.90$ \ \ & \ \ $ 9.00\cdot 10^{-3}$ \ \ &\ \ 1.4 \ \ & \ \ 0.45 \ \  & \ \ $^{+7.0}_{-6.9}$ \ \ & \ \ $^{+0.4}_{-0.4}$ \ \ & \ \ $^{+1.2}_{-1.2}$ \ \ & \ \ $^{+3.7}_{-3.5}$ \ \ & \ \ 0.6 \ \ & \ \ 0.0 \ \ & \ \ 0.0 \ \ & \ \ 0.75 \ \ & \ \ 1.8 \ \ \\[1.5mm]
\ \ $18$ \ \ & \ \ $ 1.90- 2.08$ \ \ & \ \ $ 4.30\cdot 10^{-3}$ \ \ &\ \ 1.9 \ \ & \ \ 0.43 \ \  & \ \ $^{+7.6}_{-7.5}$ \ \ & \ \ $^{+0.3}_{-0.4}$ \ \ & \ \ $^{+1.1}_{-1.1}$ \ \ & \ \ $^{+3.6}_{-3.4}$ \ \ & \ \ 0.6 \ \ & \ \ 0.0 \ \ & \ \ 0.0 \ \ & \ \ 0.75 \ \ & \ \ 1.8 \ \ \\[1.5mm]
\ \ $19$ \ \ & \ \ $ 2.08- 2.26$ \ \ & \ \ $ 2.13\cdot 10^{-3}$ \ \ &\ \ 2.6 \ \ & \ \ 0.42 \ \  & \ \ $^{+8.4}_{-8.2}$ \ \ & \ \ $^{+0.3}_{-0.3}$ \ \ & \ \ $^{+0.9}_{-1.0}$ \ \ & \ \ $^{+3.5}_{-3.4}$ \ \ & \ \ 0.7 \ \ & \ \ 0.0 \ \ & \ \ 0.0 \ \ & \ \ 0.75 \ \ & \ \ 1.8 \ \ \\[1.5mm]
\ \ $20$ \ \ & \ \ $ 2.26- 2.48$ \ \ & \ \ $ 9.74\cdot 10^{-4}$ \ \ &\ \ 3.5 \ \ & \ \ 0.44 \ \  & \ \ $^{+9.4}_{-9.2}$ \ \ & \ \ $^{+0.3}_{-0.2}$ \ \ & \ \ $^{+0.9}_{-0.9}$ \ \ & \ \ $^{+3.4}_{-3.3}$ \ \ & \ \ 0.7 \ \ & \ \ 0.0 \ \ & \ \ 0.0 \ \ & \ \ 0.75 \ \ & \ \ 1.8 \ \ \\[1.5mm]
\ \ $21$ \ \ & \ \ $ 2.48- 2.72$ \ \ & \ \ $ 3.74\cdot 10^{-4}$ \ \ &\ \ 5.4 \ \ & \ \ 0.56 \ \  & \ \ $^{+10.7}_{-10.4}$ \ \ & \ \ $^{+0.2}_{-0.2}$ \ \ & \ \ $^{+0.8}_{-0.8}$ \ \ & \ \ $^{+3.3}_{-3.2}$ \ \ & \ \ 0.8 \ \ & \ \ 0.0 \ \ & \ \ 0.0 \ \ & \ \ 0.75 \ \ & \ \ 1.8 \ \ \\[1.5mm]
\ \ $22$ \ \ & \ \ $ 2.72- 2.98$ \ \ & \ \ $ 1.33\cdot 10^{-4}$ \ \ &\ \ 8.8 \ \ & \ \ 0.72 \ \  & \ \ $^{+12.1}_{-11.6}$ \ \ & \ \ $^{+0.3}_{-0.2}$ \ \ & \ \ $^{+0.8}_{-0.8}$ \ \ & \ \ $^{+3.2}_{-3.2}$ \ \ & \ \ 0.8 \ \ & \ \ 0.0 \ \ & \ \ 0.0 \ \ & \ \ 0.75 \ \ & \ \ 1.8 \ \ \\[1.5mm]
\ \ $23$ \ \ & \ \ $ 2.98- 3.26$ \ \ & \ \ $ 5.84\cdot 10^{-5}$ \ \ &\ \ 13 \ \ & \ \ 0.77 \ \  & \ \ $^{+13.9}_{-13.2}$ \ \ & \ \ $^{+0.5}_{-0.2}$ \ \ & \ \ $^{+0.7}_{-0.7}$ \ \ & \ \ $^{+3.2}_{-3.1}$ \ \ & \ \ 0.8 \ \ & \ \ 0.0 \ \ & \ \ 0.1 \ \ & \ \ 0.75 \ \ & \ \ 1.8 \ \ \\[1.5mm]
\ \ $24$ \ \ & \ \ $ 3.26- 3.58$ \ \ & \ \ $ 1.52\cdot 10^{-5}$ \ \ &\ \ 23 \ \ & \ \ 1.4 \ \  & \ \ $^{+17.2}_{-16.9}$ \ \ & \ \ $^{+0.5}_{-0.2}$ \ \ & \ \ $^{+0.7}_{-0.7}$ \ \ & \ \ $^{+3.2}_{-3.0}$ \ \ & \ \ 0.9 \ \ & \ \ 0.0 \ \ & \ \ 0.1 \ \ & \ \ 0.75 \ \ & \ \ 1.8 \ \ \\[1.5mm]
\ \ $25$ \ \ & \ \ $ 3.58- 4.20$ \ \ & \ \ $ 5.57\cdot 10^{-6}$ \ \ &\ \ 29 \ \ & \ \ 1.2 \ \  & \ \ $^{+26.2}_{-26.6}$ \ \ & \ \ $^{+0.7}_{-0.2}$ \ \ & \ \ $^{+0.6}_{-0.6}$ \ \ & \ \ $^{+2.6}_{-2.8}$ \ \ & \ \ 2.1 \ \ & \ \ 0.0 \ \ & \ \ 0.3 \ \ & \ \ 0.75 \ \ & \ \ 1.8 \ \ \\[1.5mm]
\hline
\hline
\end{tabular}
\caption{
   Measured double-differential \trijet{} \xs{}, $\sigma$, for $\rfour{}$~jets and $2\leq\ystar{}<4$, along with uncertainties in the measurement. 
   All uncertainties are given in \%, where 
   $\delta_\textrm{stat}^\textrm{data}$ ($\delta_\textrm{stat}^\textrm{MC}$) are the statistical uncertainties in the data (MC simulation). 
   The $\gamma$ components are the uncertainty in the jet energy calibration from the \insitu{}, the pileup, the close-by jet, and flavour components. 
   The $u$ components show the uncertainty for the jet energy and angular resolution, the unfolding, the quality selection, and the luminosity. 
   While all columns are uncorrelated with each other, the \insitu{}, pileup, and flavour uncertainties shown here are the sum in quadrature of multiple uncorrelated components. 
}
\label{tab:sysunc_r00_ystar1}
\end{table*}

\begin{table*}[!ht]\tiny\centering\begin{tabular}{@{}c@{}@{}c@{}@{}c@{}@{}c@{}@{}r@{}@{}r@{}@{}r@{}@{}r@{}@{}r@{}@{}r@{}@{}r@{}@{}r@{}@{}r@{}@{}r@{}}\hline\hline \\[-2mm]\ \ $m_{jjj}$ \ \ & \ \ $m_{jjj}$-range \ \ & \ \ $\sigma$ \ \ & \ \ $\delta_{\textrm{stat}}^{\textrm{data}}$ \ \ & \ \ $\delta_{\textrm{stat}}^{\textrm{MC}}$ \ \ & \ \ $\gamma_{\insituT}$ \ \ & \ \ $\gamma_{\textrm{pileup}}$ \ \ & \ \ $\gamma_{\textrm{close-by}}$ \ \ & \ \ $\gamma_{\textrm{flavour}}$ \ \ & \ \ $u_{\textrm{JER}}$ \ \ & \ \ $u_{\textrm{JAR}}$ \ \ & \ \ $u_{\textrm{unfold}}$ \ \ & \ \ $u_{\textrm{qual.}}$ \ \ & \ \ $u_{\textrm{lumi}}$ \ \ \\ \relax
\ \ bin \# \ \ & \ \ [TeV] \ \ & \ \ [pb/GeV] \ \ & \ \ [\%] \ \ & \ \ [\%] \ \ & \ \ [\%] \ \ & \ \ [\%] \ \ & \ \ [\%] \ \ & \ \ [\%] \ \ & \ \ [\%] \ \ & \ \ [\%] \ \ & \ \ [\%] \ \ & \ \ [\%] \ \ & \ \ [\%] \\
\hline \\[-1.5mm]
\ \ $1$ \ \ & \ \ $ 0.42- 0.46$ \ \ & \ \ 36.3 \ \ & \ \ 2.1 \ \ & \ \ 0.90 \ \  & \ \ $^{+7.0}_{-7.2}$ \ \ & \ \ $^{+0.7}_{-4.4}$ \ \ & \ \ $^{+4.0}_{-4.0}$ \ \ & \ \ $^{+6.9}_{-6.8}$ \ \ & \ \ 1.8 \ \ & \ \ 0.8 \ \ & \ \ 0.0 \ \ & \ \ 0.75 \ \ & \ \ 1.8 \ \ \\[1.5mm]
\ \ $2$ \ \ & \ \ $ 0.46- 0.50$ \ \ & \ \ 35.9 \ \ & \ \ 2.0 \ \ & \ \ 0.76 \ \  & \ \ $^{+7.0}_{-7.1}$ \ \ & \ \ $^{+0.5}_{-3.3}$ \ \ & \ \ $^{+4.0}_{-3.9}$ \ \ & \ \ $^{+6.8}_{-6.6}$ \ \ & \ \ 1.9 \ \ & \ \ 0.8 \ \ & \ \ 0.0 \ \ & \ \ 0.75 \ \ & \ \ 1.8 \ \ \\[1.5mm]
\ \ $3$ \ \ & \ \ $ 0.50- 0.54$ \ \ & \ \ 29.5 \ \ & \ \ 2.1 \ \ & \ \ 0.76 \ \  & \ \ $^{+6.9}_{-7.0}$ \ \ & \ \ $^{+0.4}_{-2.5}$ \ \ & \ \ $^{+3.9}_{-3.8}$ \ \ & \ \ $^{+6.7}_{-6.5}$ \ \ & \ \ 1.9 \ \ & \ \ 0.8 \ \ & \ \ 0.0 \ \ & \ \ 0.75 \ \ & \ \ 1.8 \ \ \\[1.5mm]
\ \ $4$ \ \ & \ \ $ 0.54- 0.60$ \ \ & \ \ 19.7 \ \ & \ \ 1.8 \ \ & \ \ 0.59 \ \  & \ \ $^{+6.9}_{-6.9}$ \ \ & \ \ $^{+0.4}_{-1.9}$ \ \ & \ \ $^{+3.8}_{-3.7}$ \ \ & \ \ $^{+6.5}_{-6.3}$ \ \ & \ \ 1.8 \ \ & \ \ 0.7 \ \ & \ \ 0.0 \ \ & \ \ 0.75 \ \ & \ \ 1.8 \ \ \\[1.5mm]
\ \ $5$ \ \ & \ \ $ 0.60- 0.66$ \ \ & \ \ 11.6 \ \ & \ \ 1.7 \ \ & \ \ 0.51 \ \  & \ \ $^{+6.8}_{-6.7}$ \ \ & \ \ $^{+0.3}_{-1.6}$ \ \ & \ \ $^{+3.7}_{-3.6}$ \ \ & \ \ $^{+6.2}_{-6.0}$ \ \ & \ \ 1.7 \ \ & \ \ 0.6 \ \ & \ \ 0.0 \ \ & \ \ 0.75 \ \ & \ \ 1.8 \ \ \\[1.5mm]
\ \ $6$ \ \ & \ \ $ 0.66- 0.72$ \ \ & \ \ 6.99 \ \ & \ \ 2.0 \ \ & \ \ 0.40 \ \  & \ \ $^{+6.7}_{-6.5}$ \ \ & \ \ $^{+0.3}_{-1.4}$ \ \ & \ \ $^{+3.6}_{-3.4}$ \ \ & \ \ $^{+6.0}_{-5.7}$ \ \ & \ \ 1.6 \ \ & \ \ 0.5 \ \ & \ \ 0.0 \ \ & \ \ 0.75 \ \ & \ \ 1.8 \ \ \\[1.5mm]
\ \ $7$ \ \ & \ \ $ 0.72- 0.78$ \ \ & \ \ 4.20 \ \ & \ \ 2.0 \ \ & \ \ 0.47 \ \  & \ \ $^{+6.5}_{-6.3}$ \ \ & \ \ $^{+0.4}_{-1.4}$ \ \ & \ \ $^{+3.5}_{-3.3}$ \ \ & \ \ $^{+5.8}_{-5.4}$ \ \ & \ \ 1.5 \ \ & \ \ 0.4 \ \ & \ \ 0.0 \ \ & \ \ 0.75 \ \ & \ \ 1.8 \ \ \\[1.5mm]
\ \ $8$ \ \ & \ \ $ 0.78- 0.86$ \ \ & \ \ 2.55 \ \ & \ \ 1.7 \ \ & \ \ 0.42 \ \  & \ \ $^{+6.4}_{-6.1}$ \ \ & \ \ $^{+0.5}_{-1.3}$ \ \ & \ \ $^{+3.4}_{-3.1}$ \ \ & \ \ $^{+5.5}_{-5.2}$ \ \ & \ \ 1.4 \ \ & \ \ 0.3 \ \ & \ \ 0.0 \ \ & \ \ 0.75 \ \ & \ \ 1.8 \ \ \\[1.5mm]
\ \ $9$ \ \ & \ \ $ 0.86- 0.94$ \ \ & \ \ 1.43 \ \ & \ \ 2.0 \ \ & \ \ 0.47 \ \  & \ \ $^{+6.3}_{-6.0}$ \ \ & \ \ $^{+0.5}_{-1.3}$ \ \ & \ \ $^{+3.2}_{-3.0}$ \ \ & \ \ $^{+5.3}_{-4.9}$ \ \ & \ \ 1.3 \ \ & \ \ 0.3 \ \ & \ \ 0.0 \ \ & \ \ 0.75 \ \ & \ \ 1.8 \ \ \\[1.5mm]
\ \ $10$ \ \ & \ \ $ 0.94- 1.02$ \ \ & \ \ $ 8.60\cdot 10^{-1}$ \ \ &\ \ 1.0 \ \ & \ \ 0.47 \ \  & \ \ $^{+6.1}_{-5.9}$ \ \ & \ \ $^{+0.6}_{-1.4}$ \ \ & \ \ $^{+3.0}_{-2.8}$ \ \ & \ \ $^{+5.0}_{-4.7}$ \ \ & \ \ 1.2 \ \ & \ \ 0.2 \ \ & \ \ 0.0 \ \ & \ \ 0.75 \ \ & \ \ 1.8 \ \ \\[1.5mm]
\ \ $11$ \ \ & \ \ $ 1.02- 1.12$ \ \ & \ \ $ 4.96\cdot 10^{-1}$ \ \ &\ \ 1.2 \ \ & \ \ 0.41 \ \  & \ \ $^{+6.0}_{-5.8}$ \ \ & \ \ $^{+0.6}_{-1.4}$ \ \ & \ \ $^{+2.8}_{-2.6}$ \ \ & \ \ $^{+4.7}_{-4.5}$ \ \ & \ \ 1.1 \ \ & \ \ 0.2 \ \ & \ \ 0.0 \ \ & \ \ 0.75 \ \ & \ \ 1.8 \ \ \\[1.5mm]
\ \ $12$ \ \ & \ \ $ 1.12- 1.22$ \ \ & \ \ $ 2.66\cdot 10^{-1}$ \ \ &\ \ 1.5 \ \ & \ \ 0.37 \ \  & \ \ $^{+5.9}_{-5.7}$ \ \ & \ \ $^{+0.7}_{-1.4}$ \ \ & \ \ $^{+2.5}_{-2.4}$ \ \ & \ \ $^{+4.4}_{-4.2}$ \ \ & \ \ 1.0 \ \ & \ \ 0.1 \ \ & \ \ 0.0 \ \ & \ \ 0.75 \ \ & \ \ 1.8 \ \ \\[1.5mm]
\ \ $13$ \ \ & \ \ $ 1.22- 1.34$ \ \ & \ \ $ 1.47\cdot 10^{-1}$ \ \ &\ \ 0.79 \ \ & \ \ 0.32 \ \  & \ \ $^{+5.9}_{-5.8}$ \ \ & \ \ $^{+0.7}_{-1.2}$ \ \ & \ \ $^{+2.2}_{-2.2}$ \ \ & \ \ $^{+4.1}_{-4.0}$ \ \ & \ \ 0.9 \ \ & \ \ 0.1 \ \ & \ \ 0.0 \ \ & \ \ 0.75 \ \ & \ \ 1.8 \ \ \\[1.5mm]
\ \ $14$ \ \ & \ \ $ 1.34- 1.46$ \ \ & \ \ $ 7.89\cdot 10^{-2}$ \ \ &\ \ 0.95 \ \ & \ \ 0.32 \ \  & \ \ $^{+6.0}_{-5.8}$ \ \ & \ \ $^{+0.6}_{-1.0}$ \ \ & \ \ $^{+2.0}_{-1.9}$ \ \ & \ \ $^{+3.9}_{-3.8}$ \ \ & \ \ 0.9 \ \ & \ \ 0.1 \ \ & \ \ 0.0 \ \ & \ \ 0.75 \ \ & \ \ 1.8 \ \ \\[1.5mm]
\ \ $15$ \ \ & \ \ $ 1.46- 1.60$ \ \ & \ \ $ 4.08\cdot 10^{-2}$ \ \ &\ \ 1.1 \ \ & \ \ 0.34 \ \  & \ \ $^{+6.2}_{-6.0}$ \ \ & \ \ $^{+0.4}_{-0.7}$ \ \ & \ \ $^{+1.7}_{-1.7}$ \ \ & \ \ $^{+3.7}_{-3.5}$ \ \ & \ \ 0.9 \ \ & \ \ 0.0 \ \ & \ \ 0.0 \ \ & \ \ 0.75 \ \ & \ \ 1.8 \ \ \\[1.5mm]
\ \ $16$ \ \ & \ \ $ 1.60- 1.74$ \ \ & \ \ $ 2.15\cdot 10^{-2}$ \ \ &\ \ 0.90 \ \ & \ \ 0.39 \ \  & \ \ $^{+6.6}_{-6.3}$ \ \ & \ \ $^{+0.3}_{-0.5}$ \ \ & \ \ $^{+1.5}_{-1.4}$ \ \ & \ \ $^{+3.5}_{-3.3}$ \ \ & \ \ 0.9 \ \ & \ \ 0.0 \ \ & \ \ 0.0 \ \ & \ \ 0.75 \ \ & \ \ 1.8 \ \ \\[1.5mm]
\ \ $17$ \ \ & \ \ $ 1.74- 1.90$ \ \ & \ \ $ 1.09\cdot 10^{-2}$ \ \ &\ \ 1.2 \ \ & \ \ 0.46 \ \  & \ \ $^{+7.1}_{-6.8}$ \ \ & \ \ $^{+0.2}_{-0.3}$ \ \ & \ \ $^{+1.3}_{-1.3}$ \ \ & \ \ $^{+3.3}_{-3.2}$ \ \ & \ \ 0.8 \ \ & \ \ 0.0 \ \ & \ \ 0.0 \ \ & \ \ 0.75 \ \ & \ \ 1.8 \ \ \\[1.5mm]
\ \ $18$ \ \ & \ \ $ 1.90- 2.08$ \ \ & \ \ $ 5.29\cdot 10^{-3}$ \ \ &\ \ 1.7 \ \ & \ \ 0.49 \ \  & \ \ $^{+7.7}_{-7.4}$ \ \ & \ \ $^{+0.2}_{-0.2}$ \ \ & \ \ $^{+1.2}_{-1.1}$ \ \ & \ \ $^{+3.2}_{-3.0}$ \ \ & \ \ 0.8 \ \ & \ \ 0.0 \ \ & \ \ 0.0 \ \ & \ \ 0.75 \ \ & \ \ 1.8 \ \ \\[1.5mm]
\ \ $19$ \ \ & \ \ $ 2.08- 2.26$ \ \ & \ \ $ 2.53\cdot 10^{-3}$ \ \ &\ \ 2.4 \ \ & \ \ 0.53 \ \  & \ \ $^{+8.5}_{-8.1}$ \ \ & \ \ $^{+0.2}_{-0.1}$ \ \ & \ \ $^{+1.1}_{-1.0}$ \ \ & \ \ $^{+3.1}_{-2.9}$ \ \ & \ \ 0.9 \ \ & \ \ 0.0 \ \ & \ \ 0.0 \ \ & \ \ 0.75 \ \ & \ \ 1.8 \ \ \\[1.5mm]
\ \ $20$ \ \ & \ \ $ 2.26- 2.48$ \ \ & \ \ $ 1.17\cdot 10^{-3}$ \ \ &\ \ 3.2 \ \ & \ \ 0.48 \ \  & \ \ $^{+9.5}_{-9.2}$ \ \ & \ \ $^{+0.1}_{-0.1}$ \ \ & \ \ $^{+1.0}_{-0.9}$ \ \ & \ \ $^{+3.0}_{-2.9}$ \ \ & \ \ 0.9 \ \ & \ \ 0.0 \ \ & \ \ 0.0 \ \ & \ \ 0.75 \ \ & \ \ 1.8 \ \ \\[1.5mm]
\ \ $21$ \ \ & \ \ $ 2.48- 2.72$ \ \ & \ \ $ 4.65\cdot 10^{-4}$ \ \ &\ \ 5.1 \ \ & \ \ 0.50 \ \  & \ \ $^{+10.7}_{-10.5}$ \ \ & \ \ $^{+0.1}_{-0.1}$ \ \ & \ \ $^{+0.9}_{-0.9}$ \ \ & \ \ $^{+2.9}_{-2.9}$ \ \ & \ \ 0.9 \ \ & \ \ 0.0 \ \ & \ \ 0.0 \ \ & \ \ 0.75 \ \ & \ \ 1.8 \ \ \\[1.5mm]
\ \ $22$ \ \ & \ \ $ 2.72- 2.98$ \ \ & \ \ $ 1.58\cdot 10^{-4}$ \ \ &\ \ 7.9 \ \ & \ \ 0.61 \ \  & \ \ $^{+12.2}_{-11.9}$ \ \ & \ \ $^{+0.1}_{-0.1}$ \ \ & \ \ $^{+0.9}_{-0.8}$ \ \ & \ \ $^{+3.0}_{-2.9}$ \ \ & \ \ 0.9 \ \ & \ \ 0.0 \ \ & \ \ 0.0 \ \ & \ \ 0.75 \ \ & \ \ 1.8 \ \ \\[1.5mm]
\ \ $23$ \ \ & \ \ $ 2.98- 3.26$ \ \ & \ \ $ 7.07\cdot 10^{-5}$ \ \ &\ \ 12 \ \ & \ \ 0.75 \ \  & \ \ $^{+13.8}_{-13.6}$ \ \ & \ \ $^{+0.1}_{-0.1}$ \ \ & \ \ $^{+0.8}_{-0.8}$ \ \ & \ \ $^{+3.0}_{-2.9}$ \ \ & \ \ 0.8 \ \ & \ \ 0.0 \ \ & \ \ 0.0 \ \ & \ \ 0.75 \ \ & \ \ 1.8 \ \ \\[1.5mm]
\ \ $24$ \ \ & \ \ $ 3.26- 3.58$ \ \ & \ \ $ 1.42\cdot 10^{-5}$ \ \ &\ \ 22 \ \ & \ \ 1.6 \ \  & \ \ $^{+17.4}_{-16.8}$ \ \ & \ \ $^{+0.1}_{-0.1}$ \ \ & \ \ $^{+0.8}_{-0.7}$ \ \ & \ \ $^{+3.1}_{-3.0}$ \ \ & \ \ 0.8 \ \ & \ \ 0.0 \ \ & \ \ 0.0 \ \ & \ \ 0.75 \ \ & \ \ 1.8 \ \ \\[1.5mm]
\ \ $25$ \ \ & \ \ $ 3.58- 4.20$ \ \ & \ \ $ 6.19\cdot 10^{-6}$ \ \ &\ \ 28 \ \ & \ \ 1.2 \ \  & \ \ $^{+28.3}_{-25.0}$ \ \ & \ \ $^{+0.1}_{-0.1}$ \ \ & \ \ $^{+0.9}_{-0.7}$ \ \ & \ \ $^{+3.6}_{-3.3}$ \ \ & \ \ 2.8 \ \ & \ \ 0.0 \ \ & \ \ 0.0 \ \ & \ \ 0.75 \ \ & \ \ 1.8 \ \ \\[1.5mm]
\hline
\hline
\end{tabular}
\caption{
   Measured double-differential \trijet{} \xs{}, $\sigma$, for $\rsix{}$~jets and $2\leq\ystar{}<4$, along with uncertainties in the measurement. 
   All uncertainties are given in \%, where 
   $\delta_\textrm{stat}^\textrm{data}$ ($\delta_\textrm{stat}^\textrm{MC}$) are the statistical uncertainties in the data (MC simulation). 
   The $\gamma$ components are the uncertainty in the jet energy calibration from the \insitu{}, the pileup, the close-by jet, and flavour components. 
   The $u$ components show the uncertainty for the jet energy and angular resolution, the unfolding, the quality selection, and the luminosity. 
   While all columns are uncorrelated with each other, the \insitu{}, pileup, and flavour uncertainties shown here are the sum in quadrature of multiple uncorrelated components. 
}
\label{tab:sysunc_r01_ystar1}
\end{table*}

\begin{table*}[!ht]\tiny\centering\begin{tabular}{@{}c@{}@{}c@{}@{}c@{}@{}c@{}@{}r@{}@{}r@{}@{}r@{}@{}r@{}@{}r@{}@{}r@{}@{}r@{}@{}r@{}@{}r@{}@{}r@{}}\hline\hline \\[-2mm]\ \ $m_{jjj}$ \ \ & \ \ $m_{jjj}$-range \ \ & \ \ $\sigma$ \ \ & \ \ $\delta_{\textrm{stat}}^{\textrm{data}}$ \ \ & \ \ $\delta_{\textrm{stat}}^{\textrm{MC}}$ \ \ & \ \ $\gamma_{\insituT}$ \ \ & \ \ $\gamma_{\textrm{pileup}}$ \ \ & \ \ $\gamma_{\textrm{close-by}}$ \ \ & \ \ $\gamma_{\textrm{flavour}}$ \ \ & \ \ $u_{\textrm{JER}}$ \ \ & \ \ $u_{\textrm{JAR}}$ \ \ & \ \ $u_{\textrm{unfold}}$ \ \ & \ \ $u_{\textrm{qual.}}$ \ \ & \ \ $u_{\textrm{lumi}}$ \ \ \\ \relax
\ \ bin \# \ \ & \ \ [TeV] \ \ & \ \ [pb/GeV] \ \ & \ \ [\%] \ \ & \ \ [\%] \ \ & \ \ [\%] \ \ & \ \ [\%] \ \ & \ \ [\%] \ \ & \ \ [\%] \ \ & \ \ [\%] \ \ & \ \ [\%] \ \ & \ \ [\%] \ \ & \ \ [\%] \ \ & \ \ [\%] \\
\hline \\[-1.5mm]
\ \ $1$ \ \ & \ \ $ 0.54- 0.60$ \ \ & \ \ 15.0 \ \ & \ \ 1.6 \ \ & \ \ 1.0 \ \  & \ \ $^{+8.2}_{-7.6}$ \ \ & \ \ $^{+0.0}_{-4.1}$ \ \ & \ \ $^{+2.7}_{-2.5}$ \ \ & \ \ $^{+7.2}_{-6.4}$ \ \ & \ \ 0.2 \ \ & \ \ 0.1 \ \ & \ \ 0.0 \ \ & \ \ 0.75 \ \ & \ \ 1.8 \ \ \\[1.5mm]
\ \ $2$ \ \ & \ \ $ 0.60- 0.66$ \ \ & \ \ 15.0 \ \ & \ \ 1.6 \ \ & \ \ 0.81 \ \  & \ \ $^{+8.2}_{-7.8}$ \ \ & \ \ $^{+0.0}_{-3.2}$ \ \ & \ \ $^{+2.7}_{-2.6}$ \ \ & \ \ $^{+6.8}_{-6.3}$ \ \ & \ \ 0.4 \ \ & \ \ 0.1 \ \ & \ \ 0.0 \ \ & \ \ 0.75 \ \ & \ \ 1.8 \ \ \\[1.5mm]
\ \ $3$ \ \ & \ \ $ 0.66- 0.72$ \ \ & \ \ 12.7 \ \ & \ \ 1.3 \ \ & \ \ 0.74 \ \  & \ \ $^{+8.2}_{-8.0}$ \ \ & \ \ $^{+0.1}_{-2.4}$ \ \ & \ \ $^{+2.7}_{-2.6}$ \ \ & \ \ $^{+6.5}_{-6.2}$ \ \ & \ \ 0.6 \ \ & \ \ 0.1 \ \ & \ \ 0.0 \ \ & \ \ 0.75 \ \ & \ \ 1.8 \ \ \\[1.5mm]
\ \ $4$ \ \ & \ \ $ 0.72- 0.80$ \ \ & \ \ 8.89 \ \ & \ \ 0.98 \ \ & \ \ 0.62 \ \  & \ \ $^{+8.3}_{-8.2}$ \ \ & \ \ $^{+0.2}_{-1.8}$ \ \ & \ \ $^{+2.7}_{-2.7}$ \ \ & \ \ $^{+6.3}_{-6.1}$ \ \ & \ \ 0.7 \ \ & \ \ 0.1 \ \ & \ \ 0.0 \ \ & \ \ 0.75 \ \ & \ \ 1.8 \ \ \\[1.5mm]
\ \ $5$ \ \ & \ \ $ 0.80- 0.88$ \ \ & \ \ 5.46 \ \ & \ \ 1.2 \ \ & \ \ 0.57 \ \  & \ \ $^{+8.4}_{-8.3}$ \ \ & \ \ $^{+0.2}_{-1.4}$ \ \ & \ \ $^{+2.7}_{-2.7}$ \ \ & \ \ $^{+6.1}_{-5.9}$ \ \ & \ \ 0.8 \ \ & \ \ 0.1 \ \ & \ \ 0.0 \ \ & \ \ 0.75 \ \ & \ \ 1.8 \ \ \\[1.5mm]
\ \ $6$ \ \ & \ \ $ 0.88- 0.96$ \ \ & \ \ 3.28 \ \ & \ \ 1.5 \ \ & \ \ 0.57 \ \  & \ \ $^{+8.5}_{-8.3}$ \ \ & \ \ $^{+0.2}_{-1.3}$ \ \ & \ \ $^{+2.7}_{-2.7}$ \ \ & \ \ $^{+5.9}_{-5.7}$ \ \ & \ \ 0.8 \ \ & \ \ 0.1 \ \ & \ \ 0.0 \ \ & \ \ 0.75 \ \ & \ \ 1.8 \ \ \\[1.5mm]
\ \ $7$ \ \ & \ \ $ 0.96- 1.06$ \ \ & \ \ 1.82 \ \ & \ \ 1.8 \ \ & \ \ 0.52 \ \  & \ \ $^{+8.5}_{-8.3}$ \ \ & \ \ $^{+0.3}_{-1.3}$ \ \ & \ \ $^{+2.8}_{-2.7}$ \ \ & \ \ $^{+5.7}_{-5.4}$ \ \ & \ \ 0.8 \ \ & \ \ 0.1 \ \ & \ \ 0.0 \ \ & \ \ 0.75 \ \ & \ \ 1.8 \ \ \\[1.5mm]
\ \ $8$ \ \ & \ \ $ 1.06- 1.16$ \ \ & \ \ $ 9.98\cdot 10^{-1}$ \ \ &\ \ 1.4 \ \ & \ \ 0.61 \ \  & \ \ $^{+8.6}_{-8.2}$ \ \ & \ \ $^{+0.5}_{-1.3}$ \ \ & \ \ $^{+2.8}_{-2.7}$ \ \ & \ \ $^{+5.5}_{-5.2}$ \ \ & \ \ 0.8 \ \ & \ \ 0.1 \ \ & \ \ 0.0 \ \ & \ \ 0.75 \ \ & \ \ 1.8 \ \ \\[1.5mm]
\ \ $9$ \ \ & \ \ $ 1.16- 1.26$ \ \ & \ \ $ 5.84\cdot 10^{-1}$ \ \ &\ \ 1.1 \ \ & \ \ 0.65 \ \  & \ \ $^{+8.5}_{-8.2}$ \ \ & \ \ $^{+0.6}_{-1.3}$ \ \ & \ \ $^{+2.8}_{-2.7}$ \ \ & \ \ $^{+5.2}_{-4.9}$ \ \ & \ \ 0.8 \ \ & \ \ 0.1 \ \ & \ \ 0.0 \ \ & \ \ 0.75 \ \ & \ \ 1.8 \ \ \\[1.5mm]
\ \ $10$ \ \ & \ \ $ 1.26- 1.38$ \ \ & \ \ $ 3.31\cdot 10^{-1}$ \ \ &\ \ 1.4 \ \ & \ \ 0.66 \ \  & \ \ $^{+8.5}_{-8.2}$ \ \ & \ \ $^{+0.7}_{-1.3}$ \ \ & \ \ $^{+2.8}_{-2.6}$ \ \ & \ \ $^{+5.0}_{-4.8}$ \ \ & \ \ 0.8 \ \ & \ \ 0.1 \ \ & \ \ 0.0 \ \ & \ \ 0.75 \ \ & \ \ 1.8 \ \ \\[1.5mm]
\ \ $11$ \ \ & \ \ $ 1.38- 1.50$ \ \ & \ \ $ 1.81\cdot 10^{-1}$ \ \ &\ \ 1.3 \ \ & \ \ 0.64 \ \  & \ \ $^{+8.6}_{-8.2}$ \ \ & \ \ $^{+0.7}_{-1.3}$ \ \ & \ \ $^{+2.7}_{-2.6}$ \ \ & \ \ $^{+4.8}_{-4.6}$ \ \ & \ \ 0.8 \ \ & \ \ 0.1 \ \ & \ \ 0.0 \ \ & \ \ 0.75 \ \ & \ \ 1.8 \ \ \\[1.5mm]
\ \ $12$ \ \ & \ \ $ 1.50- 1.62$ \ \ & \ \ $ 9.89\cdot 10^{-2}$ \ \ &\ \ 0.92 \ \ & \ \ 0.66 \ \  & \ \ $^{+8.7}_{-8.2}$ \ \ & \ \ $^{+0.7}_{-1.2}$ \ \ & \ \ $^{+2.6}_{-2.5}$ \ \ & \ \ $^{+4.7}_{-4.5}$ \ \ & \ \ 0.8 \ \ & \ \ 0.0 \ \ & \ \ 0.0 \ \ & \ \ 0.75 \ \ & \ \ 1.8 \ \ \\[1.5mm]
\ \ $13$ \ \ & \ \ $ 1.62- 1.76$ \ \ & \ \ $ 5.46\cdot 10^{-2}$ \ \ &\ \ 1.1 \ \ & \ \ 0.60 \ \  & \ \ $^{+8.9}_{-8.3}$ \ \ & \ \ $^{+0.6}_{-1.1}$ \ \ & \ \ $^{+2.5}_{-2.3}$ \ \ & \ \ $^{+4.6}_{-4.3}$ \ \ & \ \ 0.9 \ \ & \ \ 0.0 \ \ & \ \ 0.0 \ \ & \ \ 0.75 \ \ & \ \ 1.8 \ \ \\[1.5mm]
\ \ $14$ \ \ & \ \ $ 1.76- 1.90$ \ \ & \ \ $ 2.99\cdot 10^{-2}$ \ \ &\ \ 1.4 \ \ & \ \ 0.57 \ \  & \ \ $^{+9.0}_{-8.4}$ \ \ & \ \ $^{+0.4}_{-0.9}$ \ \ & \ \ $^{+2.3}_{-2.1}$ \ \ & \ \ $^{+4.5}_{-4.2}$ \ \ & \ \ 0.9 \ \ & \ \ 0.0 \ \ & \ \ 0.0 \ \ & \ \ 0.75 \ \ & \ \ 1.8 \ \ \\[1.5mm]
\ \ $15$ \ \ & \ \ $ 1.90- 2.06$ \ \ & \ \ $ 1.57\cdot 10^{-2}$ \ \ &\ \ 1.1 \ \ & \ \ 0.60 \ \  & \ \ $^{+9.2}_{-8.6}$ \ \ & \ \ $^{+0.2}_{-0.7}$ \ \ & \ \ $^{+2.1}_{-1.9}$ \ \ & \ \ $^{+4.3}_{-4.1}$ \ \ & \ \ 0.9 \ \ & \ \ 0.0 \ \ & \ \ 0.0 \ \ & \ \ 0.75 \ \ & \ \ 1.8 \ \ \\[1.5mm]
\ \ $16$ \ \ & \ \ $ 2.06- 2.22$ \ \ & \ \ $ 7.92\cdot 10^{-3}$ \ \ &\ \ 1.4 \ \ & \ \ 0.67 \ \  & \ \ $^{+9.4}_{-8.9}$ \ \ & \ \ $^{+0.2}_{-0.5}$ \ \ & \ \ $^{+1.8}_{-1.7}$ \ \ & \ \ $^{+4.2}_{-4.0}$ \ \ & \ \ 0.9 \ \ & \ \ 0.0 \ \ & \ \ 0.0 \ \ & \ \ 0.75 \ \ & \ \ 1.8 \ \ \\[1.5mm]
\ \ $17$ \ \ & \ \ $ 2.22- 2.40$ \ \ & \ \ $ 4.12\cdot 10^{-3}$ \ \ &\ \ 1.8 \ \ & \ \ 0.76 \ \  & \ \ $^{+9.8}_{-9.3}$ \ \ & \ \ $^{+0.2}_{-0.3}$ \ \ & \ \ $^{+1.6}_{-1.5}$ \ \ & \ \ $^{+4.1}_{-3.9}$ \ \ & \ \ 1.0 \ \ & \ \ 0.0 \ \ & \ \ 0.0 \ \ & \ \ 0.75 \ \ & \ \ 1.8 \ \ \\[1.5mm]
\ \ $18$ \ \ & \ \ $ 2.40- 2.58$ \ \ & \ \ $ 1.99\cdot 10^{-3}$ \ \ &\ \ 2.7 \ \ & \ \ 0.98 \ \  & \ \ $^{+10.4}_{-9.9}$ \ \ & \ \ $^{+0.3}_{-0.2}$ \ \ & \ \ $^{+1.4}_{-1.3}$ \ \ & \ \ $^{+3.9}_{-3.8}$ \ \ & \ \ 1.1 \ \ & \ \ 0.0 \ \ & \ \ 0.0 \ \ & \ \ 0.75 \ \ & \ \ 1.8 \ \ \\[1.5mm]
\ \ $19$ \ \ & \ \ $ 2.58- 2.78$ \ \ & \ \ $ 9.95\cdot 10^{-4}$ \ \ &\ \ 3.6 \ \ & \ \ 1.0 \ \  & \ \ $^{+11.1}_{-10.5}$ \ \ & \ \ $^{+0.3}_{-0.1}$ \ \ & \ \ $^{+1.3}_{-1.2}$ \ \ & \ \ $^{+3.9}_{-3.7}$ \ \ & \ \ 1.2 \ \ & \ \ 0.0 \ \ & \ \ 0.0 \ \ & \ \ 0.75 \ \ & \ \ 1.8 \ \ \\[1.5mm]
\ \ $20$ \ \ & \ \ $ 2.78- 2.98$ \ \ & \ \ $ 4.54\cdot 10^{-4}$ \ \ &\ \ 5.2 \ \ & \ \ 1.2 \ \  & \ \ $^{+12.1}_{-11.2}$ \ \ & \ \ $^{+0.3}_{-0.1}$ \ \ & \ \ $^{+1.2}_{-1.1}$ \ \ & \ \ $^{+3.8}_{-3.6}$ \ \ & \ \ 1.3 \ \ & \ \ 0.0 \ \ & \ \ 0.0 \ \ & \ \ 0.75 \ \ & \ \ 1.8 \ \ \\[1.5mm]
\ \ $21$ \ \ & \ \ $ 2.98- 3.20$ \ \ & \ \ $ 1.91\cdot 10^{-4}$ \ \ &\ \ 7.7 \ \ & \ \ 1.5 \ \  & \ \ $^{+13.0}_{-12.0}$ \ \ & \ \ $^{+0.3}_{-0.1}$ \ \ & \ \ $^{+1.1}_{-1.1}$ \ \ & \ \ $^{+3.8}_{-3.6}$ \ \ & \ \ 1.4 \ \ & \ \ 0.0 \ \ & \ \ 0.0 \ \ & \ \ 0.75 \ \ & \ \ 1.8 \ \ \\[1.5mm]
\ \ $22$ \ \ & \ \ $ 3.20- 3.42$ \ \ & \ \ $ 7.88\cdot 10^{-5}$ \ \ &\ \ 12 \ \ & \ \ 1.6 \ \  & \ \ $^{+14.0}_{-12.7}$ \ \ & \ \ $^{+0.3}_{-0.0}$ \ \ & \ \ $^{+1.0}_{-1.0}$ \ \ & \ \ $^{+3.8}_{-3.5}$ \ \ & \ \ 1.5 \ \ & \ \ 0.0 \ \ & \ \ 0.0 \ \ & \ \ 0.75 \ \ & \ \ 1.8 \ \ \\[1.5mm]
\ \ $23$ \ \ & \ \ $ 3.42- 3.66$ \ \ & \ \ $ 3.33\cdot 10^{-5}$ \ \ &\ \ 19 \ \ & \ \ 1.7 \ \  & \ \ $^{+15.0}_{-13.5}$ \ \ & \ \ $^{+0.3}_{-0.0}$ \ \ & \ \ $^{+1.0}_{-1.0}$ \ \ & \ \ $^{+3.9}_{-3.5}$ \ \ & \ \ 1.6 \ \ & \ \ 0.0 \ \ & \ \ 0.0 \ \ & \ \ 0.75 \ \ & \ \ 1.8 \ \ \\[1.5mm]
\ \ $24$ \ \ & \ \ $ 3.66- 4.70$ \ \ & \ \ $ 5.24\cdot 10^{-6}$ \ \ &\ \ 23 \ \ & \ \ 1.6 \ \  & \ \ $^{+21.4}_{-21.1}$ \ \ & \ \ $^{+0.3}_{-0.0}$ \ \ & \ \ $^{+0.8}_{-0.7}$ \ \ & \ \ $^{+3.8}_{-3.7}$ \ \ & \ \ 2.4 \ \ & \ \ 0.0 \ \ & \ \ 0.0 \ \ & \ \ 0.75 \ \ & \ \ 1.8 \ \ \\[1.5mm]
\hline
\hline
\end{tabular}
\caption{
   Measured double-differential \trijet{} \xs{}, $\sigma$, for $\rfour{}$~jets and $4\leq\ystar{}<6$, along with uncertainties in the measurement. 
   All uncertainties are given in \%, where 
   $\delta_\textrm{stat}^\textrm{data}$ ($\delta_\textrm{stat}^\textrm{MC}$) are the statistical uncertainties in the data (MC simulation). 
   The $\gamma$ components are the uncertainty in the jet energy calibration from the \insitu{}, the pileup, the close-by jet, and flavour components. 
   The $u$ components show the uncertainty for the jet energy and angular resolution, the unfolding, the quality selection, and the luminosity. 
   While all columns are uncorrelated with each other, the \insitu{}, pileup, and flavour uncertainties shown here are the sum in quadrature of multiple uncorrelated components. 
}
\label{tab:sysunc_r00_ystar2}
\end{table*}

\begin{table*}[!ht]\tiny\centering\begin{tabular}{@{}c@{}@{}c@{}@{}c@{}@{}c@{}@{}r@{}@{}r@{}@{}r@{}@{}r@{}@{}r@{}@{}r@{}@{}r@{}@{}r@{}@{}r@{}@{}r@{}}\hline\hline \\[-2mm]\ \ $m_{jjj}$ \ \ & \ \ $m_{jjj}$-range \ \ & \ \ $\sigma$ \ \ & \ \ $\delta_{\textrm{stat}}^{\textrm{data}}$ \ \ & \ \ $\delta_{\textrm{stat}}^{\textrm{MC}}$ \ \ & \ \ $\gamma_{\insituT}$ \ \ & \ \ $\gamma_{\textrm{pileup}}$ \ \ & \ \ $\gamma_{\textrm{close-by}}$ \ \ & \ \ $\gamma_{\textrm{flavour}}$ \ \ & \ \ $u_{\textrm{JER}}$ \ \ & \ \ $u_{\textrm{JAR}}$ \ \ & \ \ $u_{\textrm{unfold}}$ \ \ & \ \ $u_{\textrm{qual.}}$ \ \ & \ \ $u_{\textrm{lumi}}$ \ \ \\ \relax
\ \ bin \# \ \ & \ \ [TeV] \ \ & \ \ [pb/GeV] \ \ & \ \ [\%] \ \ & \ \ [\%] \ \ & \ \ [\%] \ \ & \ \ [\%] \ \ & \ \ [\%] \ \ & \ \ [\%] \ \ & \ \ [\%] \ \ & \ \ [\%] \ \ & \ \ [\%] \ \ & \ \ [\%] \ \ & \ \ [\%] \\
\hline \\[-1.5mm]
\ \ $1$ \ \ & \ \ $ 0.54- 0.60$ \ \ & \ \ 21.9 \ \ & \ \ 2.2 \ \ & \ \ 0.98 \ \  & \ \ $^{+7.9}_{-8.1}$ \ \ & \ \ $^{+0.3}_{-5.2}$ \ \ & \ \ $^{+3.6}_{-3.6}$ \ \ & \ \ $^{+7.1}_{-6.6}$ \ \ & \ \ 2.2 \ \ & \ \ 1.8 \ \ & \ \ 0.0 \ \ & \ \ 0.75 \ \ & \ \ 1.8 \ \ \\[1.5mm]
\ \ $2$ \ \ & \ \ $ 0.60- 0.66$ \ \ & \ \ 21.7 \ \ & \ \ 2.2 \ \ & \ \ 0.85 \ \  & \ \ $^{+8.0}_{-8.2}$ \ \ & \ \ $^{+0.3}_{-3.7}$ \ \ & \ \ $^{+3.6}_{-3.6}$ \ \ & \ \ $^{+6.9}_{-6.5}$ \ \ & \ \ 2.2 \ \ & \ \ 1.8 \ \ & \ \ 0.0 \ \ & \ \ 0.75 \ \ & \ \ 1.8 \ \ \\[1.5mm]
\ \ $3$ \ \ & \ \ $ 0.66- 0.72$ \ \ & \ \ 18.0 \ \ & \ \ 1.8 \ \ & \ \ 0.81 \ \  & \ \ $^{+8.2}_{-8.2}$ \ \ & \ \ $^{+0.3}_{-2.6}$ \ \ & \ \ $^{+3.6}_{-3.6}$ \ \ & \ \ $^{+6.7}_{-6.4}$ \ \ & \ \ 2.2 \ \ & \ \ 1.7 \ \ & \ \ 0.0 \ \ & \ \ 0.75 \ \ & \ \ 1.8 \ \ \\[1.5mm]
\ \ $4$ \ \ & \ \ $ 0.72- 0.80$ \ \ & \ \ 12.3 \ \ & \ \ 1.4 \ \ & \ \ 0.61 \ \  & \ \ $^{+8.2}_{-8.2}$ \ \ & \ \ $^{+0.3}_{-1.8}$ \ \ & \ \ $^{+3.6}_{-3.6}$ \ \ & \ \ $^{+6.5}_{-6.2}$ \ \ & \ \ 2.1 \ \ & \ \ 1.6 \ \ & \ \ 0.0 \ \ & \ \ 0.75 \ \ & \ \ 1.8 \ \ \\[1.5mm]
\ \ $5$ \ \ & \ \ $ 0.80- 0.88$ \ \ & \ \ 7.56 \ \ & \ \ 1.7 \ \ & \ \ 0.61 \ \  & \ \ $^{+8.3}_{-8.1}$ \ \ & \ \ $^{+0.3}_{-1.4}$ \ \ & \ \ $^{+3.6}_{-3.5}$ \ \ & \ \ $^{+6.2}_{-6.0}$ \ \ & \ \ 2.1 \ \ & \ \ 1.4 \ \ & \ \ 0.0 \ \ & \ \ 0.75 \ \ & \ \ 1.8 \ \ \\[1.5mm]
\ \ $6$ \ \ & \ \ $ 0.88- 0.96$ \ \ & \ \ 4.49 \ \ & \ \ 2.1 \ \ & \ \ 0.65 \ \  & \ \ $^{+8.2}_{-8.0}$ \ \ & \ \ $^{+0.3}_{-1.3}$ \ \ & \ \ $^{+3.5}_{-3.4}$ \ \ & \ \ $^{+6.0}_{-5.7}$ \ \ & \ \ 2.0 \ \ & \ \ 1.2 \ \ & \ \ 0.0 \ \ & \ \ 0.75 \ \ & \ \ 1.8 \ \ \\[1.5mm]
\ \ $7$ \ \ & \ \ $ 0.96- 1.06$ \ \ & \ \ 2.54 \ \ & \ \ 1.5 \ \ & \ \ 0.47 \ \  & \ \ $^{+8.2}_{-8.0}$ \ \ & \ \ $^{+0.5}_{-1.3}$ \ \ & \ \ $^{+3.3}_{-3.2}$ \ \ & \ \ $^{+5.7}_{-5.5}$ \ \ & \ \ 1.9 \ \ & \ \ 1.1 \ \ & \ \ 0.0 \ \ & \ \ 0.75 \ \ & \ \ 1.8 \ \ \\[1.5mm]
\ \ $8$ \ \ & \ \ $ 1.06- 1.16$ \ \ & \ \ 1.35 \ \ & \ \ 1.9 \ \ & \ \ 0.50 \ \  & \ \ $^{+8.1}_{-7.9}$ \ \ & \ \ $^{+0.7}_{-1.5}$ \ \ & \ \ $^{+3.2}_{-3.1}$ \ \ & \ \ $^{+5.4}_{-5.2}$ \ \ & \ \ 1.8 \ \ & \ \ 0.9 \ \ & \ \ 0.0 \ \ & \ \ 0.75 \ \ & \ \ 1.8 \ \ \\[1.5mm]
\ \ $9$ \ \ & \ \ $ 1.16- 1.26$ \ \ & \ \ $ 7.82\cdot 10^{-1}$ \ \ &\ \ 2.6 \ \ & \ \ 0.60 \ \  & \ \ $^{+8.0}_{-7.8}$ \ \ & \ \ $^{+0.8}_{-1.6}$ \ \ & \ \ $^{+3.1}_{-3.0}$ \ \ & \ \ $^{+5.2}_{-5.0}$ \ \ & \ \ 1.7 \ \ & \ \ 0.7 \ \ & \ \ 0.0 \ \ & \ \ 0.75 \ \ & \ \ 1.8 \ \ \\[1.5mm]
\ \ $10$ \ \ & \ \ $ 1.26- 1.38$ \ \ & \ \ $ 4.48\cdot 10^{-1}$ \ \ &\ \ 1.3 \ \ & \ \ 0.62 \ \  & \ \ $^{+8.0}_{-7.7}$ \ \ & \ \ $^{+0.9}_{-1.7}$ \ \ & \ \ $^{+3.0}_{-2.8}$ \ \ & \ \ $^{+5.0}_{-4.7}$ \ \ & \ \ 1.7 \ \ & \ \ 0.6 \ \ & \ \ 0.0 \ \ & \ \ 0.75 \ \ & \ \ 1.8 \ \ \\[1.5mm]
\ \ $11$ \ \ & \ \ $ 1.38- 1.50$ \ \ & \ \ $ 2.40\cdot 10^{-1}$ \ \ &\ \ 1.6 \ \ & \ \ 0.67 \ \  & \ \ $^{+8.0}_{-7.6}$ \ \ & \ \ $^{+1.0}_{-1.6}$ \ \ & \ \ $^{+2.8}_{-2.7}$ \ \ & \ \ $^{+4.8}_{-4.5}$ \ \ & \ \ 1.6 \ \ & \ \ 0.5 \ \ & \ \ 0.0 \ \ & \ \ 0.75 \ \ & \ \ 1.8 \ \ \\[1.5mm]
\ \ $12$ \ \ & \ \ $ 1.50- 1.62$ \ \ & \ \ $ 1.31\cdot 10^{-1}$ \ \ &\ \ 2.1 \ \ & \ \ 0.69 \ \  & \ \ $^{+8.1}_{-7.6}$ \ \ & \ \ $^{+1.1}_{-1.5}$ \ \ & \ \ $^{+2.6}_{-2.5}$ \ \ & \ \ $^{+4.6}_{-4.3}$ \ \ & \ \ 1.6 \ \ & \ \ 0.4 \ \ & \ \ 0.0 \ \ & \ \ 0.75 \ \ & \ \ 1.8 \ \ \\[1.5mm]
\ \ $13$ \ \ & \ \ $ 1.62- 1.76$ \ \ & \ \ $ 7.10\cdot 10^{-2}$ \ \ &\ \ 1.8 \ \ & \ \ 0.64 \ \  & \ \ $^{+8.1}_{-7.6}$ \ \ & \ \ $^{+1.1}_{-1.4}$ \ \ & \ \ $^{+2.5}_{-2.3}$ \ \ & \ \ $^{+4.5}_{-4.1}$ \ \ & \ \ 1.6 \ \ & \ \ 0.3 \ \ & \ \ 0.0 \ \ & \ \ 0.75 \ \ & \ \ 1.8 \ \ \\[1.5mm]
\ \ $14$ \ \ & \ \ $ 1.76- 1.90$ \ \ & \ \ $ 3.92\cdot 10^{-2}$ \ \ &\ \ 1.3 \ \ & \ \ 0.60 \ \  & \ \ $^{+8.2}_{-7.7}$ \ \ & \ \ $^{+1.1}_{-1.1}$ \ \ & \ \ $^{+2.2}_{-2.1}$ \ \ & \ \ $^{+4.3}_{-4.0}$ \ \ & \ \ 1.6 \ \ & \ \ 0.3 \ \ & \ \ 0.0 \ \ & \ \ 0.75 \ \ & \ \ 1.8 \ \ \\[1.5mm]
\ \ $15$ \ \ & \ \ $ 1.90- 2.06$ \ \ & \ \ $ 2.11\cdot 10^{-2}$ \ \ &\ \ 1.7 \ \ & \ \ 0.52 \ \  & \ \ $^{+8.4}_{-7.9}$ \ \ & \ \ $^{+1.0}_{-0.9}$ \ \ & \ \ $^{+2.0}_{-1.9}$ \ \ & \ \ $^{+4.1}_{-3.8}$ \ \ & \ \ 1.6 \ \ & \ \ 0.3 \ \ & \ \ 0.0 \ \ & \ \ 0.75 \ \ & \ \ 1.8 \ \ \\[1.5mm]
\ \ $16$ \ \ & \ \ $ 2.06- 2.22$ \ \ & \ \ $ 1.05\cdot 10^{-2}$ \ \ &\ \ 1.9 \ \ & \ \ 0.63 \ \  & \ \ $^{+8.6}_{-8.3}$ \ \ & \ \ $^{+0.8}_{-0.7}$ \ \ & \ \ $^{+1.8}_{-1.7}$ \ \ & \ \ $^{+3.9}_{-3.7}$ \ \ & \ \ 1.6 \ \ & \ \ 0.3 \ \ & \ \ 0.0 \ \ & \ \ 0.75 \ \ & \ \ 1.8 \ \ \\[1.5mm]
\ \ $17$ \ \ & \ \ $ 2.22- 2.40$ \ \ & \ \ $ 5.18\cdot 10^{-3}$ \ \ &\ \ 1.7 \ \ & \ \ 0.74 \ \  & \ \ $^{+9.0}_{-8.8}$ \ \ & \ \ $^{+0.6}_{-0.5}$ \ \ & \ \ $^{+1.6}_{-1.6}$ \ \ & \ \ $^{+3.7}_{-3.7}$ \ \ & \ \ 1.6 \ \ & \ \ 0.3 \ \ & \ \ 0.0 \ \ & \ \ 0.75 \ \ & \ \ 1.8 \ \ \\[1.5mm]
\ \ $18$ \ \ & \ \ $ 2.40- 2.58$ \ \ & \ \ $ 2.62\cdot 10^{-3}$ \ \ &\ \ 2.3 \ \ & \ \ 0.98 \ \  & \ \ $^{+9.5}_{-9.4}$ \ \ & \ \ $^{+0.4}_{-0.3}$ \ \ & \ \ $^{+1.4}_{-1.4}$ \ \ & \ \ $^{+3.6}_{-3.6}$ \ \ & \ \ 1.7 \ \ & \ \ 0.2 \ \ & \ \ 0.0 \ \ & \ \ 0.75 \ \ & \ \ 1.8 \ \ \\[1.5mm]
\ \ $19$ \ \ & \ \ $ 2.58- 2.78$ \ \ & \ \ $ 1.28\cdot 10^{-3}$ \ \ &\ \ 3.2 \ \ & \ \ 1.1 \ \  & \ \ $^{+10.1}_{-10.3}$ \ \ & \ \ $^{+0.3}_{-0.2}$ \ \ & \ \ $^{+1.3}_{-1.3}$ \ \ & \ \ $^{+3.5}_{-3.5}$ \ \ & \ \ 1.9 \ \ & \ \ 0.2 \ \ & \ \ 0.0 \ \ & \ \ 0.75 \ \ & \ \ 1.8 \ \ \\[1.5mm]
\ \ $20$ \ \ & \ \ $ 2.78- 2.98$ \ \ & \ \ $ 5.77\cdot 10^{-4}$ \ \ &\ \ 4.7 \ \ & \ \ 1.4 \ \  & \ \ $^{+10.9}_{-11.3}$ \ \ & \ \ $^{+0.2}_{-0.2}$ \ \ & \ \ $^{+1.2}_{-1.2}$ \ \ & \ \ $^{+3.4}_{-3.5}$ \ \ & \ \ 2.0 \ \ & \ \ 0.2 \ \ & \ \ 0.0 \ \ & \ \ 0.75 \ \ & \ \ 1.8 \ \ \\[1.5mm]
\ \ $21$ \ \ & \ \ $ 2.98- 3.20$ \ \ & \ \ $ 2.64\cdot 10^{-4}$ \ \ &\ \ 6.7 \ \ & \ \ 1.5 \ \  & \ \ $^{+11.8}_{-12.3}$ \ \ & \ \ $^{+0.1}_{-0.2}$ \ \ & \ \ $^{+1.1}_{-1.1}$ \ \ & \ \ $^{+3.4}_{-3.6}$ \ \ & \ \ 2.2 \ \ & \ \ 0.2 \ \ & \ \ 0.0 \ \ & \ \ 0.75 \ \ & \ \ 1.8 \ \ \\[1.5mm]
\ \ $22$ \ \ & \ \ $ 3.20- 3.42$ \ \ & \ \ $ 1.16\cdot 10^{-4}$ \ \ &\ \ 10 \ \ & \ \ 1.7 \ \  & \ \ $^{+12.8}_{-13.3}$ \ \ & \ \ $^{+0.1}_{-0.1}$ \ \ & \ \ $^{+1.1}_{-1.1}$ \ \ & \ \ $^{+3.4}_{-3.6}$ \ \ & \ \ 2.3 \ \ & \ \ 0.2 \ \ & \ \ 0.0 \ \ & \ \ 0.75 \ \ & \ \ 1.8 \ \ \\[1.5mm]
\ \ $23$ \ \ & \ \ $ 3.42- 3.66$ \ \ & \ \ $ 3.72\cdot 10^{-5}$ \ \ &\ \ 17 \ \ & \ \ 1.9 \ \  & \ \ $^{+13.9}_{-14.2}$ \ \ & \ \ $^{+0.1}_{-0.1}$ \ \ & \ \ $^{+1.0}_{-1.0}$ \ \ & \ \ $^{+3.5}_{-3.6}$ \ \ & \ \ 2.5 \ \ & \ \ 0.2 \ \ & \ \ 0.0 \ \ & \ \ 0.75 \ \ & \ \ 1.8 \ \ \\[1.5mm]
\ \ $24$ \ \ & \ \ $ 3.66- 4.70$ \ \ & \ \ $ 6.07\cdot 10^{-6}$ \ \ &\ \ 22 \ \ & \ \ 1.3 \ \  & \ \ $^{+24.1}_{-18.9}$ \ \ & \ \ $^{+0.1}_{-0.1}$ \ \ & \ \ $^{+0.9}_{-0.7}$ \ \ & \ \ $^{+4.9}_{-3.6}$ \ \ & \ \ 2.9 \ \ & \ \ 0.1 \ \ & \ \ 0.0 \ \ & \ \ 0.75 \ \ & \ \ 1.8 \ \ \\[1.5mm]
\hline
\hline
\end{tabular}
\caption{
   Measured double-differential \trijet{} \xs{}, $\sigma$, for $\rsix{}$~jets and $4\leq\ystar{}<6$, along with uncertainties in the measurement. 
   All uncertainties are given in \%, where 
   $\delta_\textrm{stat}^\textrm{data}$ ($\delta_\textrm{stat}^\textrm{MC}$) are the statistical uncertainties in the data (MC simulation). 
   The $\gamma$ components are the uncertainty in the jet energy calibration from the \insitu{}, the pileup, the close-by jet, and flavour components. 
   The $u$ components show the uncertainty for the jet energy and angular resolution, the unfolding, the quality selection, and the luminosity. 
   While all columns are uncorrelated with each other, the \insitu{}, pileup, and flavour uncertainties shown here are the sum in quadrature of multiple uncorrelated components. 
}
\label{tab:sysunc_r01_ystar2}
\end{table*}

\begin{table*}[!ht]\tiny\centering\begin{tabular}{@{}c@{}@{}c@{}@{}c@{}@{}c@{}@{}r@{}@{}r@{}@{}r@{}@{}r@{}@{}r@{}@{}r@{}@{}r@{}@{}r@{}@{}r@{}@{}r@{}}\hline\hline \\[-2mm]\ \ $m_{jjj}$ \ \ & \ \ $m_{jjj}$-range \ \ & \ \ $\sigma$ \ \ & \ \ $\delta_{\textrm{stat}}^{\textrm{data}}$ \ \ & \ \ $\delta_{\textrm{stat}}^{\textrm{MC}}$ \ \ & \ \ $\gamma_{\insituT}$ \ \ & \ \ $\gamma_{\textrm{pileup}}$ \ \ & \ \ $\gamma_{\textrm{close-by}}$ \ \ & \ \ $\gamma_{\textrm{flavour}}$ \ \ & \ \ $u_{\textrm{JER}}$ \ \ & \ \ $u_{\textrm{JAR}}$ \ \ & \ \ $u_{\textrm{unfold}}$ \ \ & \ \ $u_{\textrm{qual.}}$ \ \ & \ \ $u_{\textrm{lumi}}$ \ \ \\ \relax
\ \ bin \# \ \ & \ \ [TeV] \ \ & \ \ [pb/GeV] \ \ & \ \ [\%] \ \ & \ \ [\%] \ \ & \ \ [\%] \ \ & \ \ [\%] \ \ & \ \ [\%] \ \ & \ \ [\%] \ \ & \ \ [\%] \ \ & \ \ [\%] \ \ & \ \ [\%] \ \ & \ \ [\%] \ \ & \ \ [\%] \\
\hline \\[-1.5mm]
\ \ $1$ \ \ & \ \ $ 0.76- 0.84$ \ \ & \ \ 4.95 \ \ & \ \ 2.4 \ \ & \ \ 1.6 \ \  & \ \ $^{+9.8}_{-9.9}$ \ \ & \ \ $^{+0.3}_{-4.1}$ \ \ & \ \ $^{+2.5}_{-2.7}$ \ \ & \ \ $^{+6.7}_{-6.3}$ \ \ & \ \ 0.3 \ \ & \ \ 0.7 \ \ & \ \ 0.0 \ \ & \ \ 0.75 \ \ & \ \ 1.8 \ \ \\[1.5mm]
\ \ $2$ \ \ & \ \ $ 0.84- 0.94$ \ \ & \ \ 5.00 \ \ & \ \ 2.1 \ \ & \ \ 1.2 \ \  & \ \ $^{+10.2}_{-10.1}$ \ \ & \ \ $^{+0.3}_{-3.2}$ \ \ & \ \ $^{+2.5}_{-2.7}$ \ \ & \ \ $^{+6.5}_{-6.1}$ \ \ & \ \ 0.6 \ \ & \ \ 0.6 \ \ & \ \ 0.0 \ \ & \ \ 0.75 \ \ & \ \ 1.8 \ \ \\[1.5mm]
\ \ $3$ \ \ & \ \ $ 0.94- 1.04$ \ \ & \ \ 3.80 \ \ & \ \ 1.4 \ \ & \ \ 1.1 \ \  & \ \ $^{+10.7}_{-10.5}$ \ \ & \ \ $^{+0.3}_{-2.4}$ \ \ & \ \ $^{+2.6}_{-2.6}$ \ \ & \ \ $^{+6.3}_{-6.0}$ \ \ & \ \ 0.8 \ \ & \ \ 0.5 \ \ & \ \ 0.0 \ \ & \ \ 0.75 \ \ & \ \ 1.8 \ \ \\[1.5mm]
\ \ $4$ \ \ & \ \ $ 1.04- 1.14$ \ \ & \ \ 2.67 \ \ & \ \ 1.4 \ \ & \ \ 1.2 \ \  & \ \ $^{+11.4}_{-10.9}$ \ \ & \ \ $^{+0.2}_{-1.8}$ \ \ & \ \ $^{+2.6}_{-2.7}$ \ \ & \ \ $^{+6.1}_{-6.0}$ \ \ & \ \ 1.0 \ \ & \ \ 0.4 \ \ & \ \ 0.0 \ \ & \ \ 0.75 \ \ & \ \ 1.8 \ \ \\[1.5mm]
\ \ $5$ \ \ & \ \ $ 1.14- 1.26$ \ \ & \ \ 1.74 \ \ & \ \ 1.7 \ \ & \ \ 1.1 \ \  & \ \ $^{+12.2}_{-11.4}$ \ \ & \ \ $^{+0.2}_{-1.5}$ \ \ & \ \ $^{+2.7}_{-2.7}$ \ \ & \ \ $^{+6.0}_{-5.8}$ \ \ & \ \ 1.1 \ \ & \ \ 0.3 \ \ & \ \ 0.0 \ \ & \ \ 0.75 \ \ & \ \ 1.8 \ \ \\[1.5mm]
\ \ $6$ \ \ & \ \ $ 1.26- 1.38$ \ \ & \ \ $ 9.30\cdot 10^{-1}$ \ \ &\ \ 2.2 \ \ & \ \ 1.1 \ \  & \ \ $^{+12.8}_{-11.8}$ \ \ & \ \ $^{+0.1}_{-1.5}$ \ \ & \ \ $^{+2.7}_{-2.7}$ \ \ & \ \ $^{+5.8}_{-5.5}$ \ \ & \ \ 1.2 \ \ & \ \ 0.3 \ \ & \ \ 0.0 \ \ & \ \ 0.75 \ \ & \ \ 1.8 \ \ \\[1.5mm]
\ \ $7$ \ \ & \ \ $ 1.38- 1.52$ \ \ & \ \ $ 5.52\cdot 10^{-1}$ \ \ &\ \ 2.7 \ \ & \ \ 1.1 \ \  & \ \ $^{+13.4}_{-12.2}$ \ \ & \ \ $^{+0.4}_{-1.7}$ \ \ & \ \ $^{+2.8}_{-2.7}$ \ \ & \ \ $^{+5.6}_{-5.2}$ \ \ & \ \ 1.3 \ \ & \ \ 0.2 \ \ & \ \ 0.0 \ \ & \ \ 0.75 \ \ & \ \ 1.8 \ \ \\[1.5mm]
\ \ $8$ \ \ & \ \ $ 1.52- 1.66$ \ \ & \ \ $ 2.88\cdot 10^{-1}$ \ \ &\ \ 4.0 \ \ & \ \ 1.2 \ \  & \ \ $^{+13.8}_{-12.6}$ \ \ & \ \ $^{+0.6}_{-1.8}$ \ \ & \ \ $^{+2.8}_{-2.7}$ \ \ & \ \ $^{+5.4}_{-5.0}$ \ \ & \ \ 1.3 \ \ & \ \ 0.2 \ \ & \ \ 0.0 \ \ & \ \ 0.75 \ \ & \ \ 1.8 \ \ \\[1.5mm]
\ \ $9$ \ \ & \ \ $ 1.66- 1.80$ \ \ & \ \ $ 1.49\cdot 10^{-1}$ \ \ &\ \ 2.1 \ \ & \ \ 1.3 \ \  & \ \ $^{+14.2}_{-13.0}$ \ \ & \ \ $^{+0.8}_{-1.9}$ \ \ & \ \ $^{+2.8}_{-2.8}$ \ \ & \ \ $^{+5.2}_{-4.9}$ \ \ & \ \ 1.3 \ \ & \ \ 0.2 \ \ & \ \ 0.0 \ \ & \ \ 0.75 \ \ & \ \ 1.8 \ \ \\[1.5mm]
\ \ $10$ \ \ & \ \ $ 1.80- 1.94$ \ \ & \ \ $ 7.94\cdot 10^{-2}$ \ \ &\ \ 2.5 \ \ & \ \ 1.5 \ \  & \ \ $^{+14.6}_{-13.3}$ \ \ & \ \ $^{+1.0}_{-1.8}$ \ \ & \ \ $^{+2.8}_{-2.8}$ \ \ & \ \ $^{+5.1}_{-4.7}$ \ \ & \ \ 1.2 \ \ & \ \ 0.1 \ \ & \ \ 0.0 \ \ & \ \ 0.75 \ \ & \ \ 1.8 \ \ \\[1.5mm]
\ \ $11$ \ \ & \ \ $ 1.94- 2.10$ \ \ & \ \ $ 4.19\cdot 10^{-2}$ \ \ &\ \ 3.0 \ \ & \ \ 1.5 \ \  & \ \ $^{+15.0}_{-13.6}$ \ \ & \ \ $^{+1.1}_{-1.7}$ \ \ & \ \ $^{+2.8}_{-2.8}$ \ \ & \ \ $^{+5.0}_{-4.6}$ \ \ & \ \ 1.3 \ \ & \ \ 0.1 \ \ & \ \ 0.0 \ \ & \ \ 0.75 \ \ & \ \ 1.8 \ \ \\[1.5mm]
\ \ $12$ \ \ & \ \ $ 2.10- 2.26$ \ \ & \ \ $ 2.20\cdot 10^{-2}$ \ \ &\ \ 3.8 \ \ & \ \ 1.7 \ \  & \ \ $^{+15.6}_{-14.0}$ \ \ & \ \ $^{+1.2}_{-1.5}$ \ \ & \ \ $^{+2.8}_{-2.8}$ \ \ & \ \ $^{+4.9}_{-4.5}$ \ \ & \ \ 1.3 \ \ & \ \ 0.1 \ \ & \ \ 0.0 \ \ & \ \ 0.75 \ \ & \ \ 1.8 \ \ \\[1.5mm]
\ \ $13$ \ \ & \ \ $ 2.26- 2.42$ \ \ & \ \ $ 1.21\cdot 10^{-2}$ \ \ &\ \ 2.4 \ \ & \ \ 1.8 \ \  & \ \ $^{+16.0}_{-14.3}$ \ \ & \ \ $^{+1.2}_{-1.3}$ \ \ & \ \ $^{+2.8}_{-2.7}$ \ \ & \ \ $^{+4.9}_{-4.5}$ \ \ & \ \ 1.4 \ \ & \ \ 0.0 \ \ & \ \ 0.0 \ \ & \ \ 0.75 \ \ & \ \ 1.8 \ \ \\[1.5mm]
\ \ $14$ \ \ & \ \ $ 2.42- 2.58$ \ \ & \ \ $ 5.86\cdot 10^{-3}$ \ \ &\ \ 2.8 \ \ & \ \ 2.4 \ \  & \ \ $^{+16.5}_{-14.8}$ \ \ & \ \ $^{+1.1}_{-1.1}$ \ \ & \ \ $^{+2.7}_{-2.6}$ \ \ & \ \ $^{+4.9}_{-4.5}$ \ \ & \ \ 1.6 \ \ & \ \ 0.0 \ \ & \ \ 0.0 \ \ & \ \ 0.75 \ \ & \ \ 1.8 \ \ \\[1.5mm]
\ \ $15$ \ \ & \ \ $ 2.58- 2.76$ \ \ & \ \ $ 3.24\cdot 10^{-3}$ \ \ &\ \ 3.8 \ \ & \ \ 2.1 \ \  & \ \ $^{+16.9}_{-15.2}$ \ \ & \ \ $^{+1.0}_{-0.9}$ \ \ & \ \ $^{+2.6}_{-2.5}$ \ \ & \ \ $^{+4.8}_{-4.6}$ \ \ & \ \ 1.7 \ \ & \ \ 0.0 \ \ & \ \ 0.0 \ \ & \ \ 0.75 \ \ & \ \ 1.8 \ \ \\[1.5mm]
\ \ $16$ \ \ & \ \ $ 2.76- 2.94$ \ \ & \ \ $ 1.53\cdot 10^{-3}$ \ \ &\ \ 4.8 \ \ & \ \ 1.5 \ \  & \ \ $^{+17.3}_{-15.7}$ \ \ & \ \ $^{+0.8}_{-0.7}$ \ \ & \ \ $^{+2.5}_{-2.4}$ \ \ & \ \ $^{+4.7}_{-4.7}$ \ \ & \ \ 1.7 \ \ & \ \ 0.0 \ \ & \ \ 0.0 \ \ & \ \ 0.75 \ \ & \ \ 1.8 \ \ \\[1.5mm]
\ \ $17$ \ \ & \ \ $ 2.94- 3.12$ \ \ & \ \ $ 7.02\cdot 10^{-4}$ \ \ &\ \ 4.5 \ \ & \ \ 2.2 \ \  & \ \ $^{+17.8}_{-16.2}$ \ \ & \ \ $^{+0.6}_{-0.5}$ \ \ & \ \ $^{+2.4}_{-2.3}$ \ \ & \ \ $^{+4.7}_{-4.7}$ \ \ & \ \ 1.8 \ \ & \ \ 0.0 \ \ & \ \ 0.0 \ \ & \ \ 0.75 \ \ & \ \ 1.8 \ \ \\[1.5mm]
\ \ $18$ \ \ & \ \ $ 3.12- 3.44$ \ \ & \ \ $ 2.67\cdot 10^{-4}$ \ \ &\ \ 5.6 \ \ & \ \ 2.1 \ \  & \ \ $^{+18.5}_{-16.9}$ \ \ & \ \ $^{+0.3}_{-0.4}$ \ \ & \ \ $^{+2.3}_{-2.2}$ \ \ & \ \ $^{+4.6}_{-4.7}$ \ \ & \ \ 2.0 \ \ & \ \ 0.0 \ \ & \ \ 0.0 \ \ & \ \ 0.75 \ \ & \ \ 1.8 \ \ \\[1.5mm]
\ \ $19$ \ \ & \ \ $ 3.44- 3.90$ \ \ & \ \ $ 6.67\cdot 10^{-5}$ \ \ &\ \ 10 \ \ & \ \ 2.8 \ \  & \ \ $^{+19.8}_{-18.2}$ \ \ & \ \ $^{+0.0}_{-0.2}$ \ \ & \ \ $^{+2.1}_{-2.0}$ \ \ & \ \ $^{+4.4}_{-4.5}$ \ \ & \ \ 2.1 \ \ & \ \ 0.0 \ \ & \ \ 0.0 \ \ & \ \ 0.75 \ \ & \ \ 1.8 \ \ \\[1.5mm]
\ \ $20$ \ \ & \ \ $ 3.90- 4.66$ \ \ & \ \ $ 4.17\cdot 10^{-6}$ \ \ &\ \ 30 \ \ & \ \ 5.0 \ \  & \ \ $^{+24.1}_{-24.4}$ \ \ & \ \ $^{+0.0}_{-0.5}$ \ \ & \ \ $^{+1.2}_{-0.7}$ \ \ & \ \ $^{+3.2}_{-3.0}$ \ \ & \ \ 2.8 \ \ & \ \ 0.0 \ \ & \ \ 0.0 \ \ & \ \ 0.75 \ \ & \ \ 1.8 \ \ \\[1.5mm]
\hline
\hline
\end{tabular}
\caption{
   Measured double-differential \trijet{} \xs{}, $\sigma$, for $\rfour{}$~jets and $6\leq\ystar{}<8$, along with uncertainties in the measurement. 
   All uncertainties are given in \%, where 
   $\delta_\textrm{stat}^\textrm{data}$ ($\delta_\textrm{stat}^\textrm{MC}$) are the statistical uncertainties in the data (MC simulation). 
   The $\gamma$ components are the uncertainty in the jet energy calibration from the \insitu{}, the pileup, the close-by jet, and flavour components. 
   The $u$ components show the uncertainty for the jet energy and angular resolution, the unfolding, the quality selection, and the luminosity. 
   While all columns are uncorrelated with each other, the \insitu{}, pileup, and flavour uncertainties shown here are the sum in quadrature of multiple uncorrelated components. 
}
\label{tab:sysunc_r00_ystar3}
\end{table*}

\begin{table*}[!ht]\tiny\centering\begin{tabular}{@{}c@{}@{}c@{}@{}c@{}@{}c@{}@{}r@{}@{}r@{}@{}r@{}@{}r@{}@{}r@{}@{}r@{}@{}r@{}@{}r@{}@{}r@{}@{}r@{}}\hline\hline \\[-2mm]\ \ $m_{jjj}$ \ \ & \ \ $m_{jjj}$-range \ \ & \ \ $\sigma$ \ \ & \ \ $\delta_{\textrm{stat}}^{\textrm{data}}$ \ \ & \ \ $\delta_{\textrm{stat}}^{\textrm{MC}}$ \ \ & \ \ $\gamma_{\insituT}$ \ \ & \ \ $\gamma_{\textrm{pileup}}$ \ \ & \ \ $\gamma_{\textrm{close-by}}$ \ \ & \ \ $\gamma_{\textrm{flavour}}$ \ \ & \ \ $u_{\textrm{JER}}$ \ \ & \ \ $u_{\textrm{JAR}}$ \ \ & \ \ $u_{\textrm{unfold}}$ \ \ & \ \ $u_{\textrm{qual.}}$ \ \ & \ \ $u_{\textrm{lumi}}$ \ \ \\ \relax
\ \ bin \# \ \ & \ \ [TeV] \ \ & \ \ [pb/GeV] \ \ & \ \ [\%] \ \ & \ \ [\%] \ \ & \ \ [\%] \ \ & \ \ [\%] \ \ & \ \ [\%] \ \ & \ \ [\%] \ \ & \ \ [\%] \ \ & \ \ [\%] \ \ & \ \ [\%] \ \ & \ \ [\%] \ \ & \ \ [\%] \\
\hline \\[-1.5mm]
\ \ $1$ \ \ & \ \ $ 0.76- 0.84$ \ \ & \ \ 6.96 \ \ & \ \ 3.3 \ \ & \ \ 1.4 \ \  & \ \ $^{+8.9}_{-8.8}$ \ \ & \ \ $^{+0.4}_{-5.0}$ \ \ & \ \ $^{+3.7}_{-3.6}$ \ \ & \ \ $^{+6.5}_{-6.0}$ \ \ & \ \ 2.7 \ \ & \ \ 3.6 \ \ & \ \ 0.0 \ \ & \ \ 0.75 \ \ & \ \ 1.8 \ \ \\[1.5mm]
\ \ $2$ \ \ & \ \ $ 0.84- 0.94$ \ \ & \ \ 7.23 \ \ & \ \ 3.0 \ \ & \ \ 1.2 \ \  & \ \ $^{+9.6}_{-9.3}$ \ \ & \ \ $^{+0.4}_{-3.8}$ \ \ & \ \ $^{+3.8}_{-3.5}$ \ \ & \ \ $^{+6.5}_{-5.9}$ \ \ & \ \ 2.9 \ \ & \ \ 3.6 \ \ & \ \ 0.0 \ \ & \ \ 0.75 \ \ & \ \ 1.8 \ \ \\[1.5mm]
\ \ $3$ \ \ & \ \ $ 0.94- 1.04$ \ \ & \ \ 5.74 \ \ & \ \ 1.8 \ \ & \ \ 1.0 \ \  & \ \ $^{+10.3}_{-9.8}$ \ \ & \ \ $^{+0.4}_{-2.8}$ \ \ & \ \ $^{+3.8}_{-3.5}$ \ \ & \ \ $^{+6.5}_{-5.8}$ \ \ & \ \ 2.9 \ \ & \ \ 3.5 \ \ & \ \ 0.0 \ \ & \ \ 0.75 \ \ & \ \ 1.8 \ \ \\[1.5mm]
\ \ $4$ \ \ & \ \ $ 1.04- 1.14$ \ \ & \ \ 4.09 \ \ & \ \ 2.1 \ \ & \ \ 1.2 \ \  & \ \ $^{+11.1}_{-10.3}$ \ \ & \ \ $^{+0.4}_{-2.2}$ \ \ & \ \ $^{+3.8}_{-3.5}$ \ \ & \ \ $^{+6.5}_{-5.7}$ \ \ & \ \ 2.9 \ \ & \ \ 3.4 \ \ & \ \ 0.0 \ \ & \ \ 0.75 \ \ & \ \ 1.8 \ \ \\[1.5mm]
\ \ $5$ \ \ & \ \ $ 1.14- 1.26$ \ \ & \ \ 2.50 \ \ & \ \ 2.5 \ \ & \ \ 1.2 \ \  & \ \ $^{+11.7}_{-10.8}$ \ \ & \ \ $^{+0.4}_{-2.0}$ \ \ & \ \ $^{+3.8}_{-3.5}$ \ \ & \ \ $^{+6.3}_{-5.7}$ \ \ & \ \ 3.0 \ \ & \ \ 3.2 \ \ & \ \ 0.0 \ \ & \ \ 0.75 \ \ & \ \ 1.8 \ \ \\[1.5mm]
\ \ $6$ \ \ & \ \ $ 1.26- 1.38$ \ \ & \ \ 1.34 \ \ & \ \ 3.2 \ \ & \ \ 1.2 \ \  & \ \ $^{+12.3}_{-11.2}$ \ \ & \ \ $^{+0.6}_{-1.8}$ \ \ & \ \ $^{+3.8}_{-3.5}$ \ \ & \ \ $^{+6.1}_{-5.6}$ \ \ & \ \ 3.0 \ \ & \ \ 3.0 \ \ & \ \ 0.0 \ \ & \ \ 0.75 \ \ & \ \ 1.8 \ \ \\[1.5mm]
\ \ $7$ \ \ & \ \ $ 1.38- 1.52$ \ \ & \ \ $ 7.28\cdot 10^{-1}$ \ \ &\ \ 3.2 \ \ & \ \ 1.1 \ \  & \ \ $^{+12.6}_{-11.6}$ \ \ & \ \ $^{+1.0}_{-1.7}$ \ \ & \ \ $^{+3.7}_{-3.5}$ \ \ & \ \ $^{+5.9}_{-5.5}$ \ \ & \ \ 2.9 \ \ & \ \ 2.7 \ \ & \ \ 0.0 \ \ & \ \ 0.75 \ \ & \ \ 1.8 \ \ \\[1.5mm]
\ \ $8$ \ \ & \ \ $ 1.52- 1.66$ \ \ & \ \ $ 3.81\cdot 10^{-1}$ \ \ &\ \ 3.2 \ \ & \ \ 1.2 \ \  & \ \ $^{+12.7}_{-11.9}$ \ \ & \ \ $^{+1.2}_{-1.5}$ \ \ & \ \ $^{+3.6}_{-3.4}$ \ \ & \ \ $^{+5.6}_{-5.4}$ \ \ & \ \ 2.8 \ \ & \ \ 2.3 \ \ & \ \ 0.0 \ \ & \ \ 0.75 \ \ & \ \ 1.8 \ \ \\[1.5mm]
\ \ $9$ \ \ & \ \ $ 1.66- 1.80$ \ \ & \ \ $ 2.19\cdot 10^{-1}$ \ \ &\ \ 4.2 \ \ & \ \ 1.0 \ \  & \ \ $^{+12.8}_{-12.2}$ \ \ & \ \ $^{+1.4}_{-1.4}$ \ \ & \ \ $^{+3.4}_{-3.4}$ \ \ & \ \ $^{+5.3}_{-5.3}$ \ \ & \ \ 2.7 \ \ & \ \ 2.0 \ \ & \ \ 0.0 \ \ & \ \ 0.75 \ \ & \ \ 1.8 \ \ \\[1.5mm]
\ \ $10$ \ \ & \ \ $ 1.80- 1.94$ \ \ & \ \ $ 1.10\cdot 10^{-1}$ \ \ &\ \ 4.6 \ \ & \ \ 1.3 \ \  & \ \ $^{+12.8}_{-12.4}$ \ \ & \ \ $^{+1.6}_{-1.4}$ \ \ & \ \ $^{+3.2}_{-3.4}$ \ \ & \ \ $^{+5.0}_{-5.3}$ \ \ & \ \ 2.7 \ \ & \ \ 1.6 \ \ & \ \ 0.0 \ \ & \ \ 0.75 \ \ & \ \ 1.8 \ \ \\[1.5mm]
\ \ $11$ \ \ & \ \ $ 1.94- 2.10$ \ \ & \ \ $ 6.00\cdot 10^{-2}$ \ \ &\ \ 2.8 \ \ & \ \ 1.4 \ \  & \ \ $^{+13.0}_{-12.5}$ \ \ & \ \ $^{+1.7}_{-1.5}$ \ \ & \ \ $^{+3.0}_{-3.4}$ \ \ & \ \ $^{+4.8}_{-5.2}$ \ \ & \ \ 2.7 \ \ & \ \ 1.4 \ \ & \ \ 0.0 \ \ & \ \ 0.75 \ \ & \ \ 1.8 \ \ \\[1.5mm]
\ \ $12$ \ \ & \ \ $ 2.10- 2.26$ \ \ & \ \ $ 3.15\cdot 10^{-2}$ \ \ &\ \ 3.6 \ \ & \ \ 1.7 \ \  & \ \ $^{+13.2}_{-12.6}$ \ \ & \ \ $^{+1.8}_{-1.8}$ \ \ & \ \ $^{+2.9}_{-3.3}$ \ \ & \ \ $^{+4.7}_{-5.1}$ \ \ & \ \ 2.8 \ \ & \ \ 1.1 \ \ & \ \ 0.0 \ \ & \ \ 0.75 \ \ & \ \ 1.8 \ \ \\[1.5mm]
\ \ $13$ \ \ & \ \ $ 2.26- 2.42$ \ \ & \ \ $ 1.74\cdot 10^{-2}$ \ \ &\ \ 4.9 \ \ & \ \ 2.0 \ \  & \ \ $^{+13.5}_{-12.8}$ \ \ & \ \ $^{+1.9}_{-2.0}$ \ \ & \ \ $^{+2.8}_{-3.1}$ \ \ & \ \ $^{+4.7}_{-5.0}$ \ \ & \ \ 3.0 \ \ & \ \ 0.9 \ \ & \ \ 0.0 \ \ & \ \ 0.75 \ \ & \ \ 1.8 \ \ \\[1.5mm]
\ \ $14$ \ \ & \ \ $ 2.42- 2.58$ \ \ & \ \ $ 8.55\cdot 10^{-3}$ \ \ &\ \ 4.5 \ \ & \ \ 2.2 \ \  & \ \ $^{+13.9}_{-13.0}$ \ \ & \ \ $^{+2.1}_{-2.1}$ \ \ & \ \ $^{+2.7}_{-3.0}$ \ \ & \ \ $^{+4.7}_{-4.9}$ \ \ & \ \ 3.1 \ \ & \ \ 0.8 \ \ & \ \ 0.0 \ \ & \ \ 0.75 \ \ & \ \ 1.8 \ \ \\[1.5mm]
\ \ $15$ \ \ & \ \ $ 2.58- 2.76$ \ \ & \ \ $ 4.40\cdot 10^{-3}$ \ \ &\ \ 3.5 \ \ & \ \ 1.9 \ \  & \ \ $^{+14.3}_{-13.3}$ \ \ & \ \ $^{+2.3}_{-2.2}$ \ \ & \ \ $^{+2.6}_{-2.8}$ \ \ & \ \ $^{+4.7}_{-4.8}$ \ \ & \ \ 3.3 \ \ & \ \ 0.8 \ \ & \ \ 0.0 \ \ & \ \ 0.75 \ \ & \ \ 1.8 \ \ \\[1.5mm]
\ \ $16$ \ \ & \ \ $ 2.76- 2.94$ \ \ & \ \ $ 2.24\cdot 10^{-3}$ \ \ &\ \ 4.8 \ \ & \ \ 1.9 \ \  & \ \ $^{+14.7}_{-13.6}$ \ \ & \ \ $^{+2.5}_{-2.1}$ \ \ & \ \ $^{+2.5}_{-2.7}$ \ \ & \ \ $^{+4.7}_{-4.8}$ \ \ & \ \ 3.4 \ \ & \ \ 0.7 \ \ & \ \ 0.0 \ \ & \ \ 0.75 \ \ & \ \ 1.8 \ \ \\[1.5mm]
\ \ $17$ \ \ & \ \ $ 2.94- 3.12$ \ \ & \ \ $ 1.09\cdot 10^{-3}$ \ \ &\ \ 6.8 \ \ & \ \ 1.9 \ \  & \ \ $^{+15.0}_{-13.9}$ \ \ & \ \ $^{+2.6}_{-2.0}$ \ \ & \ \ $^{+2.4}_{-2.5}$ \ \ & \ \ $^{+4.8}_{-4.7}$ \ \ & \ \ 3.6 \ \ & \ \ 0.7 \ \ & \ \ 0.0 \ \ & \ \ 0.75 \ \ & \ \ 1.8 \ \ \\[1.5mm]
\ \ $18$ \ \ & \ \ $ 3.12- 3.44$ \ \ & \ \ $ 4.01\cdot 10^{-4}$ \ \ &\ \ 8.8 \ \ & \ \ 1.9 \ \  & \ \ $^{+15.6}_{-14.5}$ \ \ & \ \ $^{+2.8}_{-1.9}$ \ \ & \ \ $^{+2.2}_{-2.3}$ \ \ & \ \ $^{+4.7}_{-4.6}$ \ \ & \ \ 3.9 \ \ & \ \ 0.7 \ \ & \ \ 0.0 \ \ & \ \ 0.75 \ \ & \ \ 1.8 \ \ \\[1.5mm]
\ \ $19$ \ \ & \ \ $ 3.44- 3.90$ \ \ & \ \ $ 7.76\cdot 10^{-5}$ \ \ &\ \ 12 \ \ & \ \ 2.7 \ \  & \ \ $^{+16.7}_{-15.7}$ \ \ & \ \ $^{+3.0}_{-1.7}$ \ \ & \ \ $^{+2.0}_{-2.1}$ \ \ & \ \ $^{+4.7}_{-4.3}$ \ \ & \ \ 4.6 \ \ & \ \ 0.6 \ \ & \ \ 0.0 \ \ & \ \ 0.75 \ \ & \ \ 1.8 \ \ \\[1.5mm]
\ \ $20$ \ \ & \ \ $ 3.90- 4.66$ \ \ & \ \ $ 1.17\cdot 10^{-5}$ \ \ &\ \ 19 \ \ & \ \ 4.3 \ \  & \ \ $^{+24.1}_{-21.6}$ \ \ & \ \ $^{+3.3}_{-1.4}$ \ \ & \ \ $^{+1.3}_{-1.1}$ \ \ & \ \ $^{+5.6}_{-3.1}$ \ \ & \ \ 9.7 \ \ & \ \ 0.6 \ \ & \ \ 0.0 \ \ & \ \ 0.75 \ \ & \ \ 1.8 \ \ \\[1.5mm]
\hline
\hline
\end{tabular}
\caption{
   Measured double-differential \trijet{} \xs{}, $\sigma$, for $\rsix{}$~jets and $6\leq\ystar{}<8$, along with uncertainties in the measurement. 
   All uncertainties are given in \%, where 
   $\delta_\textrm{stat}^\textrm{data}$ ($\delta_\textrm{stat}^\textrm{MC}$) are the statistical uncertainties in the data (MC simulation). 
   The $\gamma$ components are the uncertainty in the jet energy calibration from the \insitu{}, the pileup, the close-by jet, and flavour components. 
   The $u$ components show the uncertainty for the jet energy and angular resolution, the unfolding, the quality selection, and the luminosity. 
   While all columns are uncorrelated with each other, the \insitu{}, pileup, and flavour uncertainties shown here are the sum in quadrature of multiple uncorrelated components. 
}
\label{tab:sysunc_r01_ystar3}
\end{table*}

\begin{table*}[!ht]\tiny\centering\begin{tabular}{@{}c@{}@{}c@{}@{}c@{}@{}c@{}@{}r@{}@{}r@{}@{}r@{}@{}r@{}@{}r@{}@{}r@{}@{}r@{}@{}r@{}@{}r@{}@{}r@{}}\hline\hline \\[-2mm]\ \ $m_{jjj}$ \ \ & \ \ $m_{jjj}$-range \ \ & \ \ $\sigma$ \ \ & \ \ $\delta_{\textrm{stat}}^{\textrm{data}}$ \ \ & \ \ $\delta_{\textrm{stat}}^{\textrm{MC}}$ \ \ & \ \ $\gamma_{\insituT}$ \ \ & \ \ $\gamma_{\textrm{pileup}}$ \ \ & \ \ $\gamma_{\textrm{close-by}}$ \ \ & \ \ $\gamma_{\textrm{flavour}}$ \ \ & \ \ $u_{\textrm{JER}}$ \ \ & \ \ $u_{\textrm{JAR}}$ \ \ & \ \ $u_{\textrm{unfold}}$ \ \ & \ \ $u_{\textrm{qual.}}$ \ \ & \ \ $u_{\textrm{lumi}}$ \ \ \\ \relax
\ \ bin \# \ \ & \ \ [TeV] \ \ & \ \ [pb/GeV] \ \ & \ \ [\%] \ \ & \ \ [\%] \ \ & \ \ [\%] \ \ & \ \ [\%] \ \ & \ \ [\%] \ \ & \ \ [\%] \ \ & \ \ [\%] \ \ & \ \ [\%] \ \ & \ \ [\%] \ \ & \ \ [\%] \ \ & \ \ [\%] \\
\hline \\[-1.5mm]
\ \ $1$ \ \ & \ \ $ 1.18- 1.30$ \ \ & \ \ $ 8.88\cdot 10^{-1}$ \ \ &\ \ 3.3 \ \ & \ \ 2.8 \ \  & \ \ $^{+14.4}_{-13.3}$ \ \ & \ \ $^{+0.2}_{-2.7}$ \ \ & \ \ $^{+2.2}_{-2.1}$ \ \ & \ \ $^{+5.8}_{-5.2}$ \ \ & \ \ 1.0 \ \ & \ \ 0.8 \ \ & \ \ 0.0 \ \ & \ \ 0.75 \ \ & \ \ 1.8 \ \ \\[1.5mm]
\ \ $2$ \ \ & \ \ $ 1.30- 1.44$ \ \ & \ \ $ 8.13\cdot 10^{-1}$ \ \ &\ \ 2.6 \ \ & \ \ 2.3 \ \  & \ \ $^{+14.6}_{-14.4}$ \ \ & \ \ $^{+0.1}_{-2.3}$ \ \ & \ \ $^{+2.3}_{-2.3}$ \ \ & \ \ $^{+5.5}_{-5.5}$ \ \ & \ \ 1.0 \ \ & \ \ 0.8 \ \ & \ \ 0.0 \ \ & \ \ 0.75 \ \ & \ \ 1.8 \ \ \\[1.5mm]
\ \ $3$ \ \ & \ \ $ 1.44- 1.58$ \ \ & \ \ $ 5.67\cdot 10^{-1}$ \ \ &\ \ 2.9 \ \ & \ \ 2.4 \ \  & \ \ $^{+15.4}_{-15.6}$ \ \ & \ \ $^{+0.1}_{-2.0}$ \ \ & \ \ $^{+2.4}_{-2.5}$ \ \ & \ \ $^{+5.4}_{-5.7}$ \ \ & \ \ 1.0 \ \ & \ \ 0.8 \ \ & \ \ 0.0 \ \ & \ \ 0.75 \ \ & \ \ 1.8 \ \ \\[1.5mm]
\ \ $4$ \ \ & \ \ $ 1.58- 1.74$ \ \ & \ \ $ 3.67\cdot 10^{-1}$ \ \ &\ \ 3.4 \ \ & \ \ 2.6 \ \  & \ \ $^{+16.7}_{-16.8}$ \ \ & \ \ $^{+0.2}_{-1.8}$ \ \ & \ \ $^{+2.5}_{-2.6}$ \ \ & \ \ $^{+5.3}_{-5.6}$ \ \ & \ \ 1.0 \ \ & \ \ 0.7 \ \ & \ \ 0.0 \ \ & \ \ 0.75 \ \ & \ \ 1.8 \ \ \\[1.5mm]
\ \ $5$ \ \ & \ \ $ 1.74- 1.92$ \ \ & \ \ $ 2.04\cdot 10^{-1}$ \ \ &\ \ 4.2 \ \ & \ \ 2.8 \ \  & \ \ $^{+18.5}_{-17.9}$ \ \ & \ \ $^{+0.4}_{-1.7}$ \ \ & \ \ $^{+2.6}_{-2.6}$ \ \ & \ \ $^{+5.3}_{-5.4}$ \ \ & \ \ 1.0 \ \ & \ \ 0.6 \ \ & \ \ 0.0 \ \ & \ \ 0.75 \ \ & \ \ 1.8 \ \ \\[1.5mm]
\ \ $6$ \ \ & \ \ $ 1.92- 2.12$ \ \ & \ \ $ 1.04\cdot 10^{-1}$ \ \ &\ \ 5.5 \ \ & \ \ 2.5 \ \  & \ \ $^{+20.6}_{-19.1}$ \ \ & \ \ $^{+0.5}_{-1.7}$ \ \ & \ \ $^{+2.7}_{-2.7}$ \ \ & \ \ $^{+5.3}_{-5.1}$ \ \ & \ \ 1.0 \ \ & \ \ 0.4 \ \ & \ \ 0.0 \ \ & \ \ 0.75 \ \ & \ \ 1.8 \ \ \\[1.5mm]
\ \ $7$ \ \ & \ \ $ 2.12- 2.32$ \ \ & \ \ $ 4.48\cdot 10^{-2}$ \ \ &\ \ 8.0 \ \ & \ \ 3.8 \ \  & \ \ $^{+22.6}_{-20.2}$ \ \ & \ \ $^{+0.4}_{-1.7}$ \ \ & \ \ $^{+2.8}_{-2.8}$ \ \ & \ \ $^{+5.4}_{-4.9}$ \ \ & \ \ 1.1 \ \ & \ \ 0.3 \ \ & \ \ 0.0 \ \ & \ \ 0.75 \ \ & \ \ 1.8 \ \ \\[1.5mm]
\ \ $8$ \ \ & \ \ $ 2.32- 2.72$ \ \ & \ \ $ 1.67\cdot 10^{-2}$ \ \ &\ \ 8.7 \ \ & \ \ 1.9 \ \  & \ \ $^{+25.6}_{-21.4}$ \ \ & \ \ $^{+0.1}_{-1.7}$ \ \ & \ \ $^{+3.0}_{-3.0}$ \ \ & \ \ $^{+5.4}_{-4.4}$ \ \ & \ \ 1.6 \ \ & \ \ 0.2 \ \ & \ \ 0.0 \ \ & \ \ 0.75 \ \ & \ \ 1.8 \ \ \\[1.5mm]
\ \ $9$ \ \ & \ \ $ 2.72- 3.14$ \ \ & \ \ $ 3.52\cdot 10^{-3}$ \ \ &\ \ 7.9 \ \ & \ \ 3.3 \ \  & \ \ $^{+30.2}_{-23.6}$ \ \ & \ \ $^{+0.2}_{-1.7}$ \ \ & \ \ $^{+3.4}_{-3.2}$ \ \ & \ \ $^{+5.1}_{-3.6}$ \ \ & \ \ 2.7 \ \ & \ \ 0.1 \ \ & \ \ 0.0 \ \ & \ \ 0.75 \ \ & \ \ 1.8 \ \ \\[1.5mm]
\ \ $10$ \ \ & \ \ $ 3.14- 3.58$ \ \ & \ \ $ 6.24\cdot 10^{-4}$ \ \ &\ \ 18 \ \ & \ \ 6.3 \ \  & \ \ $^{+34.7}_{-27.2}$ \ \ & \ \ $^{+0.2}_{-1.8}$ \ \ & \ \ $^{+3.6}_{-3.5}$ \ \ & \ \ $^{+4.8}_{-3.2}$ \ \ & \ \ 3.7 \ \ & \ \ 0.1 \ \ & \ \ 0.0 \ \ & \ \ 0.75 \ \ & \ \ 1.8 \ \ \\[1.5mm]
\ \ $11$ \ \ & \ \ $ 3.58- 4.18$ \ \ & \ \ $ 1.03\cdot 10^{-4}$ \ \ &\ \ 32 \ \ & \ \ 14 \ \  & \ \ $^{+39.9}_{-32.1}$ \ \ & \ \ $^{+0.2}_{-1.8}$ \ \ & \ \ $^{+3.9}_{-3.8}$ \ \ & \ \ $^{+4.6}_{-3.6}$ \ \ & \ \ 4.4 \ \ & \ \ 0.1 \ \ & \ \ 0.1 \ \ & \ \ 0.75 \ \ & \ \ 1.8 \ \ \\[1.5mm]
\ \ $12$ \ \ & \ \ $ 4.18- 5.50$ \ \ & \ \ $ 3.03\cdot 10^{-6}$ \ \ &\ \ 40 \ \ & \ \ 14 \ \  & \ \ $^{+58.5}_{-42.4}$ \ \ & \ \ $^{+0.2}_{-1.9}$ \ \ & \ \ $^{+5.4}_{-4.1}$ \ \ & \ \ $^{+4.3}_{-8.2}$ \ \ & \ \ 5.0 \ \ & \ \ 0.1 \ \ & \ \ 0.7 \ \ & \ \ 0.75 \ \ & \ \ 1.8 \ \ \\[1.5mm]
\hline
\hline
\end{tabular}
\caption{
   Measured double-differential \trijet{} \xs{}, $\sigma$, for $\rfour{}$~jets and $8\leq\ystar{}<10$, along with uncertainties in the measurement. 
   All uncertainties are given in \%, where 
   $\delta_\textrm{stat}^\textrm{data}$ ($\delta_\textrm{stat}^\textrm{MC}$) are the statistical uncertainties in the data (MC simulation). 
   The $\gamma$ components are the uncertainty in the jet energy calibration from the \insitu{}, the pileup, the close-by jet, and flavour components. 
   The $u$ components show the uncertainty for the jet energy and angular resolution, the unfolding, the quality selection, and the luminosity. 
   While all columns are uncorrelated with each other, the \insitu{}, pileup, and flavour uncertainties shown here are the sum in quadrature of multiple uncorrelated components. 
}
\label{tab:sysunc_r00_ystar4}
\end{table*}

\begin{table*}[!ht]\tiny\centering\begin{tabular}{@{}c@{}@{}c@{}@{}c@{}@{}c@{}@{}r@{}@{}r@{}@{}r@{}@{}r@{}@{}r@{}@{}r@{}@{}r@{}@{}r@{}@{}r@{}@{}r@{}}\hline\hline \\[-2mm]\ \ $m_{jjj}$ \ \ & \ \ $m_{jjj}$-range \ \ & \ \ $\sigma$ \ \ & \ \ $\delta_{\textrm{stat}}^{\textrm{data}}$ \ \ & \ \ $\delta_{\textrm{stat}}^{\textrm{MC}}$ \ \ & \ \ $\gamma_{\insituT}$ \ \ & \ \ $\gamma_{\textrm{pileup}}$ \ \ & \ \ $\gamma_{\textrm{close-by}}$ \ \ & \ \ $\gamma_{\textrm{flavour}}$ \ \ & \ \ $u_{\textrm{JER}}$ \ \ & \ \ $u_{\textrm{JAR}}$ \ \ & \ \ $u_{\textrm{unfold}}$ \ \ & \ \ $u_{\textrm{qual.}}$ \ \ & \ \ $u_{\textrm{lumi}}$ \ \ \\ \relax
\ \ bin \# \ \ & \ \ [TeV] \ \ & \ \ [pb/GeV] \ \ & \ \ [\%] \ \ & \ \ [\%] \ \ & \ \ [\%] \ \ & \ \ [\%] \ \ & \ \ [\%] \ \ & \ \ [\%] \ \ & \ \ [\%] \ \ & \ \ [\%] \ \ & \ \ [\%] \ \ & \ \ [\%] \ \ & \ \ [\%] \\
\hline \\[-1.5mm]
\ \ $1$ \ \ & \ \ $ 1.18- 1.30$ \ \ & \ \ 1.46 \ \ & \ \ 3.8 \ \ & \ \ 2.3 \ \  & \ \ $^{+13.6}_{-13.5}$ \ \ & \ \ $^{+0.5}_{-5.0}$ \ \ & \ \ $^{+4.1}_{-3.7}$ \ \ & \ \ $^{+6.0}_{-5.9}$ \ \ & \ \ 3.2 \ \ & \ \ 6.2 \ \ & \ \ 0.0 \ \ & \ \ 0.75 \ \ & \ \ 1.8 \ \ \\[1.5mm]
\ \ $2$ \ \ & \ \ $ 1.30- 1.44$ \ \ & \ \ 1.21 \ \ & \ \ 3.4 \ \ & \ \ 2.1 \ \  & \ \ $^{+14.3}_{-13.7}$ \ \ & \ \ $^{+0.5}_{-3.8}$ \ \ & \ \ $^{+4.2}_{-3.7}$ \ \ & \ \ $^{+6.0}_{-5.8}$ \ \ & \ \ 3.6 \ \ & \ \ 6.5 \ \ & \ \ 0.0 \ \ & \ \ 0.75 \ \ & \ \ 1.8 \ \ \\[1.5mm]
\ \ $3$ \ \ & \ \ $ 1.44- 1.58$ \ \ & \ \ $ 8.88\cdot 10^{-1}$ \ \ &\ \ 3.9 \ \ & \ \ 2.3 \ \  & \ \ $^{+15.0}_{-14.0}$ \ \ & \ \ $^{+0.6}_{-2.9}$ \ \ & \ \ $^{+4.1}_{-3.8}$ \ \ & \ \ $^{+5.8}_{-5.6}$ \ \ & \ \ 4.0 \ \ & \ \ 6.8 \ \ & \ \ 0.0 \ \ & \ \ 0.75 \ \ & \ \ 1.8 \ \ \\[1.5mm]
\ \ $4$ \ \ & \ \ $ 1.58- 1.74$ \ \ & \ \ $ 5.94\cdot 10^{-1}$ \ \ &\ \ 4.5 \ \ & \ \ 2.4 \ \  & \ \ $^{+15.9}_{-14.6}$ \ \ & \ \ $^{+0.8}_{-2.2}$ \ \ & \ \ $^{+4.0}_{-3.8}$ \ \ & \ \ $^{+5.6}_{-5.4}$ \ \ & \ \ 4.2 \ \ & \ \ 6.8 \ \ & \ \ 0.0 \ \ & \ \ 0.75 \ \ & \ \ 1.8 \ \ \\[1.5mm]
\ \ $5$ \ \ & \ \ $ 1.74- 1.92$ \ \ & \ \ $ 3.44\cdot 10^{-1}$ \ \ &\ \ 5.5 \ \ & \ \ 2.4 \ \  & \ \ $^{+17.3}_{-15.4}$ \ \ & \ \ $^{+1.0}_{-1.8}$ \ \ & \ \ $^{+4.0}_{-3.9}$ \ \ & \ \ $^{+5.6}_{-5.4}$ \ \ & \ \ 4.3 \ \ & \ \ 6.7 \ \ & \ \ 0.0 \ \ & \ \ 0.75 \ \ & \ \ 1.8 \ \ \\[1.5mm]
\ \ $6$ \ \ & \ \ $ 1.92- 2.12$ \ \ & \ \ $ 1.63\cdot 10^{-1}$ \ \ &\ \ 7.2 \ \ & \ \ 3.0 \ \  & \ \ $^{+19.0}_{-16.3}$ \ \ & \ \ $^{+1.0}_{-1.6}$ \ \ & \ \ $^{+4.1}_{-3.9}$ \ \ & \ \ $^{+5.7}_{-5.4}$ \ \ & \ \ 4.3 \ \ & \ \ 6.4 \ \ & \ \ 0.0 \ \ & \ \ 0.75 \ \ & \ \ 1.8 \ \ \\[1.5mm]
\ \ $7$ \ \ & \ \ $ 2.12- 2.32$ \ \ & \ \ $ 6.64\cdot 10^{-2}$ \ \ &\ \ 6.5 \ \ & \ \ 2.9 \ \  & \ \ $^{+20.7}_{-17.1}$ \ \ & \ \ $^{+0.7}_{-1.5}$ \ \ & \ \ $^{+4.1}_{-3.9}$ \ \ & \ \ $^{+5.9}_{-5.4}$ \ \ & \ \ 4.5 \ \ & \ \ 6.2 \ \ & \ \ 0.0 \ \ & \ \ 0.75 \ \ & \ \ 1.8 \ \ \\[1.5mm]
\ \ $8$ \ \ & \ \ $ 2.32- 2.72$ \ \ & \ \ $ 2.59\cdot 10^{-2}$ \ \ &\ \ 8.1 \ \ & \ \ 1.7 \ \  & \ \ $^{+22.6}_{-18.1}$ \ \ & \ \ $^{+0.4}_{-1.4}$ \ \ & \ \ $^{+3.8}_{-3.9}$ \ \ & \ \ $^{+5.9}_{-5.4}$ \ \ & \ \ 4.9 \ \ & \ \ 5.6 \ \ & \ \ 0.0 \ \ & \ \ 0.75 \ \ & \ \ 1.8 \ \ \\[1.5mm]
\ \ $9$ \ \ & \ \ $ 2.72- 3.14$ \ \ & \ \ $ 4.95\cdot 10^{-3}$ \ \ &\ \ 16 \ \ & \ \ 3.3 \ \  & \ \ $^{+24.8}_{-19.7}$ \ \ & \ \ $^{+0.1}_{-1.3}$ \ \ & \ \ $^{+3.7}_{-4.1}$ \ \ & \ \ $^{+5.8}_{-5.9}$ \ \ & \ \ 5.7 \ \ & \ \ 4.3 \ \ & \ \ 0.0 \ \ & \ \ 0.75 \ \ & \ \ 1.8 \ \ \\[1.5mm]
\ \ $10$ \ \ & \ \ $ 3.14- 3.58$ \ \ & \ \ $ 1.12\cdot 10^{-3}$ \ \ &\ \ 14 \ \ & \ \ 5.7 \ \  & \ \ $^{+27.7}_{-22.0}$ \ \ & \ \ $^{+0.0}_{-0.8}$ \ \ & \ \ $^{+3.7}_{-4.3}$ \ \ & \ \ $^{+5.7}_{-6.1}$ \ \ & \ \ 6.6 \ \ & \ \ 3.2 \ \ & \ \ 0.0 \ \ & \ \ 0.75 \ \ & \ \ 1.8 \ \ \\[1.5mm]
\ \ $11$ \ \ & \ \ $ 3.58- 4.18$ \ \ & \ \ $ 1.44\cdot 10^{-4}$ \ \ &\ \ 33 \ \ & \ \ 9.7 \ \  & \ \ $^{+30.8}_{-25.0}$ \ \ & \ \ $^{+0.0}_{-0.2}$ \ \ & \ \ $^{+3.4}_{-4.0}$ \ \ & \ \ $^{+5.7}_{-5.5}$ \ \ & \ \ 7.3 \ \ & \ \ 2.5 \ \ & \ \ 0.1 \ \ & \ \ 0.75 \ \ & \ \ 1.8 \ \ \\[1.5mm]
\ \ $12$ \ \ & \ \ $ 4.18- 5.50$ \ \ & \ \ $ 5.14\cdot 10^{-6}$ \ \ &\ \ 43 \ \ & \ \ 9.8 \ \  & \ \ $^{+36.0}_{-34.1}$ \ \ & \ \ $^{+1.2}_{-0.2}$ \ \ & \ \ $^{+1.4}_{-2.8}$ \ \ & \ \ $^{+8.6}_{-2.8}$ \ \ & \ \ 6.5 \ \ & \ \ 1.9 \ \ & \ \ 0.8 \ \ & \ \ 0.75 \ \ & \ \ 1.8 \ \ \\[1.5mm]
\hline
\hline
\end{tabular}
\caption{
   Measured double-differential \trijet{} \xs{}, $\sigma$, for $\rsix{}$~jets and $8\leq\ystar{}<10$, along with uncertainties in the measurement. 
   All uncertainties are given in \%, where 
   $\delta_\textrm{stat}^\textrm{data}$ ($\delta_\textrm{stat}^\textrm{MC}$) are the statistical uncertainties in the data (MC simulation). 
   The $\gamma$ components are the uncertainty in the jet energy calibration from the \insitu{}, the pileup, the close-by jet, and flavour components. 
   The $u$ components show the uncertainty for the jet energy and angular resolution, the unfolding, the quality selection, and the luminosity. 
   While all columns are uncorrelated with each other, the \insitu{}, pileup, and flavour uncertainties shown here are the sum in quadrature of multiple uncorrelated components. 
}
\label{tab:sysunc_r01_ystar4}
\end{table*}

\clearpage

\begin{flushleft}
{\Large The ATLAS Collaboration}

\bigskip

G.~Aad$^{\rm 84}$,
B.~Abbott$^{\rm 112}$,
J.~Abdallah$^{\rm 152}$,
S.~Abdel~Khalek$^{\rm 116}$,
O.~Abdinov$^{\rm 11}$,
R.~Aben$^{\rm 106}$,
B.~Abi$^{\rm 113}$,
M.~Abolins$^{\rm 89}$,
O.S.~AbouZeid$^{\rm 159}$,
H.~Abramowicz$^{\rm 154}$
% ...
\bigskip
\\
% ...
$^{11}$ Institute of Physics, Azerbaijan Academy of Sciences, Baku, Azerbaijan\\
% ...
$^{84}$ CPPM, Aix-Marseille Universit{\'e} and CNRS/IN2P3, Marseille, France\\
% ...
$^{89}$ Department of Physics and Astronomy, Michigan State University, East Lansing MI, United States of America\\
% ...
$^{106}$ Nikhef National Institute for Subatomic Physics and University of Amsterdam, Amsterdam, Netherlands\\
% ...
$^{112}$ Homer L. Dodge Department of Physics and Astronomy, University of Oklahoma, Norman OK, United States of America\\
$^{113}$ Department of Physics, Oklahoma State University, Stillwater OK, United States of America\\
% ...
$^{116}$ LAL, Universit{\'e} Paris-Sud and CNRS/IN2P3, Orsay, France\\
% ...
$^{152}$ Institute of Physics, Academia Sinica, Taipei, Taiwan\\
% ...
$^{154}$ Raymond and Beverly Sackler School of Physics and Astronomy, Tel Aviv University, Tel Aviv, Israel\\
% ...
$^{159}$ Department of Physics, University of Toronto, Toronto ON, Canada\\
% ...
$^{*}$ Deceased
\end{flushleft}


\end{document}